\documentclass{article}
\usepackage[a4paper, total={17cm, 26.7cm}]{geometry}
\usepackage{graphicx} 
\usepackage{hyperref}
\usepackage{authblk}
\usepackage{xspace}
\usepackage{tabularx}
\usepackage{multirow}
\usepackage[separate-uncertainty,bracket-numbers = false, table-align-uncertainty = true, table-align-exponent=false, exponent-product=\times]{siunitx}
\usepackage{subfiles}
\usepackage{subcaption}

\usepackage{comment}
\usepackage{placeins}

\usepackage{todonotes}

\usepackage{amsmath}
\usepackage{amssymb}
\usepackage{chemformula}
\usepackage{chemfig}
\usepackage{esvect}
\usepackage{gensymb}

\title{nuSCOPE \\ A short-baseline neutrino beam at CERN \\ for high-precision cross-section measurements}

\date{}

\begin{document}
\author[a,b]{F.~Acerbi}
\author[ac]{C. Andreopoulos}
\author[n]{I. Angelis}
\author[r]{A.~Baratto Roldan} 
\author[e,p]{L.~Bomben}
\author[e]{M.~Bonesini}
\author[e,f]{F.~Bramati}
\author[e,f]{A.~Branca}
\author[e,f]{C.~Brizzolari}
\author[f]{G.~Brunetti}
\author[u]{M.~Buizza~Avanzini}
\author[e,p]{S.~Capelli}
\author[e,f]{M.~Capitani}
\author[d,g]{S.~Carturan}
\author[h]{M.G.~Catanesi}
\author[i]{S.~Cecchini}
\author[r]{N.~Charitonidis}
\author[i]{F.~Cindolo}
\author[aa]{J.~Cogan}
\author[d]{G.~Cogo}
\author[c,d]{G.~Collazuol}
\author[c]{D.~D'Ago}
\author[c]{F.~Dal Corso}
\author[j,k]{G.~De~Rosa}
\author[r]{S.~Dolan}
\author[e,f]{A.~Falcone}
\author[c,d]{M.~Feltre}
\author[a]{A.~Gola}
\author[e,f]{D.~Guffanti}
\author[m]{L.~Hali\'{c}}
\author[c,d]{F.~Iacob}
\author[r]{M.A. Jebramcik}
\author[l]{C.~Jollet}
\author[w]{A.~Kallitsopoulou}
\author[m]{B.~Kli\u{c}ek}
\author[v]{A.~Lai}
\author[n]{Ch. Lampoudis}
\author[r]{F. Lanni}
\author[c,d]{M.~Laveder}
\author[w]{P.~Legou}
\author[c]{S.~Levorato}
\author[c,d]{A.~Longhin}
\author[o]{L.~Ludovici}
\author[h,q]{L.~Magaletti}
\author[i]{G.~Mandrioli}
\author[i]{A.~Margotti}
\author[y,z]{V.~Mascagna}
\author[c,d]{M.~Mattiazzi}
\author[i]{N.~Mauri}
\author[l]{J.~McElwee}
\author[e,f]{L.~Meazza}
\author[l]{A.~Meregaglia}
\author[c]{M.~Mezzetto}
\author[r]{L.~Munteanu}
\author[t]{A.~Paoloni}
\author[r]{M.~Pari}
\author[w]{T.~Papaevangelou}
\author[e,f,r]{E.G.~Parozzi}
\author[i,s]{L.~Pasqualini}
\author[a]{G.~Paternoster}
\author[i]{L.~Patrizii}
\author[aa]{M.~Perrin-Terrin}
\author[aa,ab]{L.~Petit}
\author[i]{M.~Pozzato}
\author[e,p]{M.~Prest}
\author[c]{F.~Pupilli}
\author[h]{E.~Radicioni}
\author[r]{F. Resnati}
\author[j,k]{A.C.~Ruggeri}
\author[ab]{A.~Roueff}
\author[e,p]{G.~Saibene}
\author[n]{D. Sampsonidis}
\author[e,f]{A.~Scanu}
\author[c,d]{C.~Scian}
\author[i]{G.~Sirri}
\author[e,f]{R.~Speziali}
\author[m]{M.~Stip\u{c}evi\'{c}}
\author[i]{M.~Tenti}
\author[e,f]{F.~Terranova}
\author[e,f]{M.~Torti}
\author[n]{S. E. Tzamarias}
\author[e]{E.~Vallazza}
\author[aa]{C.~Vall\'{e}e}
\author[t]{L.~Votano}

\affil[a]{Fondazione Bruno Kessler (FBK),  Via Sommarive 18 - 38123 Povo (TN), IT}
\affil[b]{INFN-TIFPA, Università di Trento, Via Sommarive 14 - 38123 Povo (TN), IT}
\affil[c]{INFN Sezione di Padova, via Marzolo 8 - 35131 Padova, IT}
\affil[d]{Università di Padova, via Marzolo 8 - 35131 Padova, IT}
\affil[e]{INFN Sezione di Milano-Bicocca, Piazza della Scienza 3 - 20133 Milano, IT}
\affil[f]{Università di Milano-Bicocca, Piazza della Scienza 3 - 20133 Milano, IT}
\affil[g]{INFN, Laboratori Nazionali di Legnaro, Viale dell'Università 2 - 35020 Legnaro (PD), IT}
\affil[h]{INFN Sezione di Bari, Via Giovanni Amendola 173 - 70126 Bari, IT}
\affil[i]{INFN Sezione di Bologna, viale Berti-Pichat 6/2 - 40127 Bologna, IT}
\affil[j]{INFN, Sezione di Napoli, Strada Comunale Cinthia - 80126 Napoli, IT}
\affil[k]{Università ``Federico II'' di Napoli, Corso Umberto I 40 - 80138 Napoli, IT}
\affil[l]{LP2I Bordeaux, Universitè de Bordeaux, CNRS/IN2P3, 33175 Gradignan, FR}
\affil[m]{Center of Excellence for Advanced Materials and Sensing Devices, Ruder Boskovic Institute, HR-10000 Zagreb, HR}
\affil[n]{Aristotle University of Thessaloniki. Thessaloniki 541 24, GR.}
\affil[o]{INFN Sezione di Roma 1, Piazzale A. Moro 2, 00185 Rome, IT}
\affil[p]{Università degli Studi dell'Insubria, Via Valleggio 11 - 22100 Como, IT}
\affil[q]{Università degli Studi di Bari, Via Giovanni Amendola 173 - 70126 Bari, IT}
\affil[r]{CERN, Esplanade des particules - 1211 Genève 23, CH}
\affil[s]{Università degli Studi di Bologna, viale Berti-Pichat 6/2 - 40127 Bologna, IT}
\affil[t]{INFN, Laboratori Nazionali di Frascati, via Fermi 40 - 00044 Frascati (Rome), Italy}
\affil[u]{Ecole Polytechnique, IN2P3-CNRS, Laboratoire Leprince-Ringuet, Palaiseau, France}
\affil[v]{Istituto Nazionale Fisica Nucleare, Sezione di Cagliari, Cagliari, Italy}
\affil[w]{CEA, Centre de Saclay, Irfu/SPP, F-91191 Gif-sur-Yvette, FR.}
\affil[y]{DII, Universit\`a degli Studi di Brescia, via Branze 38, Brescia, IT.}
\affil[z]{INFN, Sezione di Pavia, via Bassi 6, Pavia, IT.}
\affil[aa]{Aix Marseille Univ, CNRS/IN2P3, CPPM, Marseille, France}
\affil[ab]{Université de Toulon, Aix Marseille Univ, CNRS, IM2NP, Toulon, France}
\affil[ac]{University of Liverpool, Dept. of Physics, Liverpool L69 7ZE, UK}


\def\numu        {\ensuremath{\nu_\mu}\xspace}
\def\nue        {\ensuremath{\nu_e}\xspace}

\def\pidecay        {\ensuremath{\pi_{\mu\nu}}\xspace}
\def\kdecay        {\ensuremath{K_{\mu\nu}}\xspace}

\def\cczeropi        {\ensuremath{\text{CC}0\pi}\xspace}
\def\cconepi        {\ensuremath{\text{CC}1\pi}\xspace}
\def\ccNpi        {\ensuremath{\text{CC}N\pi}\xspace}
\def\ncpizero      {\ensuremath{\text{NC}\pi^0}\xspace}
\def\ncpizeronop      {\ensuremath{\text{NC}\pi^0+0p}\xspace}
\def\ncpizeroonep      {\ensuremath{\text{NC}\pi^0+1p}\xspace}

\def\argenie
{\texttt{AR23\_20i\_00\_000}\xspace}
\def\susagenie
{\texttt{G21\_11a\_00\_000}\xspace}
\def\genietune
{\texttt{G18\_10a\_02\_11b}\xspace}

\def\enureco
{\ensuremath{E_{\nu}^{\text{reco}}}\xspace}
\def\enutrue
{\ensuremath{E_{\nu}^{\text{true}}}\xspace}
\def\enu        {\ensuremath{E_\nu}\xspace}
\def\pmu        {\ensuremath{p_\mu}\xspace}
\def\cosmu        {\ensuremath{\cos\theta_\mu}\xspace}
\def\eavail       {\ensuremath{E_{\textrm{avail}}}\xspace}
\def\qthree      {\ensuremath{q_3}\xspace}
\def\etransfer      {\ensuremath{\omega}\xspace}
\def\wexp     {\ensuremath{W}\xspace}
\def\nuemuxsec
{\ensuremath{\sigma(\nue)/\sigma(\numu)}\xspace}
\def\sigmanue
{\ensuremath{\sigma(\nue)}\xspace}
\def\sigmanumu
{\ensuremath{\sigma(\numu)}\xspace}

\maketitle
\newpage
\begin{abstract}
    A new generation of neutrino cross-section experiments at the GeV scale is crucial in the precision era of oscillation physics and lepton flavor studies. In this document, we present a novel neutrino beam design that leverages the experience and R\&D achievements of the NP06/ENUBET and NuTag Collaborations and explore its potential implementation at CERN. This beam enables flux monitoring at the percent level and provides a neutrino energy measurement independent of final state particle reconstruction at the neutrino detector. As a result, it eliminates the two primary sources of systematic uncertainty in cross-section measurements: flux normalization and energy bias caused by nuclear effects. We provide a detailed description of the beam technology and instrumentation, along with an overview of its physics potential, with particular emphasis on cross-sections relevant to DUNE and Hyper-Kamiokande.   
\end{abstract}

\tableofcontents

\section{A high precision neutrino beam at CERN}

Neutrino physics entered its precision era in 2012 when accelerator and reactor experiments demonstrated that all mixing angles, including $\theta_{13}$, are large \cite{pdg}. This groundbreaking discovery paved the way for a comprehensive exploration of the lepton Yukawa sector of the Standard Model. Oscillation experiments can now probe the neutrino mass hierarchy, the mixing angles, and the Dirac CP-violating phase, providing a nearly complete picture of neutrino properties. Two remaining pieces~—the determination of the absolute neutrino mass and the Dirac/Majorana nature—~require dedicated measurements beyond oscillation experiments.
A key finding of the 2020 European Strategy for Particle Physics \cite{Ellis:2691414} was the recognition that the ambitious goals of neutrino physics are critically hindered by the limited understanding of standard neutrino interactions, particularly neutrino cross-sections at the GeV scale. This concern has grown over the past five years, emphasizing the need for a new generation of experiments. These experiments must achieve percent-level precision in cross-section measurements—an order of magnitude improvement over current facilities—and are now considered a top priority in the field. 

Cross-section uncertainties already represent the largest source of systematic error in long-baseline experiments such as T2K, where statistical errors still contribute significantly. However, as DUNE \cite{DUNE:2020lwj} and Hyper-Kamiokande \cite{Hyper-Kamiokande:2018ofw} enter their data-taking phases, their unprecedented beam power and detector mass will push statistical uncertainties far below the systematic budget. In this scenario, cross-section uncertainties will become the dominant factor limiting the physics reach of these experiments.
Given the extraordinary complexity and cost of long-baseline experiments, strategic investment in reducing the final systematic budget is essential. The construction of high-precision cross-section experiments, operating concurrently with DUNE and Hyper-Kamiokande, offers the most robust and cost-effective pathway to achieve the ultimate sensitivity to the lepton Yukawa sector.

The evaluation of neutrino cross-sections in the energy range relevant for DUNE and Hyper-Kamiokande (1–5 GeV) and beyond presents significant theoretical challenges \cite{Athar:2020kqn, NuSTEC:2017hzk}. This may seem surprising, as the electroweak interactions of neutrinos with nuclei are perturbatively well-understood at the parton level. However, the complexity arises from the subsequent interactions of the parton with spectator quarks and gluons within the nucleon, the hadronization process, and the interactions of final-state hadrons within nuclear matter.

Over the past two decades, several data-driven models have been developed to address these processes. However, achieving a precision below 50\% in cross-section calculations introduces entirely new levels of complexity. At present, state-of-the-art generators cannot achieve better than 30\% agreement with experimental data for most inclusive, single, and double-differential cross-sections. The precision for exclusive channels is often even worse, highlighting the limitations of existing theoretical and computational frameworks. These limitations can be mitigated when the projectile is a charged lepton, such as an electron or a muon. In these cases, experimentalists can select the incoming lepton with high precision, achieving accuracies in the flux, flavor 
($e, \mu$), and energy well below a few percent. This enables a detailed study of lepton scattering processes, with a robust a priori evaluation of the four-momentum transfer and a precise knowledge of the incoming lepton flux \cite{Amaro:2019zos}.
In contrast, until now, none of these control measurements have been feasible for neutrino scattering. At present, the incoming neutrino flux is typically known with a precision of about 10\%, and the energy of the neutrinos can only be inferred indirectly from the four-momentum of the final-state particles. However, the detection efficiency of final-state hadrons—particularly neutral hadrons—is limited, and nuclear reinteractions further distort the kinematics of these particles. As a result, the inferred energy of the incoming neutrinos is strongly biased, significantly contributing to the uncertainty of the cross-section measurement.
This uncertainty is further exacerbated by normalization uncertainties stemming from the limited knowledge of the incoming neutrino flux. 

Consequently, despite considerable progress in neutrino scattering measurements achieved through the near detectors of T2K and NO$\nu$A, as well as dedicated cross-section experiments such as SciBooNE, MINERvA, LArIAT, and MicroBooNE \cite{Katori:2016yel}, the precision of these measurements remains insufficient for the demands of high-precision neutrino oscillation physics.

The 2020 European Strategy for Particle Physics identified promising candidates for a new generation of high-precision neutrino beams capable of addressing the limitations of current cross-section measurements. It encouraged further R\&D efforts to bring these innovative approaches to maturity. Over the last five years, this goal has been achieved by one of the leading candidates: the monitored neutrino beam developed by the ENUBET Collaboration.

ENUBET was initially conceived in 2016 to tackle a specific challenge: the creation of a narrow-band neutrino beam whose electron neutrino flux is known with an unprecedented precision of 1\% \cite{Longhin:2014yta}. This idea stemmed from the realization that -- even today -- the most advanced cross-section measurements are limited by flux uncertainties, which are typically around 10\%.
In its original proposal, ENUBET focused on producing electron neutrinos via the three-body decay of kaons ($K_{e3} \equiv K^+  \rightarrow \pi^0 e^+ \nu_e$) and optimizing the beam to have $K_{e3}$ as the dominant source of $\nu_e$.  The innovative aspect of ENUBET’s approach was its ability to identify positrons at the single-particle level by detecting large-angle positrons that impinge on the walls of the decay tunnel. Since there is a one-to-one correspondence between the positrons in the decay tunnel and the corresponding electron neutrino, this method provided a direct handle on the flux normalization, paving the way for precision measurements at a level unattainable by conventional neutrino beams.
In 2021, the ENUBET Collaboration achieved a groundbreaking advancement that significantly expanded the experiment's physics potential while simplifying its implementation. The collaboration successfully designed a beamline capable of producing a moderate-intensity neutrino flux from 
$\pi$ and $K$ decays without relying on magnetic horns \cite{ENUBET:2023hgu}. Remarkably, this beamline generated sufficient flux to enable 
$\nu_\mu$ and $\nu_e$ cross-section measurements with statistical uncertainties below 1\%.
The elimination of magnetic horns brought substantial benefits. Without the need to address the Joule heating constraints of magnetic horns, primary protons could be delivered to the target in long extractions (lasting several seconds) rather than the brief (10 
$\mu$s to 1 ms) bursts that are typically required. These longer extractions reduced the instantaneous rate of charged leptons in the decay tunnel by several orders of magnitude, allowing the use of conventional, cost-effective, and moderately granular detectors to sustain operations.
The magnetic horn was replaced by a series of quadrupoles and dipoles, which focus and sign-select secondary mesons, a technique reminiscent of the narrow-band beams developed in the 1970s \cite{kopp2006}. However, ENUBET’s approach leverages the higher intensities of modern proton accelerators, the larger numerical apertures of its beamline components, and the fast response of its decay tunnel instrumentation to achieve a flux intensity comparable to that of contemporary neutrino beams used in cross-section measurements.
Furthermore, ENUBET’s horn-less design enables precise monitoring of neutrino fluxes by directly detecting charged leptons produced in the decay tunnel or the hadron dump: muons from pion and kaon decays, and positrons from kaon decays. This capability allows a horn-less monitored neutrino beam employing the ENUBET technique to monitor all neutrino flavors originating from the beamline with an unprecedented precision of 1\%.

The introduction of long extractions has unlocked new possibilities, capitalized upon by the NuTag Collaboration starting in 2022 \cite{Perrin-Terrin:2021jtl,Baratto-Roldan:2024bxk}. The NuTag proposal builds upon recent advancements in fast silicon tracking technologies, which have been successfully demonstrated in kaon physics by the NA62 Gigatracker and further enhanced by the technological innovations driven by LHC upgrade programs \cite{AglieriRinella:2019eri,Cadeddu:2024qrz}.
A monitored neutrino beam utilizing fast ($<$100 ps), radiation-hard silicon trackers placed along the beamline can track both the parent meson and its daughter leptons, extending ENUBET’s capabilities in two ways. First, lepton monitoring with advanced tracking significantly improves the determination of the neutrino flux by increasing the number of monitored neutrinos and enhancing the spectral characterization of the flux. This method further eliminates dependencies on Monte Carlo simulations of the beamline, offering a direct and precise flux measurement.
Second, if the neutrino detector possesses sufficient time resolution ($<$1 ns), each observed neutrino can be uniquely correlated with its parent meson and associated charged lepton. This operation mode, referred to as a ``tagged neutrino beam,'' has been a long-standing goal in accelerator neutrino physics but was historically unattainable due to technological limitations \cite{Hand1969,Pontevcorvo1979}. These limitations are being lifted, and NA62 reported the detection of a first tagged neutrino candidate~\cite{hep-ph_NA62Collaboration_2024}.
A tagged neutrino beam would constitute a groundbreaking advancement in experimental physics, enabling the direct measurement of neutrino energy on an event-by-event basis using the two-body kinematics of the parent meson decay.

In 2023, the convergence of the ENUBET and NuTag approaches revealed the potential to achieve neutrino-nucleus cross-section measurements with a precision comparable to electron-nucleus cross-sections. Recognizing this opportunity, the two collaborations joined forces to explore the feasibility of implementing such a facility at CERN \cite{PBC2024}. The beamline design was optimized within the framework of Physics Beyond Colliders (PBC) at CERN, culminating in the design presented in this document.
This document outlines the results of these efforts and provides the foundation for the proposal of a new high-precision, short-baseline neutrino beam at CERN, tentatively named SBN@CERN. In June 2025, an international collaboration was established, and the facility was officially named {\bf nuSCOPE} -- neutrino SPS COmplex for Precision Experiments.  This facility aims to deliver neutrino cross-section measurements at the percent level in the energy range relevant for DUNE and Hyper-Kamiokande. While alternative implementations of this innovative technology outside CERN are under consideration, CERN's unparalleled infrastructure, proton driver, and existing neutrino detectors position CERN as the most cost-effective venue for realizing this project in compliance with the CERN fixed-target program for the coming decade.

The nuSCOPE beamline is detailed in Sec. \ref{sec:beamline_design}, with a focus on its implementation at CERN, utilizing the SPS as the proton driver. Section \ref{sec:instrumentation} outlines the beamline instrumentation, which monitors and tracks mesons and charged leptons, while Sec. \ref{sec:neutrino_detectors} discusses the neutrino detectors that leverage the advancements of the CERN Neutrino Platform.
Section \ref{sec:flux_characterization} summarizes the technique that enables a sub-percent precision measurement of the neutrino flux using lepton monitoring and tracking. Section \ref{sec:energy} presents the two methods by which nuSCOPE can measure the neutrino energy without relying on neutrino interactions: the Narrow Band Off-Axis technique developed by ENUBET and the SBND Collaboration at Fermilab \cite{delTutto2021SBND,Jones:2024qco}, which achieves a precision of approximately $\mathcal{O}(10)$\%, and the tagging method, which can reach the ultimate precision of $\sim$1\%.
The expected performance of the facility for a range of reference cross-section measurements is detailed in Sec.~\ref{sec:cross-section}.

\section{Beamline design}
\label{sec:beamline_design}

\begin{figure}[htb]
    \centering
    \includegraphics[width=\textwidth]{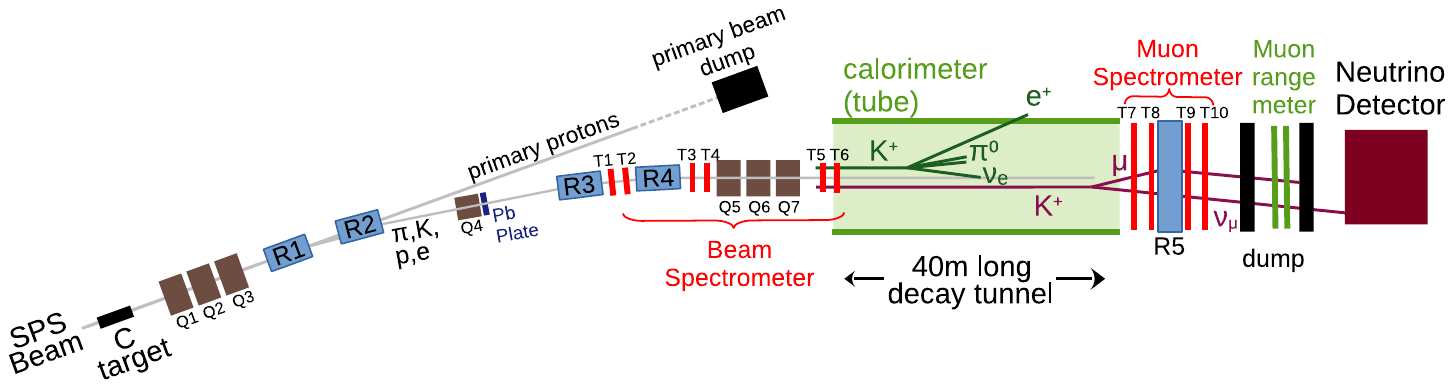}
    \caption{Layout of the experimental setup of the nuSCOPE beamline. The beamline (not to scale) combines both the ENUBET (in green) and the NuTag (in red) beam instrumentation.  The first part of the beamline contains the beamline magnets (blue: rectangular bending magnets; brown: quadrupole magnets).}
    \label{fig:layout}
\end{figure}

The nuSCOPE beamline aims at producing kaons and pions with a central momentum range of $p=\SI{8.5}{GeV/c}$.  The beamline, shown in Fig.~\ref{fig:layout}, was optimized to produce (ideally) a monochromatic meson beam that remains parallel throughout a 40\,m-long decay tunnel. The neutrino detectors that are under consideration are discussed in Sec. \ref{sec:neutrino_detectors}. In the following, it is assumed to be a \SI{500}{ton} liquid-argon detector located \SI{25}{m} downstream of the decay tunnel. It has a transverse size of $4\times 4$ m$^2$ and a longitudinal depth of roughly \SI{23}{m} (see Sec. \ref{sec:detector_parameters}). The meson production is achieved via a thin graphite target that is impinged by the $p=\SI{400}{GeV}$ proton beam of CERN's SPS. 
The initial three quadrupoles downstream of the target provide the necessary transverse phase advance that allows as many particles as possible to pass through the momentum selection of the beamline. The momentum selection section features bending magnets with small vertical gaps that deflect the secondary beam by 18$^\circ$. This section removes particles with momenta that are outside the desired momentum range -- usually a $\pm \SI{10}{\%}$ relative momentum range around the chosen beamline momentum $p$.  The last quadrupole triplet produces a parallel beam in the \SI{40}{m}-long decay tunnel. 
The beamline instrumentation offers two key features. The calorimeters inside the decay tunnel and the instrumented hadron dump (Sec. \ref{sec:instrumentation}) monitor charged lepton production, enabling a high-precision measurement of the $\nu_e$ and $\nu_\mu$ fluxes. Lepton monitoring is performed using the methods and devices developed by ENUBET. Meson reconstruction and tagging are carried out by four silicon pixel detectors positioned on either side of the fourth bending magnet. These detectors provide momentum measurements, while two additional detectors located just upstream of the decay tunnel determine the particle trajectory. Four more detectors, placed on either side of an additional bending magnet downstream of the decay tunnel, measure the momentum of the daughter muon. Meson and lepton tagging are thus performed using the methods and devices developed by NuTag.

\begin{table}[htb]
    \centering
    \begin{tabular}{l|c}
            \hline \hline
        \textbf{Parameter} & \textbf{Value} \\
        \hline 
        Primary proton momentum (GeV/c) & $400$  \\
        Beamline meson momentum (GeV/c) & max. $8.5$ \\
        Proton-beam spill duration & slow ($4.8$\,s to $9.6$\,s) \\
        Spill intensity (protons/spill) & $\SI{1E13}{}$ \\
        Event rate (THz) & $1$ -- $2$ \\
        Instantaneous power on target (W) & $170$ -- $340$ \\
        ($K^+$, $\pi^+$) yield per proton & ($\SI{1.3e-3}{}$, $\SI{1.9e-2}{}$) \\
        ($K^+$, $\pi^+$) rate (GHz) & max. ($\SI{2.7}{}$, $\SI{40}{}$) \\
        Annual proton intensity (protons/year) & $2.1$--$\SI{3.2e18}{}$ \\
        Total proton requirement (protons) & $\SI{1.4e19}{}$ \\
        \hline \hline
    \end{tabular}
   \caption{Beamline parameters and specifications of the newly optimized nuSCOPE beamline at SPS energies. Meson yields and rates refer to the location at the beginning of the decay tunnel. The total proton requirement targets a \SI{1}{\%} statistical uncertainty on the inclusive $\nu_e$ cross-section.  The annual proton intensity assumes a spill duration of \SI{4.8}{ s}.} 
    \label{tab:specs}
\end{table}

At CERN, the beamline can only be effectively driven by the SPS accelerator with a proton-beam energy in the range up to $\SI{400}{GeV/c}$. The preliminary beamline specifications are listed in Tab. \ref{tab:specs}. At PS energies, the meson yield is highly suppressed making this option not viable.  If implemented at CERN, the nuSCOPE beamline will be confronted with the issue of having a limited number of protons-on-target (PoT) available since other experiments are also dependent on a high annual proton flux from the SPS. For this reason, the beamline design has been optimized to minimize the required number of PoTs.

\subsection{Beamline optimization}
\label{sec:beamline_optimization}

The starting point of the present beamline design was an update of the ENUBET beamline baseline design \cite{ENUBET:2023hgu} by the Conventional Beams Working Group (CBWG) of the PBC initiative at CERN \cite{Parozzi:2882547}. This CBWG design roughly doubled the transmission of the $K^+$ and $\pi^+$ compared to the ENUBET baseline design. Within the scope of the present study, a further optimization of the beamline has been performed and has developed into the new beamline design which is presented hereafter.

The study has shown that a piece-wise optimization of the beamline, e.g., the optimization of the production target and the beamline optics broken up into two separate processes, leads to sub-optimal results. The production target and the beamline optics have to be matched to each other very carefully and both have to be optimized in a self-consistent manner.
The pixel detectors that are introduced for particle tagging feature a maximum read-out rate of   $\mathcal{O}(\SI{100}{MHz/mm^2})$ (see Sec. \ref{sec:silicon_detectors}) and the optimization is aimed at $\mathcal{O}(\SI{10}{MHz/mm^2})$ to suppress pile-up and the detector throughput. A challenging aspect that is introduced by the silicon detectors is the mitigation of the large number of positrons that are produced in the graphite target. To counter these, a thin Pb plate was inserted into the beamline right after the Q4 quadrupole to degrade their energy outside the beamline acceptance.  Without such a positron countermeasure, the particle flux of the monitors would exceed $ \SI{100}{MHz/mm^2}$ with spill intensities of a few $10^{12}\,$protons, thus compromising momentum measurement on a particle-by-particle basis. As in K12/NA62 in CERN's North Area \cite{Hahn:1404985}, it may be useful to adjust the thickness of the positron-absorbing Pb plate (using a motorized wedge with variable thickness) to effectively tune the flux on the pixel detectors.

\begin{figure}[htb]
    \centering
    \includegraphics[width=0.6\textwidth]{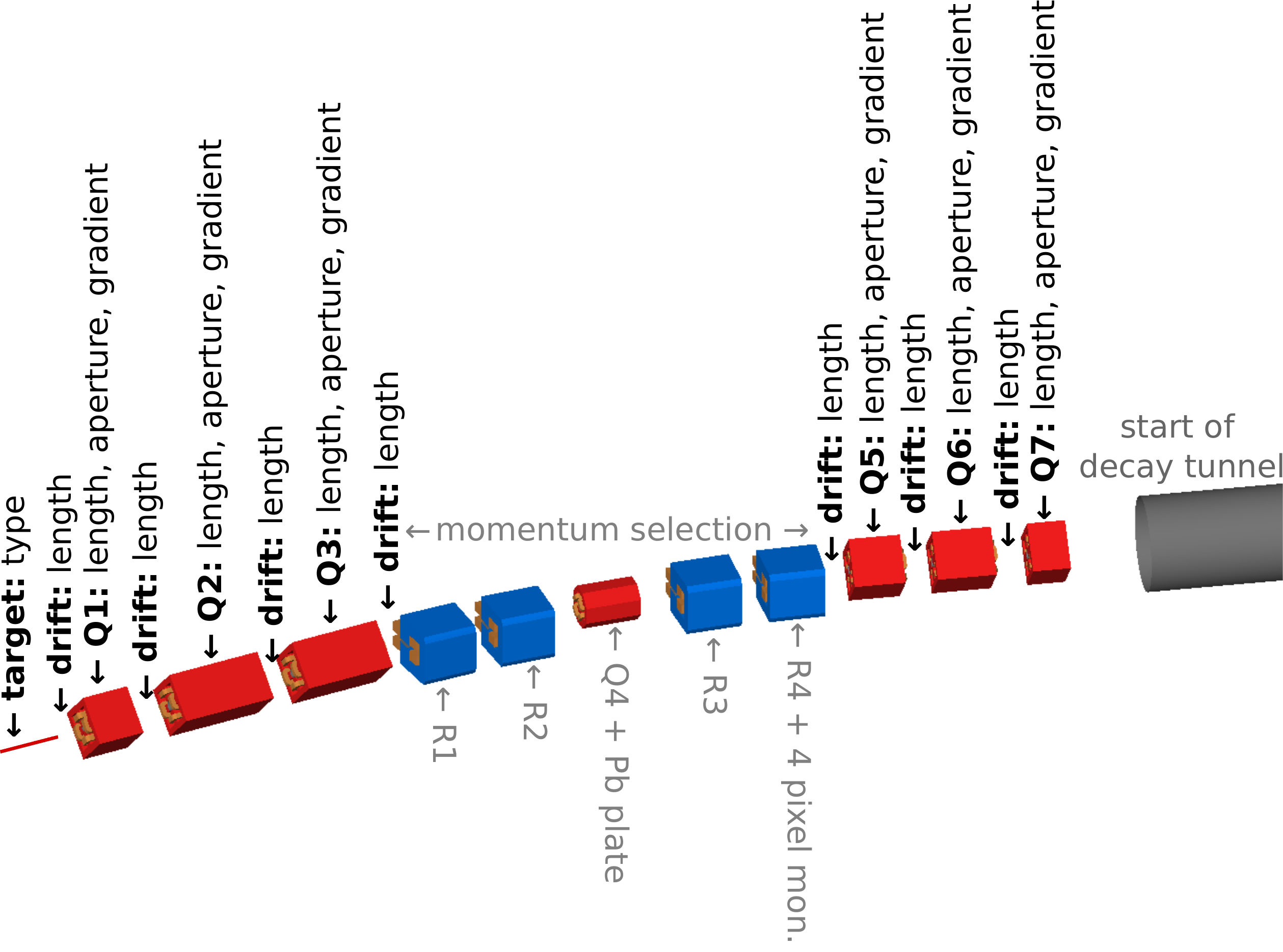}
    \caption{The figure shows the new layout of the nuSCOPE beamline with the parameters that can be changed by the MOGA optimisation algorithm printed in black on top while the parameters that are fixed and are not part of the optimization process are printed in grey.
    }
    \label{fig:sbn_drawing}
\end{figure}

With these changes, the beamline is optimised using a multi-objective genetic algorithm (MOGA). The algorithm has the capability of optimising 26 beamline parameters simultaneously. These parameters are (see Fig.~\ref{fig:sbn_drawing}):
\begin{itemize}
    \item Drift spaces: The drift spaces in the upstream and downstream quadrupole triplets are optimised (7 free parameters)
    \item Quadrupole parameters: For each quadrupole in both triplets, the aperture, length, and gradient are optimised (18 free parameters)
    \item Production target: the optimiser is capable of changing the graphite production target and choosing another target type from a pre-calculated list that contains a total of 18 different targets (one free parameter). The list includes the T2K, CNGS, ENUBET, and NuMi targets and variations of these \cite{Simos2019,densham2010design,cngs}. The list of targets covers target lengths from \SI{0.7}{m} to \SI{1.35}{m}, target radii from \SI{2.5}{mm} to \SI{30}{mm} and graphite densities from $\SI{1.70}{g/cm^3}$ to $\SI{2.26}{g/cm^3}$.
\end{itemize}

The optimisation was carried out using the NSGA-III \cite{nsgaiii}, SMS-EMOA \cite{Beume2007}, and AGE-MOEA \cite{Panichella2019} implementations within the PyMoo Python library \cite{pymoo}. The optimised beamline is verified by a full start-to-end BDSIM simulation \cite{Nevay:2018zhp}. The results of the respective BDSIM simulation are listed in Tab.~\ref{tab:sbn_performance}. The table shows the distinct improvement of the beamline performance in comparison to the  ENUBET baseline design. These improvements are achieved although obstacles like a Pb plate and silicon pixel detectors have been inserted into the beamline.
\begin{table}[htb]
    \centering
    \begin{tabular}{lcc}
    \hline \hline
      \textbf{Particle yield}  &  \textbf{{ENUBET design}} &   \textbf{Optimized nuSCOPE design}  \\ \hline
      $K^+/$PoT ($10^{-4}$)  & $3.6 $        & $12.6 $   \\
      $\pi^+/$PoT ($10^{-2}$)& $0.4$    &   $1.9 $ \\ \hline \hline
    \end{tabular}
    \caption{Comparison of the optimised nuSCOPE beamline design with the initial ENUBET baseline design \cite{ENUBET:2023hgu} at $p=8.5\,$GeV/c in the $\Delta p/p \in \left[-10\%; 10\% \right]$ momentum selection.}
    \label{tab:sbn_performance}
\end{table}
The main reasons for the improvements in terms of transmission are the following.
Firstly, the optimal target has been found to be a variation of the CNGS target with a length of \SI{1.30}{ m}, a radius of \SI{3}{ mm}, and a density of $\SI{2.26}{g/cm^2}$. Note that this particular graphite density may not be achieved for this particular application. Hence, a reduced (however still aggressive) density of $\SI{2.0}{g/cm^3}$ has been used throughout all simulations.
Secondly, the beamline has become shorter by more than 5 m. Hence, fewer $K^+$ mesons ($\pi^+$ to a less significant degree) decay in the beamline before they reach the decay tunnel. The most significant improvement, however, stems from a modified beamline acceptance that captures more of the $K^+$ and $\pi^+$ mesons that emerge from the production target.

\begin{figure}[ht]
    \centering
    \includegraphics[width=0.55\textwidth]{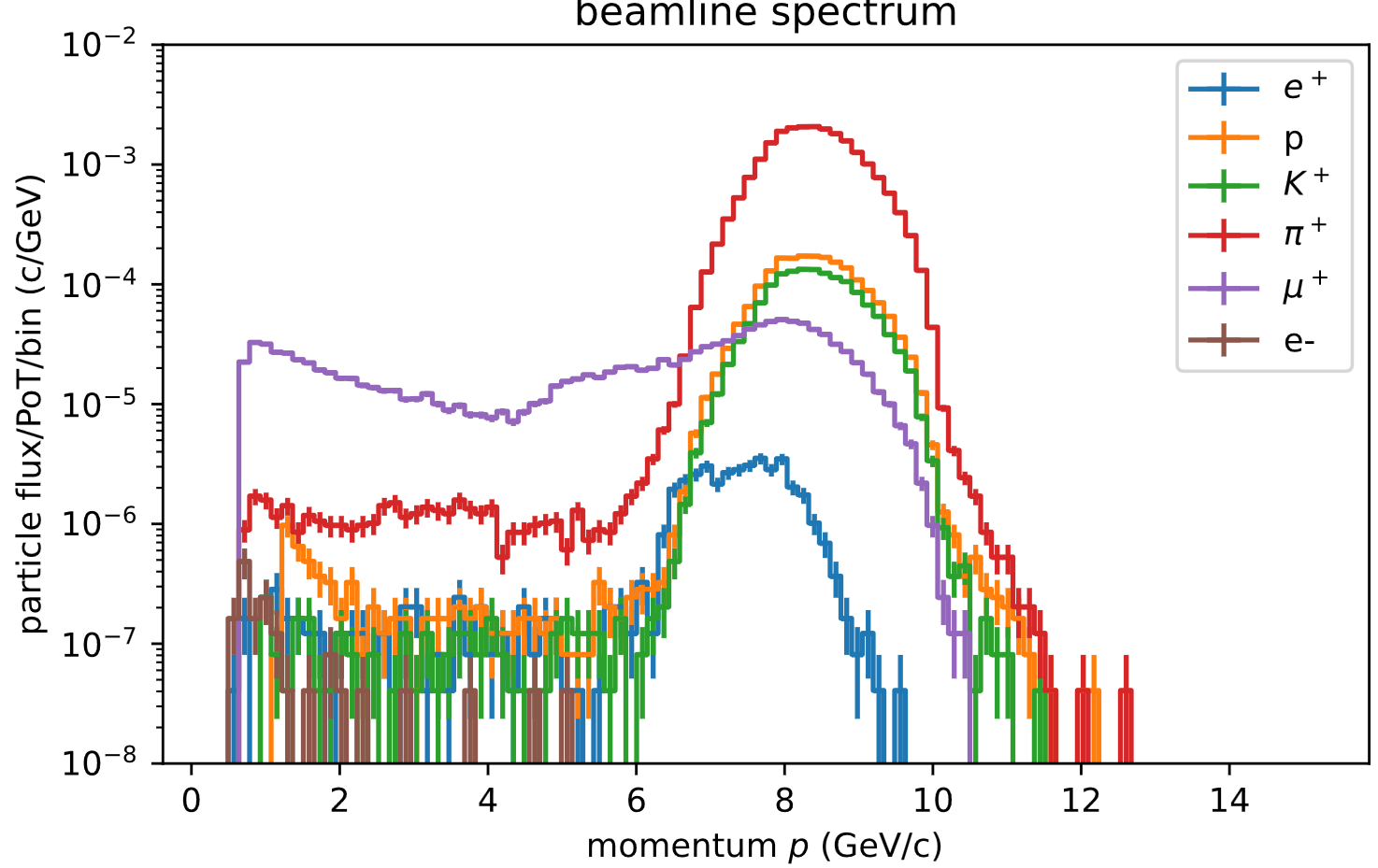}
    \includegraphics[width=0.44\textwidth]{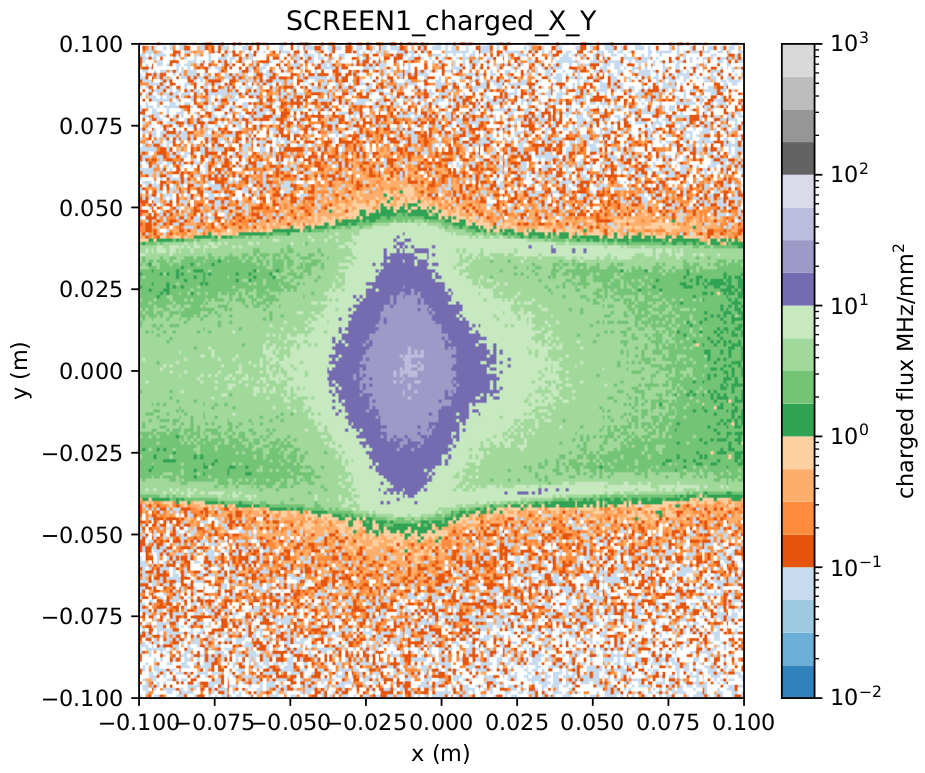}
    \caption{Left: Beamline spectrum after the optimisation process. Compared to any previous design, the positron transmission is strongly suppressed due to the Pb plate followed by two bending magnets. The study of the background level of low-energy leptons is still ongoing. Right: Flux of charged particles on the first pixel detector at the R4 bending magnet of the beamline with a spill intensity of $\SI{1e13}{PoT}$ within a spill length of $\SI{4.8}{s}$.}
    \label{fig:sbn_spectra}
\end{figure}

The spectrum of the particles transmitted by the optimized beamline at the entrance of the decay tunnel is shown in Fig.~\ref{fig:sbn_spectra} (left) with the beamline momentum being set to $p = \SI{8.5}{GeV/c}$. In the $\pm \SI{10}{\%}$ momentum range, the $K^+$ yield is at $\SI{12.6e-4}{}\,K^+/\mathrm{PoT}$ while the  $\pi^+$ yield is at $\SI{1.9e-2}{\pi^+/PoT}$.
A critical parameter for the beamline is the particle rate at the pixel detectors that are used for the momentum reconstruction at the R4 bending magnet (see Fig.~\ref{fig:layout}). In the proposed beamline design, the flux on the first pixel detector upstream of the R4 bending magnet is shown in Fig.~\ref{fig:sbn_spectra} (right). With a spill intensity of $\SI{1e13}{PoT}$ within $\SI{4.8}{s}$, the particle flux is in the range of 10 -- $\SI{40}{ MHz/mm^2}$ -- a flux that is  fully acceptable. A longer spill duration of $\SI{9.6}{s}$ for future operation of the SPS is currently under consideration. Any increase in the SPS spill duration while keeping the number of extracted protons constant would lead to an additional reduction of the flux on the pixel detectors.

\subsection{Radiation-protection considerations}
\label{sec:doses}

nuSCOPE is a proposed beamline featuring novel instrumentation for flux monitoring and tagging. However, radiation effects must be carefully considered, as high radiation doses may impact beamline components, silicon taggers, and beamline instrumentation. This challenge has already been addressed by the ENUBET R\&D and is further mitigated by the reduction of PoT achieved through the optimization described in Sec. \ref{sec:beamline_optimization}. Concerning the ambient dose equivalent (H*(10)) studies are currently ongoing. 

Non-ionizing doses in the decay tunnel must be estimated to assess the potential damage to the photosensors over the entire duration of the run. In addition, ionizing doses are key to evaluating the damage to the beamline components and, in particular, to the first quadrupole located downstream of the target as well as the potential doses to personnel. These studies were performed using FLUKA ~\cite{Ahdida2022,Battistoni2015,Roesler2001}.

\begin{figure*}[htb!]
    \centering
    \includegraphics[width=15cm]{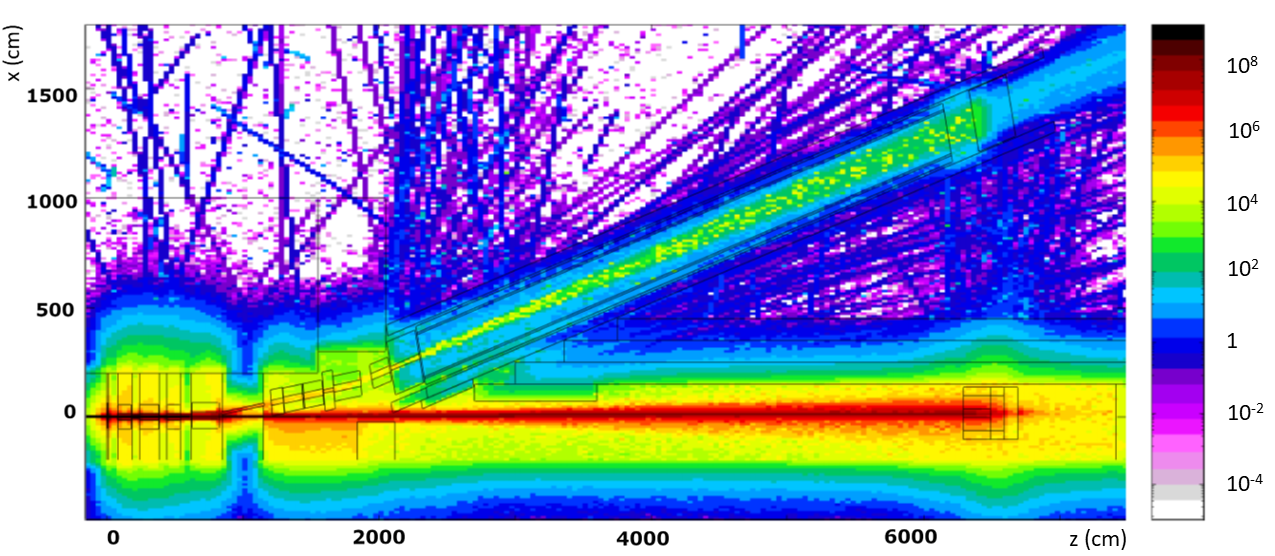}
 \includegraphics[width=15cm]{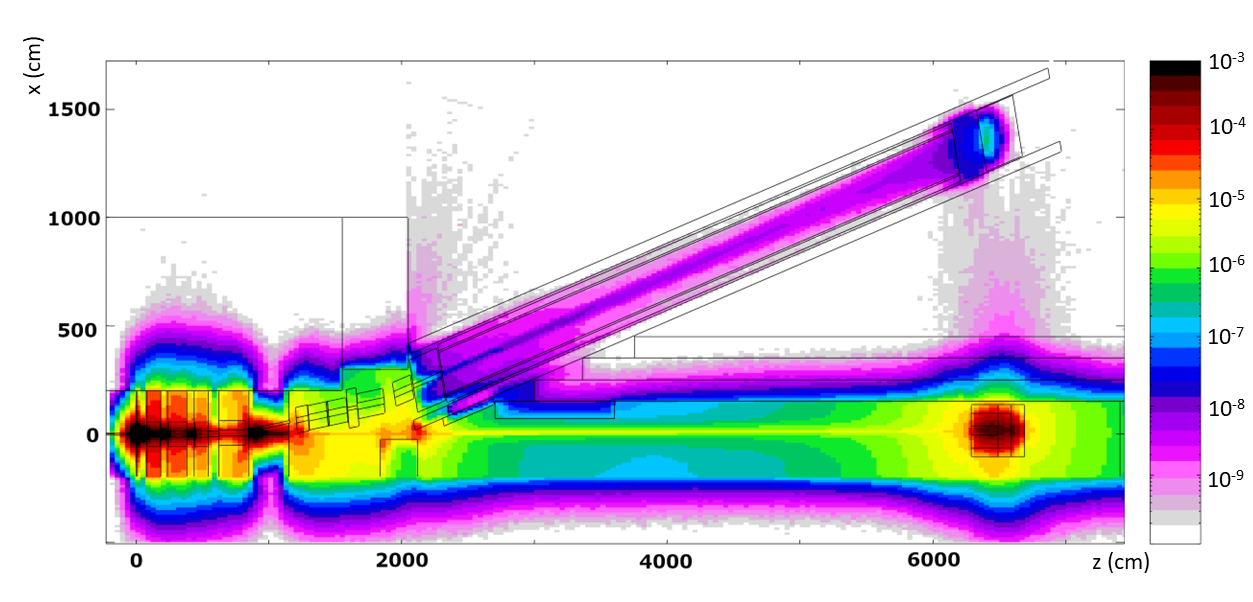}
    \caption{Top: dose map in Gy for 10$^{20}$~PoT. The first quadrupole in the map is located between $z\simeq 200$ and 500~cm. Bottom: 1-MeV-eq neutron fluences.}
 \label{fig:mapdose}
\end{figure*}

The map of the accumulated ionizing dose in Gy obtained with FLUKA for 10$^{20}$~PoT is shown in Fig.~\ref{fig:mapdose} (top plot). The dose at the hottest point of the quadrupole closest to the target is 
$<300$ kGy for 10$^{20}$~PoT. The maps in the proximity of the target prove that conventional magnets can be operated without risk in the proposed facility for the entire duration of the data taking.
Neutron fluences (non-ionizing doses) are shown in Fig.~\ref{fig:mapdose} (bottom plot) in units of neutrons/cm$^2$/primary proton.
This study, in particular, validates the choice of reading the ENUBET instrumentation—specifically, the modular sampling calorimeter described in Sec. \ref{sec:instrumentation}—using SiPMs located outside the calorimeter and protected by a neutron shield.

Although radiation damage is not critical for the nuSCOPE components up to $10^{20}$ PoT, shielding for radiation protection within the CERN complex remains an important task. The required shielding depends on the final location of nuSCOPE and is currently under evaluation. Location options are discussed in the next section.

\subsection{Possible implementation at CERN}

\begin{figure}[htb]
    \centering
    \includegraphics[width=0.7\linewidth]{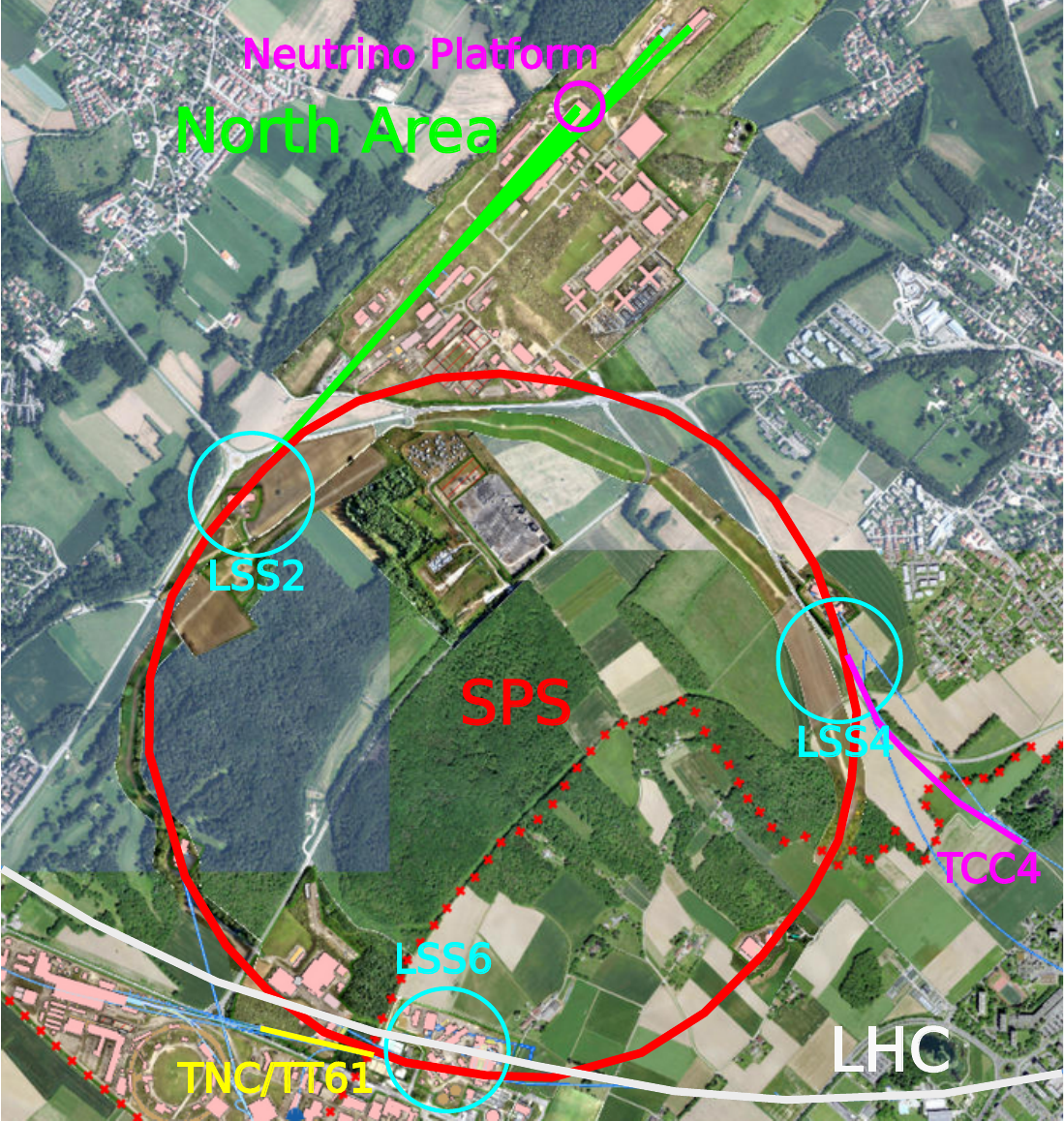}
    \caption{Top view of the CERN SPS. The three experimental areas at the SPS are located at the three long straight sections LSS2, LSS4 and LSS6 of the SPS. The Neutrino Platform that houses the ProtoDUNEs are located at the ends of the H2 and H4 beamlines in the North Area.}
    \label{fig:cern_map}
\end{figure}

The feasibility study for a potential placement of the nuSCOPE beamline on the CERN campus is currently ongoing and is not final \cite{jebramcik_sx}; however, two approaches have been identified as promising and could lead to the successful implementation of the beamline.

CERN's North Area (see Fig. \ref{fig:cern_map}) is usually the prime location for fixed-target experiments at CERN due to the availability of a slow extraction from the SPS (multiple seconds long spill duration) at the long straight section 2 (LSS2)  and of the existing infrastructure and beamlines. The former CERN Neutrinos to Gran Sasso (CNGS) experiment, however, was located deep underground downstream of LSS4 in TCC4 to be able to point the neutrino beam towards Gran Sasso, Italy. The ideal location in CERN's North Area could be the ECN3 cavern due to its depth and radiation-protection limits. However, following the recent Research Board decision at CERN, the SHiP \cite{Alekhin:2015byh} experiment will occupy that cavern starting from LHC Run 4. The beamline cannot be placed inside the surface halls EHN1 and EHN2 in CERN's North Area due to radiation constraints. The expected spill intensity exceeds \SI{1E13}{PoT/spill} and the secondary beam is of the order of \SI{1E11}{particles/spill}, while the radiation protection (RP) constraints for EHN1 and EHN2 (supervised areas) are two to four orders of magnitude lower than these values even in a future EHN2 scenario with additional shielding that is currently proposed in the context of the AMBER experiment \cite{Adams:2018pwt}. As a result, implementing such a beamline in EHN1 or EHN2 would be highly challenging without extensive shielding modifications, which would also affect the operation of other North Area beamlines. Additionally, the beamline's bending angle makes it difficult to implement in experimental halls like EHN1 and EHN2, as these halls are predominantly long rather than wide. 
A placement below EHN1 is also excluded due to the technical galleries below the building and the challenges of excavating underneath an existing building in the vicinity of the Lion river. As a result, the existing ProtoDUNEs at the North Area's Neutrino Platform (see Fig. \ref{fig:cern_map}) cannot be used in their current location for the nuSCOPE beamline, as the neutrino detector must be placed no more than \SI{50}{m} from the end of the decay tunnel.\footnote{A beamline bypassing EHN1 to approach the ProtoDUNEs has been excluded due to the vicinity of the fence and the SPS beam rigidity that results in a large bending radius.}
Hence, building extensions and an experimental hall for the neutrino detectors would be necessary in the North Area.

\subsubsection{Implementation at ECN4} \label{sec:ecn4}
\begin{figure}[htb!]
    \centering
    \includegraphics[width=0.72\linewidth]{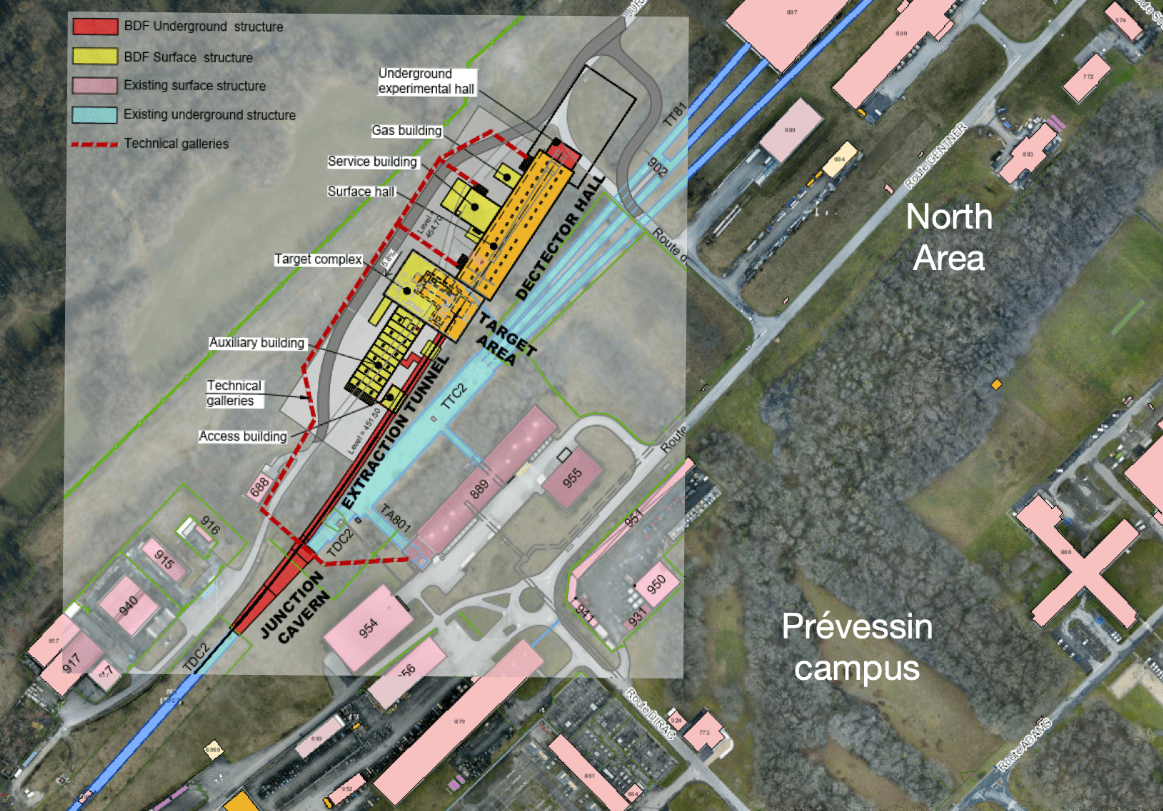}
    \caption{Schematic drawing of ECN4 solution studied as a possibility to house the BDF/SHiP facility. From \cite{Ahdida2847433}.}
    \label{fig:ship_sketch}
\end{figure}

CERN's Accelerator and Technology Sector has produced a feasibility study on a possible realization of a new underground cavern (designated ``ECN4'') as an option for the proposed BDF/SHiP facility (see Fig.~\ref{fig:ship_sketch}). This facility would be constructed on the Jura side of TCC2 -- the underground cavern that houses the targets for the EHN1 beamlines (see Fig. \ref{fig:cern_map}).
The implementation of the nuSCOPE beamline has similar requirements compared to BDF/SHiP; however, a crucial difference is the required PoT (overall and annually) as well as the spill intensity. The envisaged spill intensity of $\SI{1e13}{PoT/spill}$ is a factor of four smaller than the one envisaged by SHiP, which is $\SI{4e13}{PoT/spill}$. In addition, the nuSCOPE beamline would consume up to a maximum of 2 --\SI{3e18}{PoT/year}, a value approximately ten times smaller than the BDF/SHiP requirement. This lower intensity allows for a reduced amount of shielding, cooling, and dilution measures that are required by BDF/SHiP, and offers greater flexibility in the location of the experimental hall. Simultaneously, key components of the initial ECN4 for the BDF/SHiP proposal are nearly identical in the case of nuSCOPE:
\begin{itemize}
    \item \textbf{Junction cavern:} The BDF/SHiP study had proposed replacing the first splitter triplet (MSSB.2117) with a laminated in-vacuum beam splitter \cite{Ahdida:2703984}, enabling the establishment of a dedicated cycle in which the new ECN4 facility receives the entire SPS spill intensity. Such a device would require extensive R\&D and appears unlikely at the time of writing this document. If achieved, the beam would reach the target via a new junction cavern pointing towards the Jura side. A dedicated cycle is clearly the ideal solution; however, it may create proton competition with other experiments. The installation of an additional splitter appears not to be possible due to the optics layout of the transfer line. Such a splitter would provide the possibility of having a shared cycle (the SPS spill intensity is split among all experiments in the North Area with the exception of BDF/SHiP). An alternative solution could be the implementation of a P42-like equivalent at the T2 target. This would allow the beamline to be part of a shared cycle by receiving the full T2 intensity. This option is promising but requires in-depth analysis.
    
    \item  \textbf{Transfer tunnel}: The nuSCOPE beamline also requires an extraction tunnel downstream of a junction cavern, which must also be constructed. The tunnel is comparable in length to that of the ECN4 proposal for BDF/SHiP; however, the entire beamline and decay tunnel would be placed fully underground once sufficient separation from the junction cavern is reached. Since the beamline only requires a shielding that is comparable to K12/NA62  due to a similar target and PoT requirements, the beamline should have sufficiently small transverse dimensions to achieve a manageable excavation process.
    
    \item \textbf{Target complex}: Unlike BDF/SHiP, nuSCOPE does not require a complex target system due to its significantly lower intensity and different operational needs. The proposed graphite target is expected to have a simple design similar to that of CNGS and may not require replacement, as CNGS operated with a single graphite target throughout its lifetime. A beam dump for the primary protons that do not interact with the target is required, nevertheless. Here, a device like the XTAX currently used in the K12 beamline for the NA62 experiment is proposed, i.e., a \SI{3.2}{m} long element with different layers of Al, Fe, and Cu.

    \item \textbf{Auxiliary building:} An access building is ideally not required since the beamline would stop after roughly \SI{100}{m} to \SI{150}{m} downstream of the junction cavern. The neutrinos would then penetrate rocks and soil to reach the neutrino detectors in their dedicated experimental area.

    \item \textbf{Experimental Hall}: In contrast to the BDF/SHiP proposal, the experimental area is not connected to the target area via a tunnel or beamline to the underground beamline. A surface building comparable to the Neutrino Platform (extension of EHN1 - including a trench for the liquid-argon detector) is sufficient to house the neutrino detectors. If ECN4 was about to be chosen for this project, either the ProtoDUNEs would have to be moved (the Neutrino Platform is too far away from TCC2) or new liquid argon and/or water Cherenkov detectors would have to be constructed ---see Sec. \ref{sec:neutrino_detectors}. Since the proposed secondary beamline naturally features an 18° bending angle due to the double-bend achromat setup, its layout can be slightly adjusted to direct the decay tunnel (and neutrino beam) upward at any angle smaller than 18°. This configuration allows neutrinos to reach detectors in a dedicated surface-level experimental hall, enabling placement much closer than 100 m from the end of the decay tunnel.
    
\end{itemize}

\subsubsection{Implementation at SPS LSS6}  \label{sec:slow_extraction}
Apart from the North Area Extraction in LSS2, the two other long straight sections LSS4 and LSS6 (see Fig. \ref{fig:cern_map}) also feature extraction septa followed by transfer tunnels that serve the LHC Beam 2 and TCC4 (AWAKE) in the case of LSS4 and LHC Beam 1 and TCC6/TNC (HiRadMat) in the case of LSS6. These locations are deep underground and can be considered to house the proposed nuSCOPE facility. The main disadvantage compared to a location close to LSS2, however, is the missing ZS septa that are necessary for a slow extracted beam. Given the presence of AWAKE downstream in TCC4 and the depth of TCC4, placing the beamline downstream of LSS6 is considered more advantageous.

A conceptual analysis  \cite{pbc_rep_sx} was performed to investigate the possibility of re-installing a slow extraction setup at SPS LSS6. At that location, the fast extraction to the LHC but also to the HiRadMat facility (TNC tunnel) is performed by a combination of fast kickers (MKE) and magnetic septa (MST and MSE-type septa). The authors have been studying two approaches to revive the slow extraction towards LSS6. The most promising approach appears to be a so-called  ``non-local'' extraction setup. In this scenario, the electrostatic ZS septa are not located in LSS6 but upstream in a different LSS. As a result, two orbit bumps are required; i.e., at the locations of both the electrostatic septa upstream of LSS6 and the magnetic septa in LSS6. The extracted beam oscillates through a considerable part of the SPS ring until it is fully extracted by the magnetic septa. This type of solution is not particularly novel: the PS successfully employs a similar non-local extraction scheme \cite{django}. 
The non-local solution has the significant advantage over a local solution within LSS6 in that it does not require any modifications to the fast extraction for the LHC, while still allowing the successful extraction of the maximum energy  $p\geq \SI{400}{ GeV/c}$,  
 as discussed in detail in~\cite{pbc_rep_sx}. The study has identified two $>\SI{20}{m}$ long drift spaces in LSS4 and LSS5 that provide sufficient space for five ZS wire-septum modules identical to the ones in LSS2 that are used for the extraction towards the North Area. These locations also have the ideal phase advance with respect to the MST septa in LSS6. 

\begin{figure}[htb]
    \centering
    \includegraphics[width=0.49\linewidth]{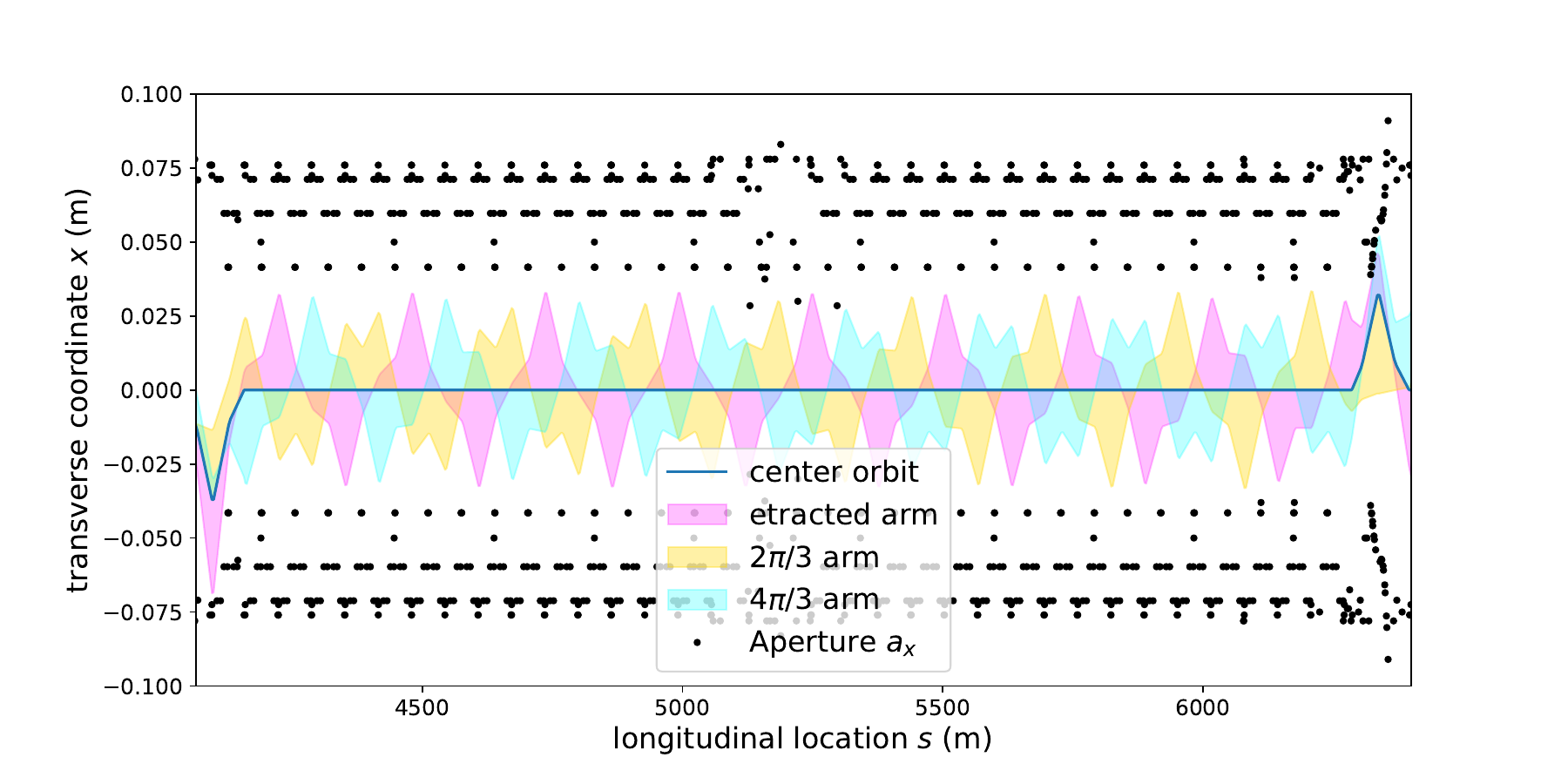}
    \includegraphics[width=0.49\linewidth]{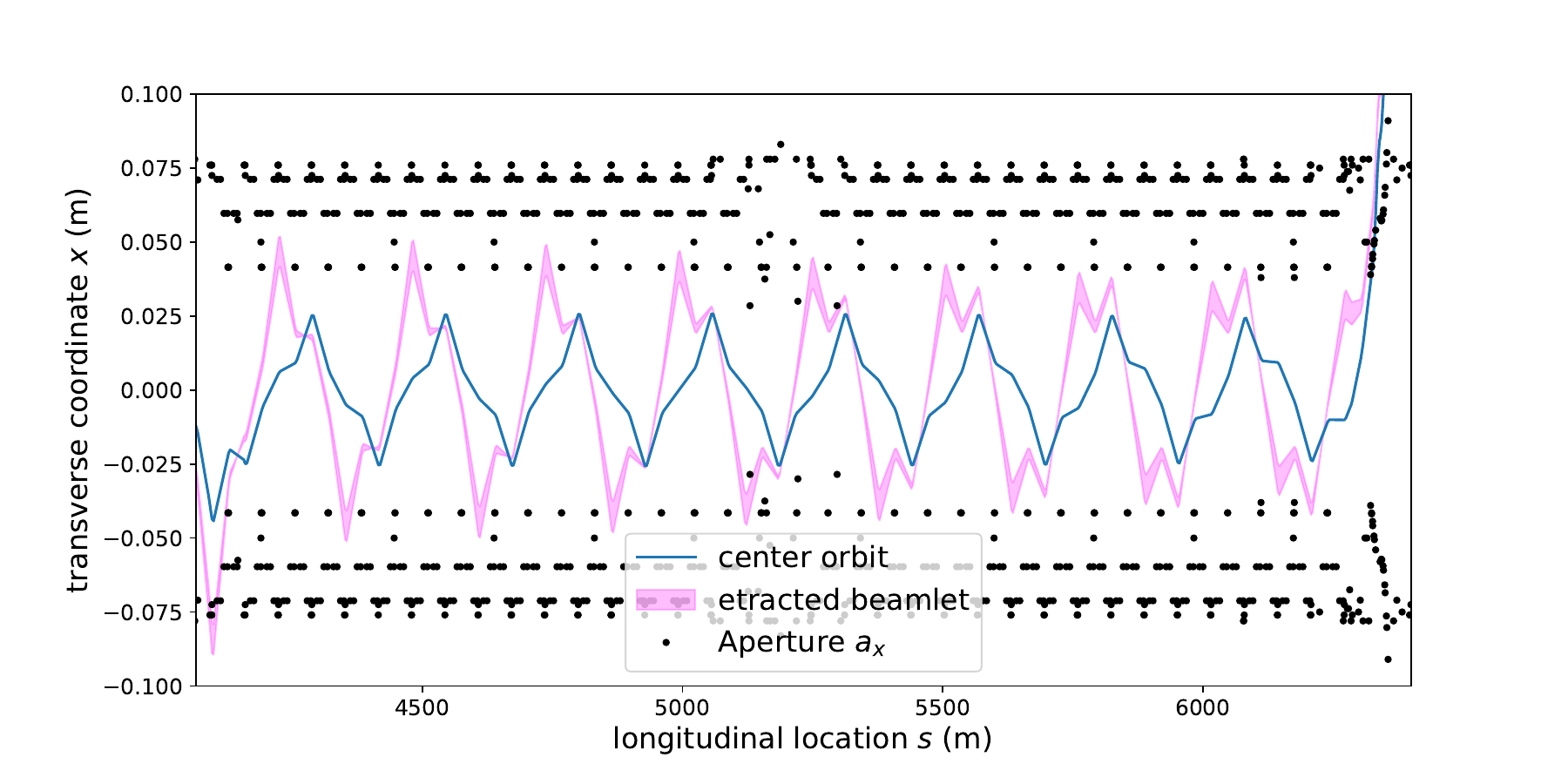}
    \caption{
    Left: Oscillation of the three stored arms of the stored beam on resonance. One can see the two local orbit bumps in LSS4 and LSS6 in dark blue. Right: The oscillation of the extracted part of the beam (right) from LSS4 up to LSS6 in the SPS.  Figures taken from~\cite{pbc_rep_sx}.}
    \label{fig:5}
\end{figure}

In \cite{pbc_rep_sx}, the authors have performed a study for the placement of the five necessary ZS modules at the aforementioned location in LSS4 as an example. An alternative location could be in LSS5; however, the feasibility of placing the ZS septa downstream of the beam dump has to be investigated in the future. Figure~\ref{fig:5} shows the behavior of the on-resonance beam as well as the extracted beam fraction in case the ZS are placed in LSS4 to then extract in LSS6. 
If the non-local solution is pursued, some modifications of the SPS magnetic lattice have to be performed. A key aspect of this setup still has to be studied thoroughly -- that is, the simultaneous extraction to LSS2 (North Area) and LSS6 (location of the nuSCOPE beamline). If accomplished, the simultaneous extraction would allow the beamline to be operated without competing for protons with other projects. Significant beam simulation and optics optimization studies have to be performed to ensure correct tune and resonance-driving term matching in the SPS before such an operation mode can be deemed feasible. 

\begin{figure}
    \centering
    \includegraphics[width=0.95\linewidth]{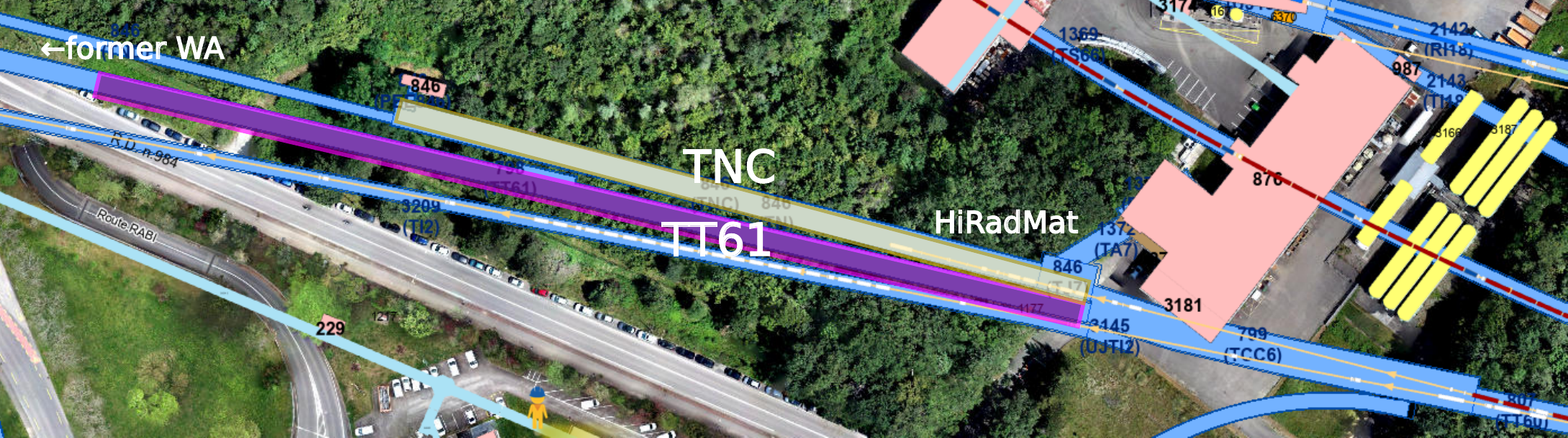}
    \caption{View of the TNC (yellow frame) and TT61 (purple frame) tunnels downstream of the extraction location in SPS LSS6. The first part of the TNC tunnel is occupied by the HiRadMat test facility, while the downstream tunnel (approximately \SI{120}{m}) can be used for the beamline. The TT61 tunnel -- although slightly narrower than the TNC tunnel -- is currently not occupied.}
    \label{fig:tnc_tunnel}
\end{figure}

A potential placement of the nuSCOPE beamline in the TNC or TT61 tunnel has the advantage of the primary production target being deep underground. The TNC tunnel with a width of \SI{5.6}{m} is slightly wider than the TT61 tunnel (\SI{4}{m} wide). The former TCC6 cavern (see Fig.~\ref{fig:tnc_tunnel}; cavern upstream of TNC and TT61) housed the T1 target for the former West Area, receiving large numbers of protons on a yearly basis (more than $\SI{1e19}{PoT/year}$). HiRadMat also features a beam dump in the TNC tunnel; however, here it is worth noticing that HiRadMat can only receive $\SI{1e16}{PoT/year}$ due to RP constraints.
Nevertheless, the placement of the beamline in either TNC or TT61 should be possible in terms of RP considerations with sufficient shielding. The exact placement of the beamline in and around the TNC and TT61 tunnels can be adjusted using multiple parameters; still, the angle of the double-bend achromat in the beamline can only be slightly reduced from its design value of 18\,\textdegree{} to maintain the required particle background suppression rate in the decay tunnel (see Sec. \ref{sec:neutrino_monitoring_wall}).

\begin{figure}[htb]
    \centering
    \includegraphics[width=0.95\linewidth]{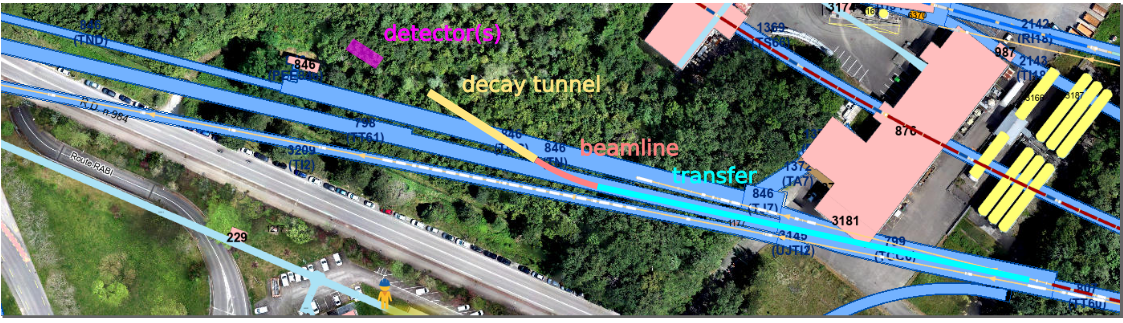}
    \caption{In the current beamline design and pending further optimization, the beamline cannot be operated successfully at a reduced bending angle. For this reason, the beamline is placed with a straight transfer and requires the excavation for the neutrino detector and the final part of the decay tunnel.}
    \label{fig:example_placement}
\end{figure}

Figure~\ref{fig:example_placement} gives an example of a potential placement of the beamline in TNC/TT61. This is currently considered the least challenging scenario. In this case, the transfer line follows the TT61 slope directly, while the secondary beamline, with an 18° bending angle, is maintained. It is worth noticing that the HiRadMat facility does not need to be decommissioned and may stay operational. The exact shielding necessary and possible operational interlock implications to ensure the safe operation of HiRadMat need to be addressed in a future study. It is also important to note that the neutrino detector(s) have to be placed underground in this scenario. In all cases,  excavation for the proposed decay tunnel as well as the neutrino detectors will be required independently of the exact placement of the beamline in TT61/TNC. This excavation has been estimated to be on the scale of $\geq\SI{2000}{m^3}$. 
Further investigation of all possible configurations concluded that the removal of HiRadMat would not offer any advantage to the project, since in that case the neutrino detector pit would get very close to the TI2 tunnel that contains the injection line to the LHC. For this reason, it was decided not to pursue the HiRadMat removal option any further. 

Key components and considerations to accomplish a successful implementation of the beamline downstream of SPS LSS6 are: 
\begin{itemize}
    \item \textbf{Slow extraction and transfer line:} As explained in Sec.~\ref{sec:slow_extraction}, the scheme that is proposed here requires the placement of the ZS septa in LSS4 of the SPS and other SPS modifications (additional bumper magnets, operation with two orbit bumps, etc.). The beam would then be guided into TCC6, similar to the beam directed to LHC Beam 1 or HiRadMat. With the placement of the beamline in the TT61 tunnel, the transport from TCC6 into the TT61 tunnel is straight and therefore not challenging. A FODO cell structure with a final focus quadrupole triplet can be envisaged.
    \item \textbf{Target complex:} A beamline in either TT61 or TNC requires a K12-like target complex and a beam dump like the existing XTAX, which today is widely used in TCC2 for the North Area slowly extracted beams. The placement of the dump may require excavation. It is of utmost importance to shield the decay tunnel from the beam-dump location to avoid background radiation influencing particle detection in the decay tunnel.
    \item \textbf{Secondary beamline:} The beamline requires excavation for the decay tunnel (see Fig. \ref{fig:example_placement} bottom) that passes over the top of the TNC tunnel (vertical separation). 
    \item \textbf{Experimental Area:} The experimental area has to be excavated as shown in Fig.~\ref{fig:example_placement}. A beam dump for the secondary particles between the end of the decay tunnel and the neutrino detector is required, as well. The setup that is proposed here envisages a short distance of \SI{25}{m} between the decay tunnel and the detectors to enhance the number of interacting neutrinos in the detectors.
\end{itemize}    

\subsection{Proton sharing at the SPS}

In the assumed scenario, the beamline is considered to be operated within a shared cycle with the other North Area experiments (SFTPRO cycle). This assumption is independent of the beamline placement, and a placement in TT61/TNC is also possible. A neutrino detector with a \SI{500}{ton} liquid argon mass is assumed to be located at a distance of \SI{25}{m} from the end of the decay tunnel, requiring $1.4 \times 10^{19}$ PoT to collect up to $10^4 \ \nu_e$ ($10^6 \ \nu_\mu$) charged-current (CC) events. The assumed SPS Flattop duration is $9.6\,$s, delivering $1.0 \times 10^{13}\,$PoT per spill to the nuSCOPE beamline in a shared cycle. This corresponds to $\SI{25}{\%}$ of the TCC2 intensity per spill and, ultimately, per year. ECN3 (BDF/SHiP) receives the maximum requested $4.0 \times 10^{19}\,$PoT/year. The nuSCOPE runtime for $10^4 \nu_e$ CC events in the 500 ton detector is then between \SI{5.7}{years} and \SI{7.3}{years} with a yearly PoT consumption of \SI{1.8e18}{PoT} to \SI{2.4e18}{PoT}.  After five years, the statistical uncertainty on the inclusive  $\nu_e$ cross-section would be better than $\SI{1.2}{\%}$ in all cases. 

In case the SPS Flattop duration would remain at the current value of \SI{4.8}{s} (see~\cite{Ahdida2847433}), the nuSCOPE beamline would still take roughly \SI{25}{\%} of the TCC2 intensity (\SI{1E13}{PoT/spill}) and would reach the physics goal of $10^4\,\nu_e$ within \SI{4.3}{years} and \SI{6.5}{years}. A statistical uncertainty of 1\% on the inclusive  $\nu_e$ cross-section is therefore likely to be achievable within five years. 
It is important to mention the possibility of an enhanced spill number for TCC2 in case the SHiP spill intensity is increased to $\SI{5e13}{PoT/spill}$. This would increase the number of spills for the North Area (SHiP excluded) by roughly $\SI{25}{\%}$ and would shorten the experiment's required runtime. 

A second slow-extraction setup for extraction at either LSS4 or LSS6 of the SPS could provide the necessary flexibility to achieve simultaneous extractions to nuSCOPE as well as the North Area (SFTPRO and SHiP cycles). This possibility is described in more detail in the corresponding PBC note, which analyzes a potential second slow-extraction setup in the SPS \cite{jebramcik_sx}. It would thus be possible to extract a small fraction of the beam during a SHiP cycle (at the percent level) in a simultaneous slow extraction. The cycle intensity can then be increased by that amount to still provide the target intensity to the SHiP experiment. This would allow for the reduction of the PoTs taken from the SFTPRO cycle. Such a more complicated sharing scenario will be studied further in the future.
It is important to note that the beamline may never be operated in a dedicated cycle in either scenario. This is necessary to ensure coexistence with the other experiments while minimizing competition for PoTs.

\section{Beamline instrumentation}
\label{sec:instrumentation}
The neutrino beamline is instrumented with three sets of detectors as shown in \autoref{fig:layout}. The flux of outgoing neutrinos originating from kaon decays and reaching the detectors can be monitored by the instrumentation in the tunnel walls, as discussed in Sec. \ref{sec:neutrino_monitoring_wall}. The $\nu_\mu$ flux from pion decays is primarily monitored by the instrumented hadron dump (Sec. \ref{sec:instrumented_hadron_dump}). Neutrinos can be tagged at the single-particle level by tracking their parent mesons and daughter muons using the silicon planes described in Sec. \ref{sec:silicon_detectors}. 

\subsection{Neutrino monitoring in the tunnel wall}
\label{sec:neutrino_monitoring_wall}

\begin{figure}[!h]
    \centering
    \includegraphics[width=0.8\linewidth]{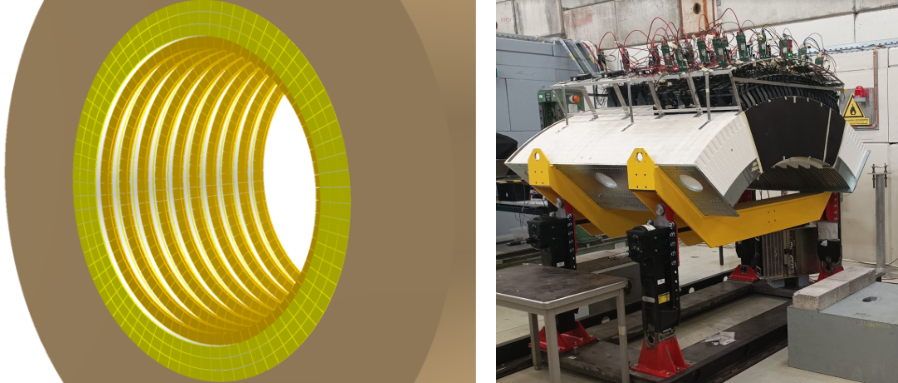}
    \caption{Left: Instrumented decay tunnel scheme. The three layers of modules of the calorimeter (light green) constitute the inner wall of the tunnel. The rings of the scintillator tiles of the photon veto (yellow) are located just below the modules. The optical fibers (not shown) bring the light to the outer part of the tunnel in the radial direction. They cross the neutron shielding (light brown) where the SiPMs (not shown) are positioned.
    Right: Prototype of a section of the instrumented decay tunnel during a testbeam at CERN-PS.}
    \label{fig:tagger}
\end{figure}

To monitor the $\nu_e$ flux and the high-energy component of the $\nu_{\mu}$ flux, charged leptons emitted at large angles from kaon decays are measured in the decay tunnel, whose
walls will be instrumented with a coarse sampling calorimeter, as a cost-effective solution for the full coverage of its 40~m length, and with a photon veto ("$t_0$~layer"), as sketched in Fig.~\ref{fig:tagger}, left.

The calorimeter is employed to achieve $e^+/\pi^+/\mu^+$ separation and segmented in the longitudinal, azimuthal and radial directions, with the basic unit, called Lateral Compact Module (LCM), made of five slabs of iron interleaved with tiles of plastic scintillator. The transverse size of the iron and scintillator tiles is $3\times3$~cm$^2$ and the thickness is 1.5 and 0.7~cm, respectively, for a total thickness of a LCM of 11~cm, corresponding to 4.3~$X_0$.
The LCMs are arranged in three radial layers, with every module covering an azimuthal angle of 31~mrad for a total of 200 LCMs per layer at a fixed longitudinal position along the tunnel.
The $t_0$~layer is composed of rings of plastic scintillator tiles placed in the innermost part of the tunnel in correspondence of the first two tiles of each LCM, to distinguish $\gamma$-initiated e.m. showers from the ones due to positrons and to provide the absolute time of each charged particle impinging on the tunnel walls.
The light is routed by a couple of WLS fibers, placed on the front face of each scintillator tile of both calorimeter and $t_0$~layer modules, towards SiPMs placed above a 30~cm thick borated polyethylene shielding that protects against neutron radiation damage of the sensors. The 10 fibers of each LCM are read out together, while each $t_0$~layer tile is read out individually by digitizers connected to the respective SiPM that sample the resulting waveforms~\cite{Acerbi:2805716}.
A large-scale prototype (1.65~m length, 90° azimuthal coverage) of the instrumented decay tunnel (Fig.~\ref{fig:tagger}, right) was constructed and exposed to mixed charged-particle beams in the T9 secondary line of the CERN-PS for validation in 2022-24~\cite{demonstrator}.
Collected data confirm the viability of the proposed solution, already tested with smaller prototypes \cite{Acerbi:2020nwd}, in terms of energy resolution ($\sim$17\% at 1~GeV) and charged particle identification for neutrino monitoring, and are used to fine-tune the detector simulation.

A full Geant4 simulation of the 40~m long tunnel has been developed, tracking all the particles at the tunnel entrance derived from the output of the BDSIM simulation of the beamline described in Sec.~\ref{sec:beamline_design}, and recording in each detector module the energy deposition from the particles impinging the tunnel walls. 
A proton-on-target intensity of 1$\times$10$^{13}$ in a 9.6~s spill is assumed for a realistic treatment of pile-up effects in the instrumentation.
This framework has been used to develop and assess the performance of the charged lepton identification algorithms in the decay tunnel~\cite{ENUBET:2023hgu}.
The first step of the reconstruction chain is the event building, in which hits in different LCMs and $t_0$~layer tiles are correlated in space and time to select the ones belonging to the same particle. The cuts are tuned to preselect signal events and to suppress the background arising from beam-halo particles and other than leptonic mesons' decay modes. Then, a multivariate analysis based on a Multilayer Perceptron Neural Network (NN) provided by the TMVA toolkit~\cite{Hocker:2007ht} is applied to discriminate the signal from the background. 
Two different event reconstruction and analysis chains, with specific cuts, NN training samples, and discriminating variables, are used for positrons and muons from kaon decays, respectively.

\begin{figure}[!h]
    \centering
    \includegraphics[width=0.55\linewidth]{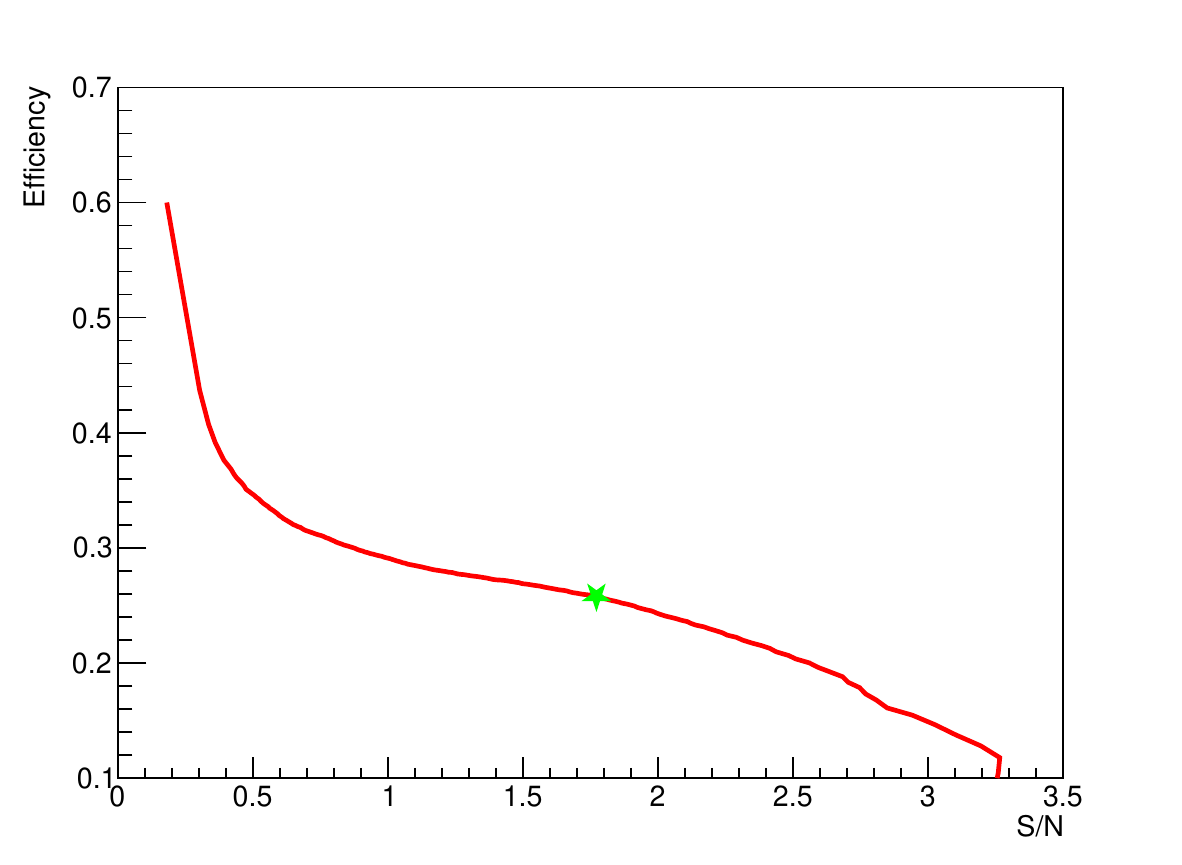}
    \caption{Signal efficiency versus signal-to-noise ratio for the $K_{e3}$ event selection. The green marker corresponds to the working point for signal selection, that is the point in the curve maximizing the product between signal efficiency and purity.}
    \label{fig:PID_positron}
\end{figure}

The $K_{e3}$ event building is triggered by the identification of an energy deposition in one LCM of the innermost calorimetric layer exceeding 28 MeV (i.e. $\sim$4 times the energy left by a m.i.p.), in order to enhance the preselection of positron candidates suppressing muons and non-interacting hadrons. Energy deposits compatible in time with the seed are sought for in an area large enough to allow for an efficient e.m./hadronic shower separation by the NN, while hits in the 9 upstream $t_0$~layer tiles are clustered for the suppression of photons originated from the interaction of stray particles with the elements of the beamline or $\pi^0$ produced in kaon decays.
The input layer of the NN is fed by 19 variables describing the energy deposition pattern in the calorimeter and the energy released in the photon veto; the event location in both the longitudinal and transverse directions is also used as a discriminating variable, as signal positrons impact the tunnel well after its entrance, whereas the halo background is primarily distributed near the tunnel entrance and along the bending plane. The ROC curve of the NN, describing the PID performance of the full positron analysis, is shown in Fig.~\ref{fig:PID_positron}. 
Each point on the curve reports the efficiency and the signal-to-noise (S/N) ratio for progressively stronger cuts on the NN classifier.
A positron selection efficiency of 25.8\% (including the $\sim$ 50\% geometrical acceptance) and a S/N of 1.8 are achieved by applying a cut on the NN classifier that maximizes their product.

\begin{figure}[!h]
    \centering
    \includegraphics[width=0.7\linewidth]{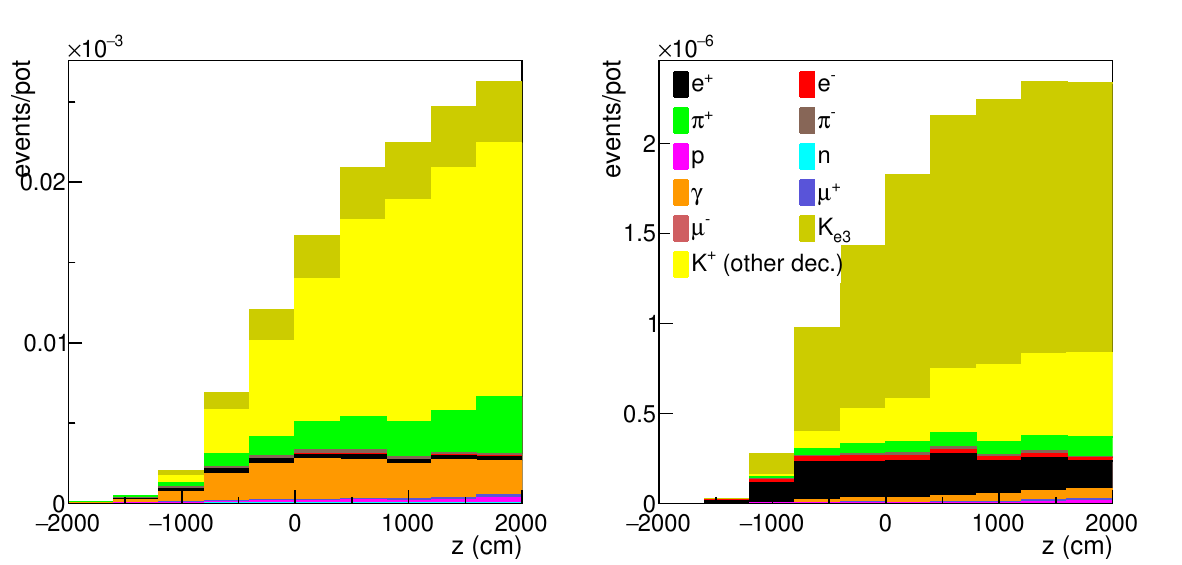}
    \caption{Longitudinal position of the reconstructed events in the $K_{e3}$ positron selection, before (left) and after (right) the cut on the NN classifier.}
    \label{fig:Z_positron}
\end{figure}

Figure~\ref{fig:Z_positron} reports the longitudinal position along the decay tunnel of reconstructed events before and after the cut on the NN classifier. The dominant background at the event building level, represented by hadronic decays of kaons (in yellow), non-collimated pions coming directly from the target or early decays in the transfer line (in green), and photons from the beamline (in orange) are efficiently suppressed by the NN classifier, while halo positrons produced in the beamline and transported to the walls of the tunnel (black) are left as the main component of the background.

\begin{figure}[!h]
    \centering
    \includegraphics[width=0.55\linewidth]{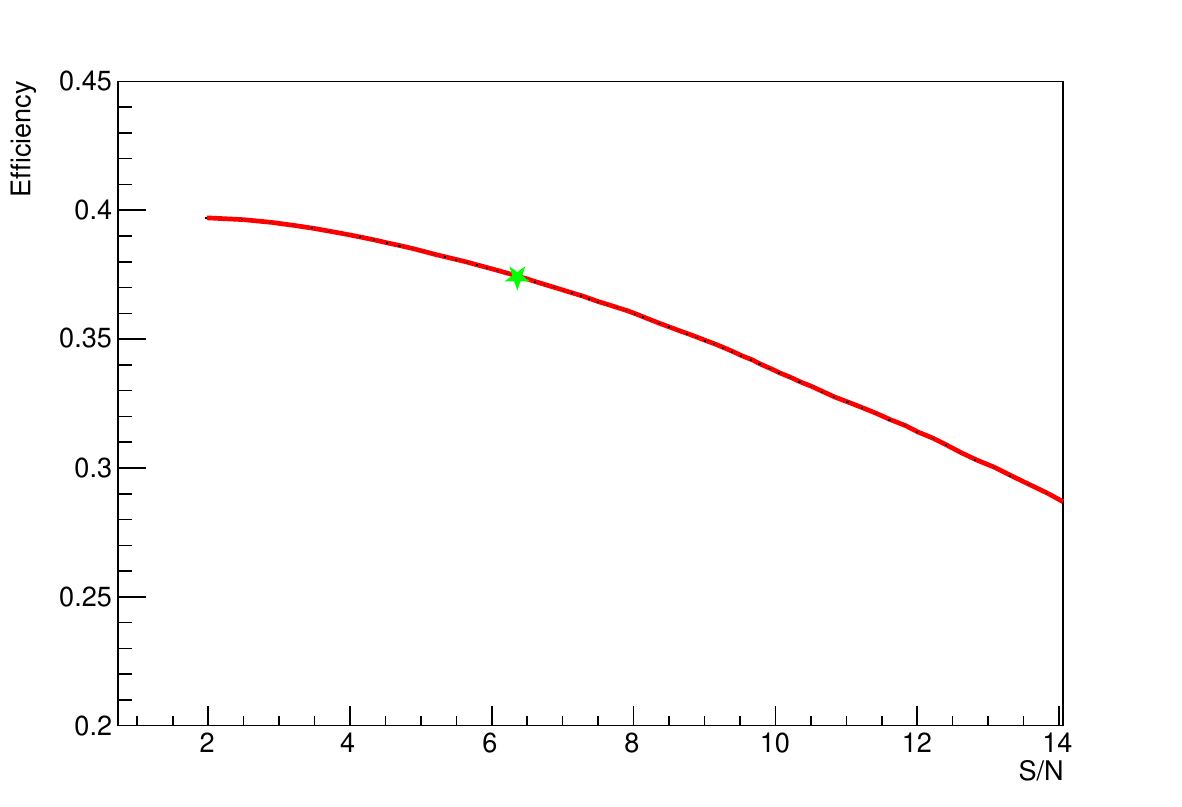}
    \caption{Signal efficiency versus signal-to-noise ratio for the $K_{\mu\nu}$ event selection. The green marker corresponds to the working point for signal selection, that is, the point in the curve maximizing the product between signal efficiency and purity.}
    \label{fig:PID_muon}
\end{figure}

Muons from kaon decays (both $K_{\mu2}$ and $K_{\mu3}$) are identified through a building algorithm that promotes as event seed an energy deposition in an LCM in the innermost layer of the calorimeter compatible with that of a m.i.p. All other LCMs and photon veto doublets, whose energy deposits are compatible in space and time with the propagation line of a muon, are grouped in a muon event candidate. The peculiar m.i.p.-like topology of a muon allows an excellent suppression of the e.m. and hadronic background already at the event builder level. A further reduction is achieved with a Neural Network that exploits variables accounting for the energy deposition pattern in the calorimeter and around the clustered track to further reduce the $\pi \rightarrow \mu$ misidentification, and topological variables to suppress the halo muon background originating in the beamline, which is mostly affecting the first half of the decay tunnel.
The ROC curve for the $K_{\mu\nu}$ event selection is shown in Fig.~\ref{fig:PID_muon}: at the working point, the signal selection efficiency, including the geometrical acceptance, amounts to 37.4\% with a $\textrm{S/N} = 6.36$.

\begin{figure}[!h]
    \centering
    \includegraphics[width=0.7\linewidth]{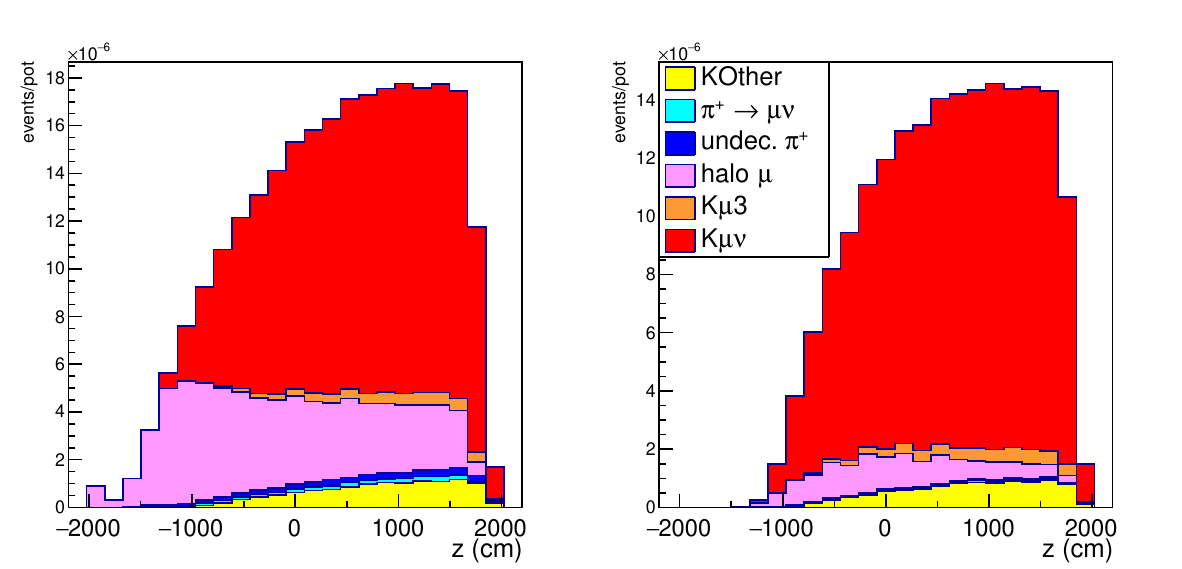}
    \caption{Longitudinal position of the reconstructed events in the $K_{\mu\nu}$ muon selection, before (left) and after (right) the cut on the NN classifier.}
    \label{fig:Z_muon}
\end{figure}

In Fig.~\ref{fig:Z_muon}, the longitudinal position of reconstructed events is shown before and after the cut on the NN classifier: the dominant background is represented by halo muons with a similar topology to muons coming from kaon decays.

\subsection{Instrumented hadron dump}
\label{sec:instrumented_hadron_dump}

The low-energy component of the $\nu_\mu$ flux is produced by $\pi^+$ decays. This flux cannot be monitored through the instrumented tunnel because the muons produced in pairs with the neutrinos in pion decays are almost collinear with the axis of the tunnel, and thus outside the geometric acceptance of the calorimeter. A constraint of this flux component can only be achieved by installing detectors downstream of the hadron dump, allowing the measurement of forward muons exiting the tunnel. This places very stringent requirements on the detector technology and the optimisation of the detector layout. Nevertheless, the narrow-band beam and the slow proton extraction time of the nuSCOPE facility are useful for identifying these muons and achieving precise control of the neutrino flux. 

We plan to install different detector layers at increasing depth, interleaved by absorbing material, so that the spatial distribution and energy of the muons can be measured by range out. The detector layers will be installed after the hadron dump, which is dimensioned to absorb the entire hadronic component of the beam from the calorimetric tunnel. By exploiting the differences in shape between the signal and background contributions, dominated by halo-muons, the rate of muons from pion decays can be estimated from a fit to the observables. The estimated rate is a proxy for the low-energy neutrino flux.  

A preliminary configuration of the muon detection system has already been studied. The hadron dump is a 2 m long iron slab. The first muon station is placed after the dump, followed by another iron absorber of the same length as the hadron dump. Seven additional muon stations follow the first station, interspersed with 0.5 m long iron absorbers. In this configuration, the muon spectrum can be measured by exploiting their range out through the absorbers. The total number of muon stations was chosen so that no muons are left after the last station.

\begin{figure}[!htp]
	\centering
	\begin{subfigure}{0.4\textwidth}
		\centering
		\includegraphics[width=\textwidth]{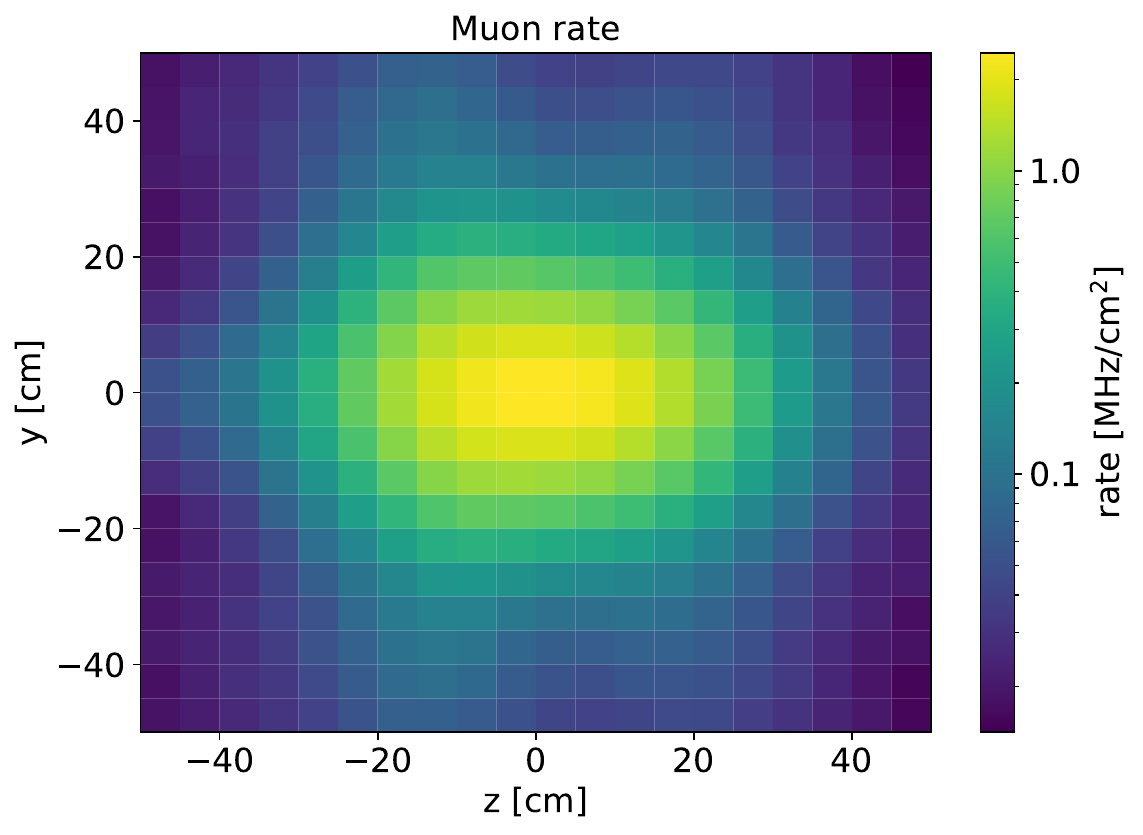}
	\end{subfigure}
        \begin{subfigure}{0.4\textwidth}
        \includegraphics[width=\textwidth]{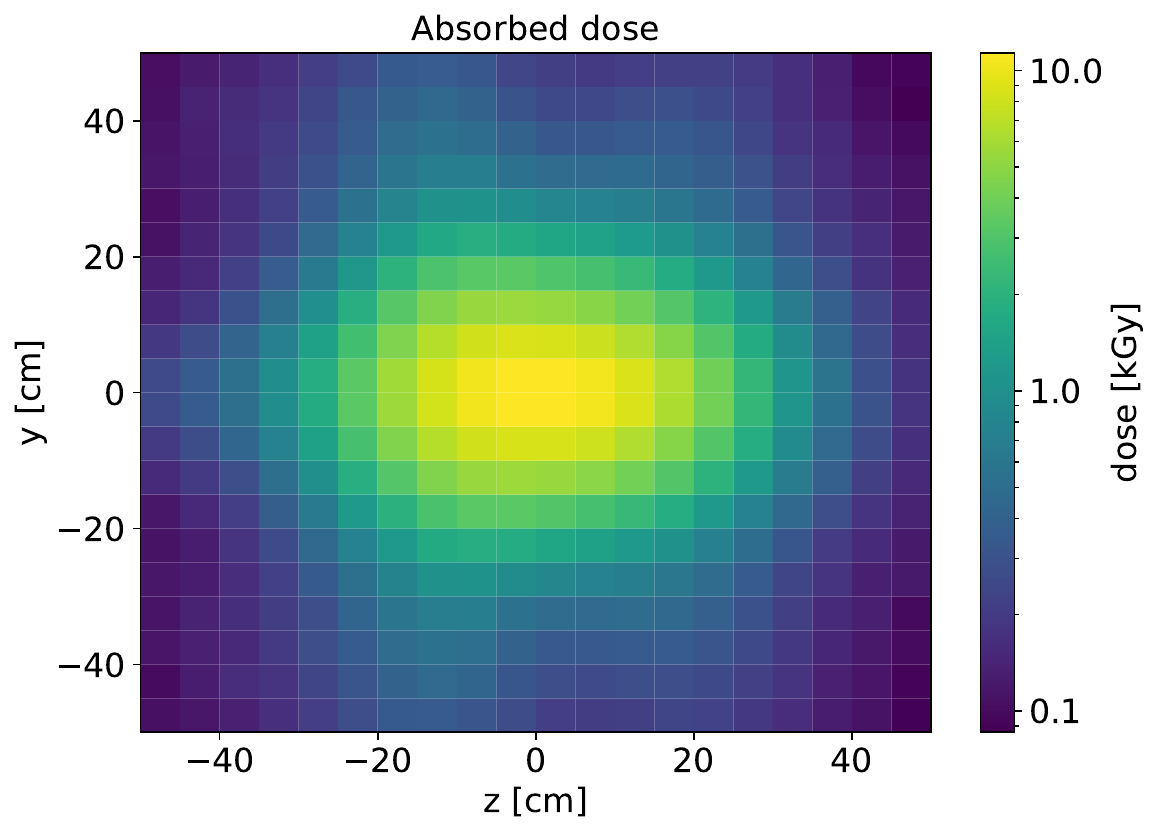}
        \end{subfigure}    
    \caption{Muon rate (left) and absorbed dose (right) on the first muon detector station in the instrumented hadron-dump. Each bin in the distribution corresponds to an area of 5$\times$5~cm$^2$.}
    \label{fig:hdump}
\end{figure}

The GEANT4 simulation of the muon system shows that the muon rate at the first muon station, where the highest rates are observed, has a maximum value of 2.5 MHz/cm$^2$, corresponding to the bulk of the muon beam, as shown in Fig.~\ref{fig:hdump}. The evaluated rate is within the specifications of a high-granularity fast tracking detector such as the MicroMegas (MM) or picosecond MM \cite{Kallitsopoulou:2024dmk}. In Fig.~\ref{fig:hdump}, the expected absorbed dose is shown, whose peak value amounts to 10~kGy. Furthermore, the simulation shows that at the first muon station, the pion punch-through is 0.5~$\%$ and the neutron fluence is of the order of 10$^{13}$ 1 MeV n$_{eq}$/cm$^2$ integrated over the entire data acquisition period. The absorbed dose and neutron fluence can be further reduced by tuning the length of the hadron-dump with a dedicated optimization study using FLUKA to better estimate their values. Tests with prototype detectors will be carried out in 2025 to evaluate the final detector technology.

\subsection{Spectrometers for neutrino tagging}
\label{sec:silicon_detectors}
\sisetup{parse-numbers=false}

In order to kinematically reconstruct the neutrinos produced in $K^+\to\mu^+\nu_\mu$ and $\pi^+\to\mu^+\nu_\mu$ decays, the beamline is instrumented with two spectrometers measuring the momentum, direction, time, and position of charged beam particles and their decay products. These devices, referred to as the \textit{beam} and \textit{muon} spectrometers, are positioned upstream and downstream of the decay tunnel, respectively.
Each spectrometer consists of a dipole magnet installed between two pairs of silicon pixel tracking planes.

\begin{figure}[!hb]
    \centering
    \includegraphics[width=0.9\linewidth]{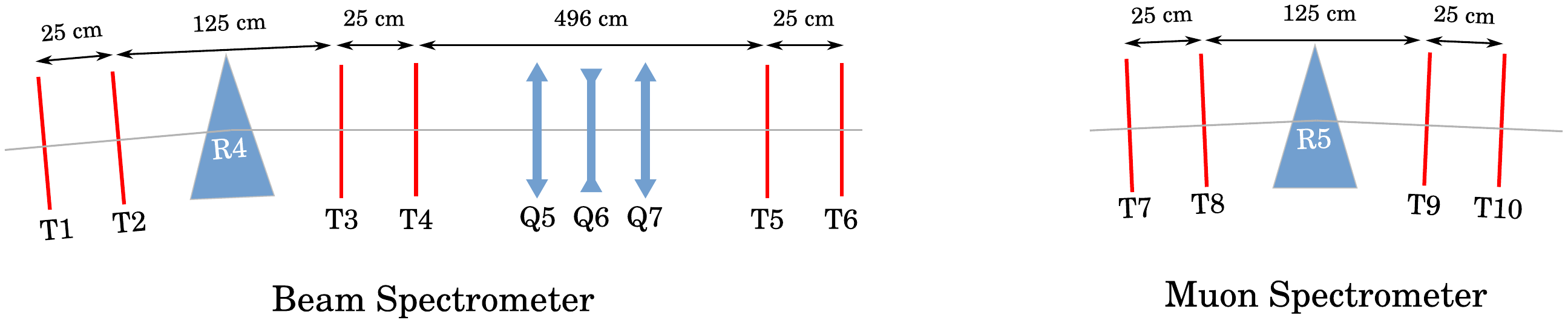}
    \caption{Layout of the beam (left) and muon (right) spectrometers.}
    \label{fig:spectrometers}
\end{figure}

To minimize the length of the region upstream of the decay tunnel, where $K^+$ and $\pi^+$ decays cannot be reconstructed, the beam spectrometer exploits the last dipole magnet (R4) of the frontend beamline. Since the field map of the quadrupole triplet (Q5-7) downstream of R4 may vary, the beam spectrometer includes two additional tracking planes (T5-6), positioned at the decay tunnel entrance, to precisely measure the beam particle direction.
The spectrometer reference layouts are shown in \autoref{fig:spectrometers}. 

The spectrometer's momentum and angular resolutions are essential for the tagging technique, as they allow for the correct association of the neutrino interaction with its parent decay. With a minor adjustment of the tracking plane spacing, the contribution of the tracking plane's spatial resolution to the spectrometer's momentum and angular resolutions can be made smaller than that from multiple Coulomb scattering. A material budget per plane of approximately \SI{1}{\%} of a radiation length is sufficient to meet the requirements of the tagging technique. Simulations used for this study assumed a material budget of \SI{0.5}{\%} of a radiation length.

\begin{figure}
    \centering
    \includegraphics[width=0.32\linewidth]{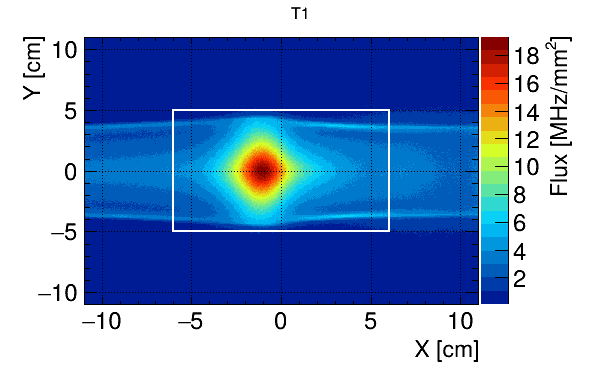}
    \includegraphics[width=0.32\linewidth]{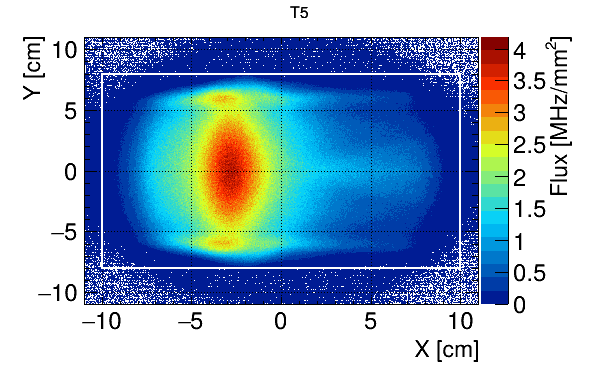}
    \includegraphics[width=0.32\linewidth]{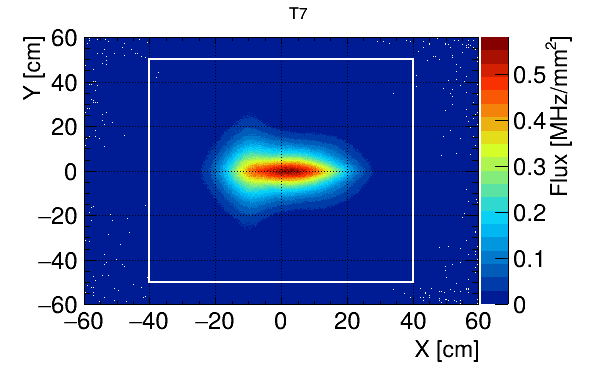}
    \caption{Spatial distribution of the charged particles in the plane transverse to the beamline at T1, T5, and T7 overlaid with the tracking plane acceptances (white line).}
    \label{fig:beam_profile}
\end{figure}

The tracking plane sizes were chosen to provide the largest ratio between the number of reconstructed $K^+\to\mu^+\nu_\mu$ and $\pi^+\to\mu^+\nu_\mu$ decays and the total tracking plane surface. The spatial distribution of the charged particle flux in the plane transverse to the beamline is shown for T1, T5, and T7 in \autoref{fig:beam_profile}. Tracking planes for the muon spectrometer are notably larger than for the beam spectrometer, due to beam divergence and the muon emission angle. The tracking plane surfaces are \SI{12\times10}{cm^2} in T1-4, \SI{20\times16}{cm^2} in T5-6 and \SI{80\times100}{cm^2} in T7-10. 
The particle flux at T1 is the highest. With \SI{9.6}{s} spills of \SI{10^{13}}{PoTs}, it reaches \SI{20}{MHz/mm^2} at the center of the tracking plane. In the muon spectrometer, the flux is significantly lower with a peak flux of \SI{0.6}{MHz/mm^2}. 

Track reconstruction will be performed in 4D, exploiting the precise measurement of the particles' position and time at the tracking planes. The integrated rate of particles on the beam spectrometer planes reaches up to \SI{45}{\giga\hertz} and \SI{20}{\giga\hertz} on the muon spectrometer ones. 
Time resolutions of \SI{\sim 40}{ps} for the beam spectrometer tracking planes and \SI{\sim 100}{ps} for the muon spectrometer ones should allow for the track and decay reconstructions. Reconstruction algorithms are currently being developed based on a global event reconstruction using Kalman filters, and will consolidate these specifications. The list of all trackers' specifications is reported in \autoref{tab:tagging-specs}.

\begin{table}[h!]
\renewcommand{\arraystretch}{1.2}
	\centering
	\begin{tabular}{r|c|c|c|c}
		\hline\hline
		                                           & \textbf{Beam}               & \textbf{Muon}     & \textbf{LHCb-VELO} & \textbf{NA62-GTK}  \\
		\textbf{Specifications [units]}           & \textbf{Spectro.}           & \textbf{Spectro.} & (2028)             & (since 2014)       \\\hline
		Peak Dose            [\unit{Mrad}]             &  $ 700 $                    & $ 60 $                & $>10^3$            & $16$               \\\hline
		Peak Fluence         [\unit{1MeV n_{eq}/cm^2}] &  $ 1\times10^{16}$          & $ 6\times10^{14}$      & $5\times10^{16}$   & $4.5\times10^{14}$ \\\hline
		Peak Rate            [\unit{MHz/mm^2}]  & $20$                 & $0.6$    & $10-100$           & $2$               \\\hline
                                               
		Time Resolution [\unit{ps}]               & $<40$                       & $<100$            & $<50$              & $<130$             \\\hline
		Pixel Pitch     [\unit{\mu m}]            & \multicolumn{2}{c|}{$300$}                     & $45$              & $300$               \\\hline
		Material Budget [\unit{X_0}]              & \multicolumn{2}{c|}{$<1\%$}                      & $0.8\%$         & 0.5\%                 \\\hline\hline
	\end{tabular}
	\caption{Tracker specifications of the beam and muon spectrometers compared to the present NA62 Gigatracker and the future upgraded LHCb VELO. Simulations of the neutrino tagging presented in this document assume a material budget of \SI{0.5}{\%} of a radiation length and a spill duration of \SI{9.6}{s}.}
	\label{tab:tagging-specs}
\end{table}

The 4D tracking approach has been successfully pioneered by the NA62 collaboration, which has developed \SI{3 \times 6}{\centi\meter\squared} silicon pixel detectors with a pitch of \SI{300}{\micro\meter} and a time resolution of \SI{130}{\pico\second}~\cite{AglieriRinella:2019eri}. These detectors are capable of operating at a peak flux of  \SI{2}{MHz/mm^2}.

Tracking planes for the muon spectrometer could use this technology either in a fully silicon-based configuration or in a hybrid approach, where silicon is used for the most illuminated region while more scalable technologies -- such as Micromegas, straw tubes, or scintillating fibers -- are employed for the outer regions.

New silicon technologies are required to match the flux at the beam spectrometer. In the context of the LHC experiment upgrades for the High Luminosity phase -- particularly the second upgrade of the LHCb Vertex Locator (VELO) -- R\&D programs for high-intensity 4D trackers were initiated and later joined by the NuTag collaboration. 

Such pixel detectors consist of a silicon sensor hybridized with readout electronics made of application-specific integrated circuits (ASICs). These devices must be extremely thin, as the total material budget allows for less than \SI{500}{\micro\meter} of silicon per tracking plane. Under these constraints, thermal management of the detector is particularly challenging. The readout ASIC dissipates \SIrange{1}{3}{\watt/\centi\meter\squared}, and its power ultimately determines the time resolution.

Currently, the only technological solution capable of providing the required cooling power while meeting the stringent material budget constraints is micro-channel cooling plates, such as those developed for the NA62 GigaTracker and the LHCb VELO. These devices consist of a silicon plate a few hundred microns thick, encapsulating micro-channels through which a refrigerant circulates. The refrigerant can be either liquid or evaporative; the latter enhances cooling performance but requires the circuit to withstand pressures of a few hundred bars, necessitating thicker channel walls.

Technological solutions on pixel sensors and integrated electronics, capable of standing the challenge in terms of time resolution and radiation hardness, have been developed and are still under study by the INFN TimeSPOT (2018-2021 and follow-ups) and IGNITE (2023, ongoing) projects. 

TimeSPOT has developed 3D silicon sensors, based on the so-called trench geometry. Two batches of such 3D-trench sensors were produced by the FBK foundry (Trento, Italy), respectively in 2019 and 2021. Both productions contain several test structures with a variable number of pixels, including also matrices of relatively small size (32$\times$32 pixels), all featuring pixels of \SI{55}{\micro\meter} pitch. An additional batch has been recently delivered (end 2024) and will be tested soon. This last production contains several process improvements, aimed at a better production yield, and contains pixel matrices up to 128$\times$128 pixels, always with a \SI{55}{\micro\meter} pitch.

3D-trench sensors of the 2019 and 2021 batches were characterized both in the laboratory (infrared laser scans) and dedicated test beams at PSI-Villigen and SPS-CERN. Details on experimental methodologies and results can be found in \cite{Borgato10e16,Addison10e17}. 3D-trench sensors demonstrated outstanding performance both in time resolution and radiation resistance. They are capable of an intrinsic time resolution below \SI{10}{\pico\second} and withstand a fluence of at least \SI{10^{17}}{n_{eq}\per\centi\meter\squared}, without significant loss in timing and efficiency performance, after a moderate increase in bias voltage (from 100 V before irradiation to 250 V at \SI{10^{17}}{n_{eq}\per\centi\meter\squared}). Recent results \cite{VCI2025_sensors} show that 3D-trench sensors are still satisfactorily operating up to \SI{10^{18}}{n_{eq}\per\centi\meter\squared}, with a voltage increase above 400 V.

The characterization campaigns on the aforementioned silicon sensors have been performed using dedicated discrete-component electronics, based on high-speed (and high-consumption) Si-Ge bipolar transistors \cite{CossuLai2023}. 
After having identified a suitable pixel technology for neutrino tagging, it is necessary to develop suitable integrated electronics capable of maintaining high timing performance in the same harsh conditions (rate, fluence) and using limited power.

The same TimeSPOT initiative has started to develop integrated electronics with pixel readout in CMOS $\SI{28}{nm}$ technology since 2020. This technology is presently the only validated one \cite{Termo2024} against the high dose expected in a neutrino tagging apparatus (above 1 Grad). The TimeSPOT starting developments \cite{Cadeddu2023} are being finalized by the IGNITE project \cite{VCI2025_elec}. IGNITE is developing an ASIC to read out a 256$\times$256 pixel matrix, with a \SI{45}{\micro\meter} pitch. The IGNITE ASIC integrates one fast amplifier and one TDC per pixel. Pre-processing capabilities and large bandwidth data readout facilities (100 Gbps) are also integrated. The ASIC is being designed mainly following the LHCb VELO Upgrade2 specifications~\cite{hep-ph_LHCb_2021}, which are summarized in Table~\ref{tab:tagging-specs}. Such specifications are very close, although not identical, to the neutrino tagging requirements. 

A parallel development named LA-Picopix is presently ongoing at CERN. Both the IGNITE and the LA-Picopix developments aim to deliver their ASICs in 2026. They target the same specifications with some relevant differences. LA-Picopix will have a 50~{\textmu}m pitch, IGNITE a 45~{\textmu}m pitch. IGNITE should also be capable of a higher rate per pixel, having one TDC per pixel instead of one serving 8 channels in LA-Picopix. On the other hand, LA-Picopix aims at a traditional ASIC implementation, while IGNITE will use a 3D-integration ASIC stack to allow better power and timing reference distributions, for optimal time resolution.

High-Intensity 4D-Tracking techniques, satisfying the requirement given in Table \ref{tab:tagging-specs}, are technically extremely demanding. For this reason, having two parallel developments that face such a challenge with alternative approaches is an advantage. The decision about which approach will be chosen and/or if the two projects will merge into a single one is planned for the end of 2026, to close the implementation of the final ASIC by 2028.

\sisetup{parse-numbers=true}
\section{Neutrino detectors}
\label{sec:neutrino_detectors}

nuSCOPE aims to be the reference facility for neutrino cross-section measurements in the DUNE and Hyper-Kamiokande era. To maximize its potential, the facility requires detectors that can fully leverage the unprecedented precision of the beam while maintaining an optimal target for next-generation long-baseline cross-section studies. The two primary detectors will therefore utilize argon and water targets, eliminating systematic uncertainties associated with extrapolating nuclear effects across different atomic numbers ($Z$). The detailed detector choice will also depend on the site implementation of nuSCOPE, but the total (fiducial) detector mass is requested to be in the 500-1000 ton range, given the intensity and integrated protons of nuSCOPE. 

\subsection{Liquid-argon detector}
\label{sec:detectors_liquidargon}

The leading candidates for the argon-based detectors \cite{Bonivento:2024qpn} at nuSCOPE are the ProtoDUNEs \cite{DUNE:2021hwx} and their future upgrades. During LHC Run 4, the existing ProtoDUNEs will undergo further upgrades to support the validation of the DUNE Phase II detectors \cite{DUNE:2024wvj}. Compared to the current ProtoDUNEs, all proposed technologies share a common set of characteristics. Most importantly, they preserve the imaging and particle reconstruction capabilities of DUNE at the GeV scale, ensuring that the primary objective of Phase II—precision observation of neutrino oscillations with the LBNF beam—remains fully achievable. A key advancement in the Phase-II detectors is the enhanced performance of the photon detection system, an upgrade also planned for DUNE’s third module, which is expected to be based on APEX or POWER technologies \cite{DUNE:2024wvj,Steklain:2025flt}. This new photon detection system will play a crucial role in determining DUNE’s energy resolution at the few-MeV scale, significantly extending the experiment’s physics reach in solar and supernova neutrino studies while also providing additional sensitivity to BSM searches.
The fourth module, often referred to as the ``Module of Opportunity,'' is still in the R\&D phase. In addition to improved photon detection, it is expected to incorporate a pixelated charge readout, which would enhance spatial resolution, lower the energy threshold at the MeV scale, improve trigger efficiency by voxelizing the fiducial volume, and strengthen the identification of low-energy photons while improving radiogenic background rejection.

None of these improvements are mandatory to achieve the physics goals of nuSCOPE, as the DUNE Phase I detectors—and consequently the ProtoDUNEs—were optimized for optimal performance in the 0.5–5 GeV range. However, the capability to tag neutrinos beyond monitoring requires a time resolution better than $\SI{1}{ns}$ in the neutrino detectors. Since the time resolution for GeV-scale neutrino events critically depends on the number of detected photoelectrons, enhancing the photon detection system (PDS) becomes essential to enable neutrino tagging in liquid argon.

As a result, the ideal detector for nuSCOPE would be one of the current ProtoDUNEs equipped with an upgraded timing system derived from the Phase II upgrades. Additionally, the implementation of a dedicated cosmic ray veto system to enhance cosmic background suppression is particularly valuable, given that nuSCOPE operates with a slow proton extraction. The design of this veto system, along with the potential installation of a muon catcher for full event containment, is currently under evaluation.

\subsection{Water-based detector}

Water is the target material for Hyper-Kamiokande, making a detector based on this target an invaluable tool for achieving the nuSCOPE physics goals through precise neutrino interaction measurements in water. A widely used option, water Cherenkov detectors, offer less precise reconstruction of neutrino interactions at the GeV scale compared to liquid argon detectors, but this limitation is compensated by reduced complexity and cost.

Tests of Water Cherenkov detector performance using charged beams—mirroring the ProtoDUNE physics program—are currently ongoing at CERN with the WCTE detector \cite{WCTE_proposal}. WCTE is a 40-ton demonstrator of the Intermediate Water Cherenkov Detector (IWCD) \cite{Scott:2016kdg}, which is currently under construction in Japan. Water Cherenkov technology offers an ideal setting for nuSCOPE operations, even in tagged mode, as it can achieve sub-nanosecond time resolution at large scales with limited light coverage, due to the inherently faster propagation of Cherenkov light compared to liquid argon scintillation light.
The WCTE/IWCD technology employs multi-PMT modules instead of the classical 20-inch PMTs used in Hyper-Kamiokande, enhancing energy linearity over a broader dynamic range and thereby improving the precision of cross-section measurements. However, the 40-ton size of WCTE is insufficient to complete the full nuSCOPE physics program in water; a dedicated detector with a fiducial volume of a few hundred tons would be required.
Further enhancement of WCTE’s particle identification capability could be achieved by replacing the water target with modern water-based scintillators. This novel technology, pursued by several experimental groups, aims to overcome the limitations of Water Cherenkov detectors in medium- and large-scale experiments. It is also under consideration for the aforementioned DUNE Module of Opportunity, an initiative led by the THEIA collaborations in the US and Europe.

\subsection{Reference detector parameters}
\label{sec:detector_parameters}

The studies presented here are based on a reference liquid argon (LAr) detector with a 500-ton fiducial mass, positioned 25~m from the end of the decay tunnel. The front face measures  $4 \times4$ m$^2$, with a total length of 22.3~m.
The LAr TPC is assumed to have a time resolution improved to 300~ps compared to the current ProtoDUNE, thanks to an enhanced photon detection system as planned for DUNE Phase II upgrades.
The final detector optimization is still a work in progress and will depend on the chosen installation site for nuSCOPE.

\section{Flux characterization}
\label{sec:flux_characterization}

\subsection{Flux constraints from lepton monitoring}
\label{sec:flux_from_lepton_monitoring}

In conventional neutrino beams, the knowledge of the neutrino flux at the detector is limited to a 5-10\% level \cite{T2K:2012bge,MINERvA:2016iqn}. The dominant systematic is given by the uncertainties in the yield of mesons produced in the interaction of the primary proton beam with the target, known as the hadroproduction systematics. This limitation is overcome by the monitoring technique, where the charged leptons produced together with the neutrinos in the meson decays are measured directly in the instrumented decay tunnel. The monitoring technique allows for bypassing the hadroproduction systematics and most beam-related systematics, since the measured lepton rates are directly correlated with the neutrino flux at the detector. In this way, a high gain in the precision of the neutrino flux can be achieved.

A proof-of-concept study has been developed to assess the performance of the monitoring technique in constraining the neutrino flux. This is based on a realistic hadroproduction model fitted to the data from the NA56/SPY and NA20 experiments, which took data with 450 and $\si{400}{GeV/c}$ proton beams on target, respectively \cite{NA56SPY:1999zez,Atherton:1980vj}. The fitted parameters and corresponding covariance matrices are used to propagate the hadroproduction uncertainties to the lepton observables measured in the instrumented decay tunnel and to the neutrino flux. The propagation is done by applying the multi-universe method, and exploiting the link between the mesons exiting the target, the leptons measured in the calorimeter, and the neutrinos crossing the detector, which is provided by the simulation of the facility. The covariance matrices for the lepton observables and the neutrino flux induced by the hadroproduction uncertainty are computed from the propagation. They encode the hadroproduction uncertainty before applying any constraint from the lepton monitoring (pre-fit result). A signal plus background model is built for the charged leptons measured in the instrumented decay tunnel. For this purpose, we use the distribution of lepton observables obtained from the MC events, reweighted with the hadroproduction model. From each universe, corresponding to a possible realization of the hadroproduction values, a set of pseudo-data is generated and fitted by building an extended maximum likelihood (EML) from the model PDF. The covariance matrices for the lepton observables are used as our a priori knowledge to constrain the fit. A software toolkit, implemented in C++ and using the Minuit2 package, was developed to construct the model PDF, perform the EML fits, and to generate pseudo-data. The strong correlation between the lepton observables and the neutrino flux at the detector, together with the result of the fitted number of leptons, are used to place a constraint on the neutrino flux. The residual hadroproduction uncertainties after the fit are propagated to the neutrino flux, and the covariance matrix after the constraint from lepton monitoring is computed (post-fit result). From the post-fit covariance matrix, the residual systematic on the neutrino flux due to hadroproduction is of $\mathcal{O}(1\%)$, for both $\nu_e$ and $\nu_{\mu}$ \cite{Longhin:NBI2024}.

The same strategy can be adopted to determine the impact of the detector effects and beamline subdominant systematics on the neutrino flux. Three different types of detector effects are considered for the study: calibration uncertainty and bias of the calorimetric channels, and ageing of the scintillator material used in the instrumentation of the sampling calorimeter. A conservative value of 5\% was assumed for the calibration uncertainty and bias. The effect of scintillator ageing is parameterized by the total drop in the light yield after the full data-taking period and the error in the light yield measurement from the monitoring system. Again, conservative values of 10\% for the light yield drop and 1\% for the light monitoring error are assumed. The detector effect systematics are injected into the output simulation of the instrumented decay tunnel, by varying all energy depositions in each channel depending on the detector effect under consideration. The full analysis chain, lepton reconstruction, and PID, are applied to the simulated data, embedding the detector's systematic effects. The previously described workflow for assessing the impact of lepton monitoring on the neutrino flux is applied to this new data set, and the residual hadroproduction uncertainties, including detector systematic effects, are computed. The result for the $\nu_{\mu}$ flux, from the comparison of the post-fit neutrino flux for the hadroproduction only systematics and the detector effects systematics, shows that the detector effects on the neutrino flux are subdominant compared to the residual hadroproduction systematics. The study is currently ongoing for the $\nu_{e}$ flux.

\subsection{Flux constraints from tagging}
\sisetup{parse-numbers=false}

The goal of the tagging is to fully control the muon neutrino flux impinging on the neutrino detector by measuring each mesonic decay of the beam and associating each observed neutrino interaction with its parent mesonic decay.
The method is based on two quantities:
\begin{itemize}
\item taggable beam neutrinos: beam neutrinos within the fiducial acceptance of the neutrino detector, whose parent meson and decay muon are within the acceptance of the beam and muon spectrometers, respectively.
\item tagged neutrino interaction: neutrino interaction observed in the neutrino detector matching a single measured parent mesonic decay.\end{itemize}

The performance of the tagging has been quantitatively evaluated with a full simulation of the beam secondary particle propagation within the optimized optics presented in Sec. \ref{sec:beamline_optimization}, the reference beam and muon spectrometer trackers presented in Sec. \ref{sec:silicon_detectors}, and the reference LAr neutrino detector of Sec. \ref{sec:detector_parameters}. 

\subsubsection{Taggable beam neutrino flux}

\begin{figure}[h!]
    \centering
    \includegraphics[width=0.7\textwidth]{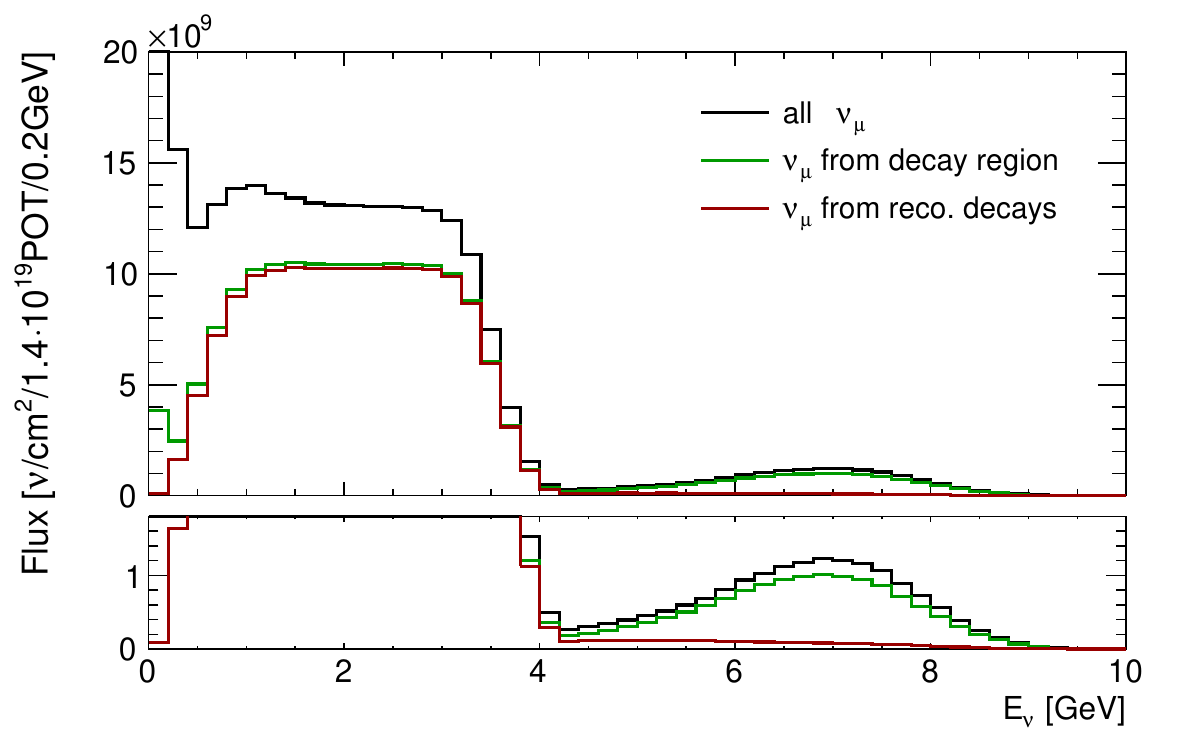}
    \caption{Beam neutrino fluxes within the fiducial acceptance of the neutrino detector, as a function of energy. Black: all neutrinos; green: neutrinos from decays within the decay tunnel; red: same as green with, in addition, the parent meson and decay muon in acceptance of the beam and muon spectrometers, respectively ("taggable beam neutrinos").} 
    \label{fig:tagging_fluxes}
\end{figure}

Since the tracking planes of the most downstream beam spectrometer and the most upstream muon spectrometer are located immediately upstream and downstream of the decay tunnel, respectively, a taggable beam neutrino must be produced within the decay tunnel. Figure~\ref{fig:tagging_fluxes} compares the flux of all neutrinos reaching the detector to the flux of neutrinos produced in the decay tunnel, as well as to the flux of those that can additionally be tagged by the beam and muon spectrometers. 

\begin{figure}[h!]
    \centering
    \includegraphics[width=0.5\textwidth]{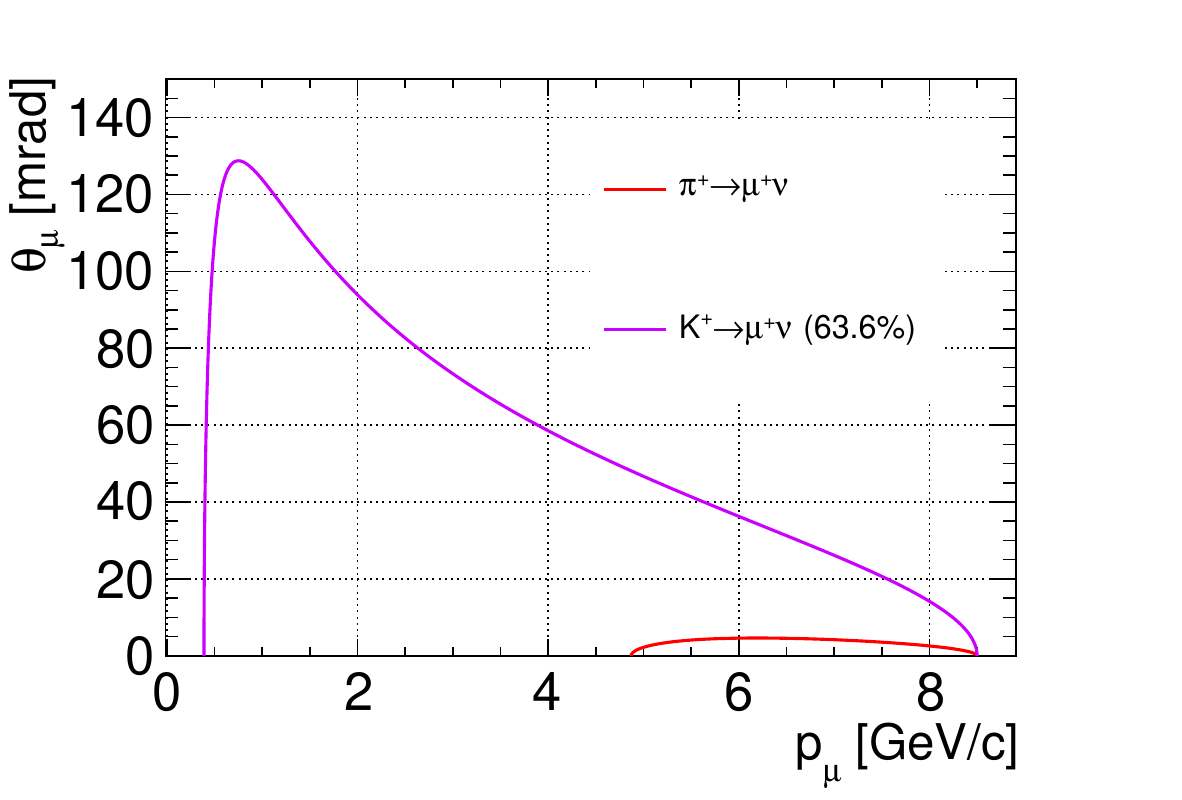}
    \caption{Correlation between the secondary muon laboratory angle $\theta_{\mu}$ and momentum $p_{\mu}$ for \SI{8.5}{GeV/c} $K$ and $\pi$-decays.}
    \label{fig:tagging_kinematics}
\end{figure}

The contribution of neutrinos produced outside the decay tunnel concentrates at low energy.
The energy spectrum exhibits two domains below and above $\sim$\SI{4.5}{GeV}, dominated by $\pi$ and $K$ decays, respectively.
The efficiency of the kinematic reconstruction of $\pi$-decays is high, $\mathcal{O}$(70\%), whereas that of $K$-decays is lower, $\mathcal{O}$(10\%). This is due to the lower acceptance of the muon spectrometer to muons from $K$-decays, which are emitted at larger angles, as shown in Fig.~\ref{fig:tagging_kinematics}.

Figure~\ref{fig:tagging_kinematics} shows that $\pi$ and $K$ parent mesons can be fully discriminated based on the measured kinematics of each decay. This enables an additional method for controlling the beam $K$ flux (and consequently the $\nu_e$ spectrum), complementing the beam tunnel calorimeters (see Sec. \ref{sec:flux_from_lepton_monitoring}). This approach depends on maintaining control over the $K$ tagging inefficiency (i.e., the difference between the green and red curves in Fig.~\ref{fig:tagging_fluxes}). The corresponding expected precision on the $\nu_e$ flux has not been quantified yet.
The red curve of Fig.~\ref{fig:tagging_fluxes} defines the input flux of taggable beam muon neutrinos that will be used for the cross-section measurements.

\subsubsection{Tagged neutrino interactions}
\label{sec:tagged_neutrino_interactions}

Tagging a neutrino interaction measured in the detector consists of looking for measured meson decays that match the interaction vertex in time and space, both within $3\sigma$, where $\sigma$ corresponds to the detector resolution. The performance of the time coincidence is dominated by the time resolution of the LAr detector, expected to be \SI{300}{ps} in the reference detector of Sec.\ref{sec:detector_parameters}, i.e. much worse than the tagging trackers' time resolution discussed in Sec. \ref{sec:silicon_detectors}. The performance of the space coincidence is determined by the precision of the neutrino direction prediction and is dominated by multiple scattering in the tagging trackers. For the $0.5\%$ radiation length tracker planes, the precision is \SI{\mathcal{O}(1)}{mrad}, corresponding to \SI{\sim40}{mm} in the LAr detector, i.e. much worse than the LAr spatial resolution. 

\begin{figure}[h!]
    \centering
    \includegraphics[width=0.47\textwidth]{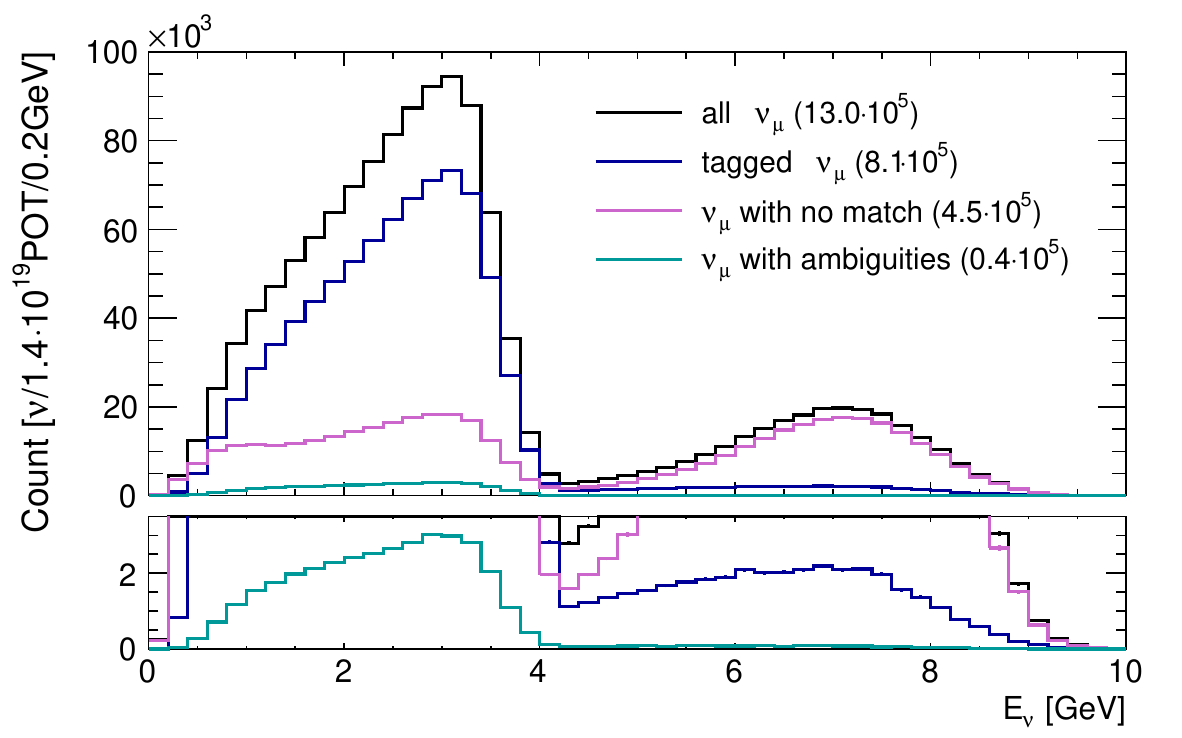}
    \includegraphics[width=0.47\textwidth]{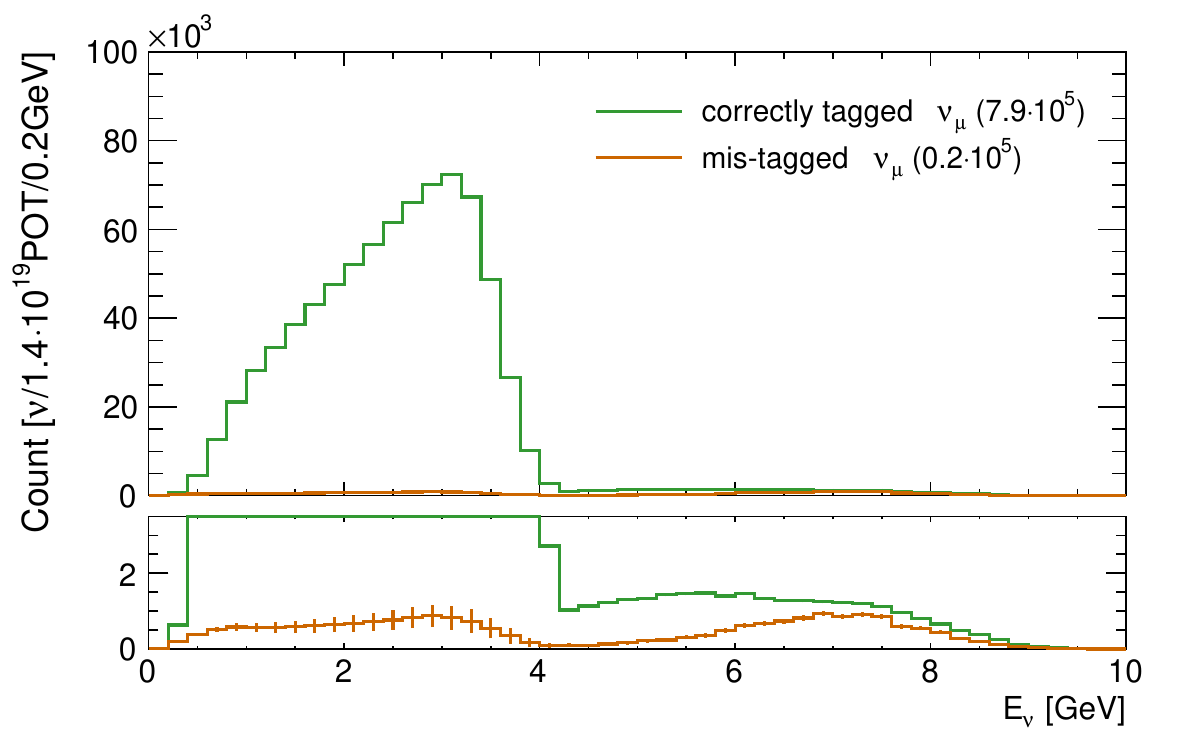}
    \caption{$\nu_\mu$ interaction rates as a function of energy. Left: The interaction sample (black) is divided into three categories: tagged neutrino interactions, i.e., matched to a single decay (dark-blue); interactions matched to no-decay (pink); interactions matched to multiple decays (light-blue). Right: The tagged sample is subdivided into two sub-samples: interactions associated with the correct meson decay (green); interactions associated with a wrong meson decay (orange). The numbers in the legend indicate the total number of interactions in each category.}
    \label{fig:tagging_efficiencies}
\end{figure}

The association of a given neutrino interaction with a measured meson decay can result in three possible outcomes: no match is found (no match case); multiple matches are found (ambiguous case); a single match is found (tagged case). In the tagged case, the associated decay can either be the correct one (correctly tagged case) or a pile-up decay (mistag case).
Fig.~\ref{fig:tagging_efficiencies} shows the distributions of these categories. 
It can be noted that for neutrinos with energies below \SI{4}{GeV} (i.e., mostly from pion decays), most of the observed neutrino interactions are associated with their correct parent meson decay, with wrong and ambiguous matches remaining at a level of a few \%. These performances significantly degrade for neutrinos with energies above \SI{4}{GeV} (i.e., mostly from kaon decays) due to the reduced acceptance of the spectrometer for large-angle muons produced in kaon decays.

 The observed neutrino interactions with a single match (dark-blue curve of Fig.~\ref{fig:tagging_efficiencies}) define the tagged neutrino interactions that will be used for the cross-section measurements.

\subsubsection{Cross-section measurement method}

Beam tagging enables the determination of $\nu_{\mu}$ cross-sections by directly comparing the number of tagged neutrino interactions with the measured input flux of taggable beam neutrinos, as defined in the previous sections. Notably, this method provides a direct measurement of the input neutrino flux, eliminating the need to rely on simulations of the hadronic components in beam production.

Some corrections are anyway needed. 
A significant number of measured neutrino interactions are expected to show no match. Most of them correspond to the non-taggable beam neutrinos that are in the acceptance of the detector, quantified by the difference between the black and red curves of Fig.~\ref{fig:tagging_fluxes}. These events can simply be ignored since they do not correspond to the input taggable beam neutrinos. However, a small correction has to be applied for the instrumental inefficiencies of the tagging trackers, which may also result in events with no match.
In addition, the contamination of wrong matches and the loss due to ambiguous matches (orange and light-blue curves in Fig.~\ref{fig:tagging_efficiencies}, respectively) must also be corrected. These corrections are performed with data-driven methods based on, e.g., variation of the matching criteria.  
Finally, as in all cross-section measurements, detector acceptances and efficiencies need to be accounted for.
The expected precision of the cross-section measurements is discussed in Sec. \ref{sec:cross-section} below.

\sisetup{parse-numbers=true}
\section{A priori measurement of the neutrino energy}
\label{sec:energy}

\subsection{The Narrow-Band Off-Axis technique}

Due to the proximity of the detector to the neutrino beam, and to the large size of the detector itself, the detector surface ($4 \times 4 ~\si{m}^2$) is exposed to a range of incoming neutrino directions.
The energy of \numu produced in $\pi^+ \to \mu^+ \nu_\mu$ and $K^+ \to \mu^+ \nu_\mu$ decays is correlated to the emission angle, due to the two-body decay kinematics. As a result, neutrinos interacting at different off-axis angles span different energy ranges.
The off-axis technique has already been exploited by a cross-section measurement from the T2K experiment~\cite{T2K:2023qjb}, and is central for the PRISM technique~\cite{nuPRISM:2014mzw}, which is planned for use in the DUNE~\cite{DUNE:2021tad} and Hyper-Kamiokande near detectors~\cite{Hyper-Kamiokande:2018ofw} as well as in the SBND experiment~\cite{delTutto2021SBND}.
A key advantage of a narrow-band beam like nuSCOPE is that the off-axis neutrino energy estimation can leverage the parent meson momentum knowledge at $\mathcal{O}$(10\%), enabling an a priori determination of the neutrino energy without relying on the reconstruction of final-state particles. This is the fundamental principle of the Narrow-Band Off-Axis (NBOA) technique, which can be applied to \numu fluxes from \pidecay and \kdecay decays.

\begin{figure}[!htp]
    \centering
    \includegraphics[width=0.5\linewidth]{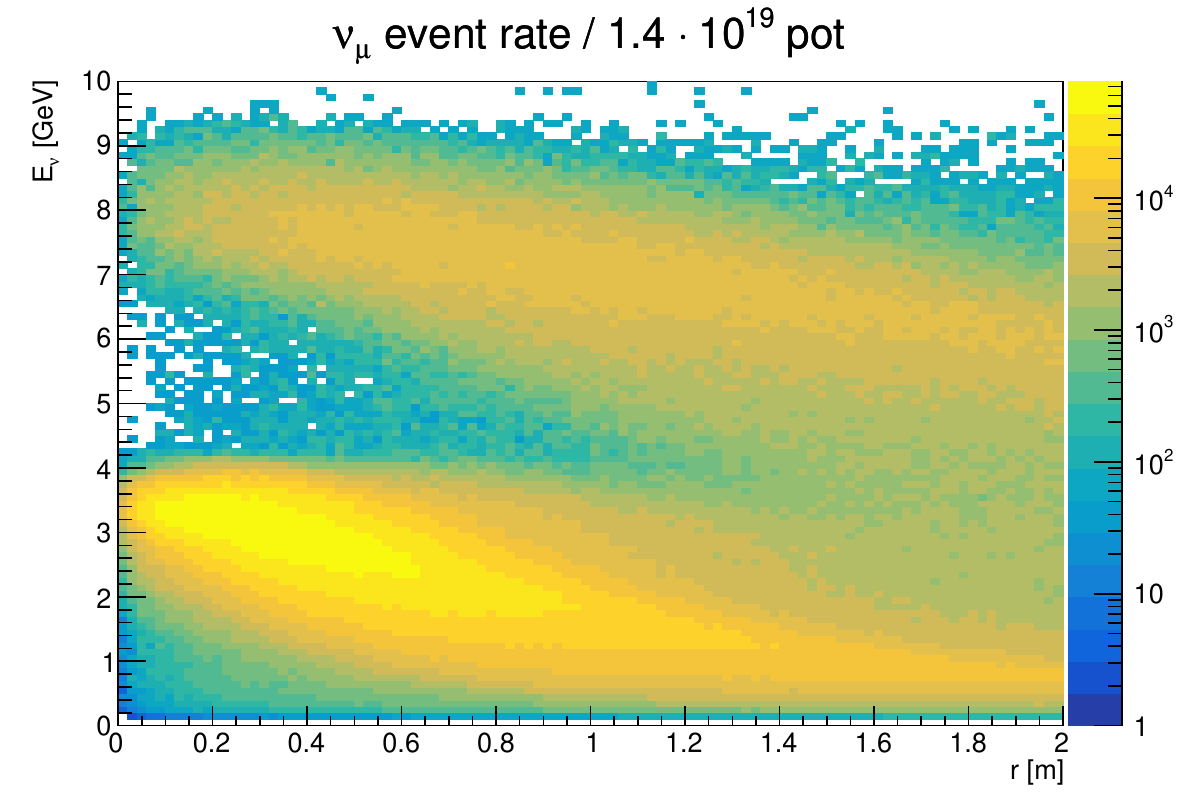}
    \caption{Distribution of neutrino energies from \pidecay and \kdecay decays as a function of the radial distance $r$ of the neutrino interaction vertex on the exposed surface of the detector.}
    \label{fig:enu_vs_r}
\end{figure}

Neutrinos interacting at different off-axis angles can be identified by defining a set of hollow cylinders within the detector volume, corresponding to different radii where interaction vertices occur. The relationship between the probed neutrino energy and the radial distance $r$ of the interaction point with respect to the center of the detector surface is shown in \autoref{fig:enu_vs_r}.

\begin{figure}[!htp]
	\centering
	\begin{subfigure}{0.49\textwidth}
		\centering
		\includegraphics[width=\textwidth]{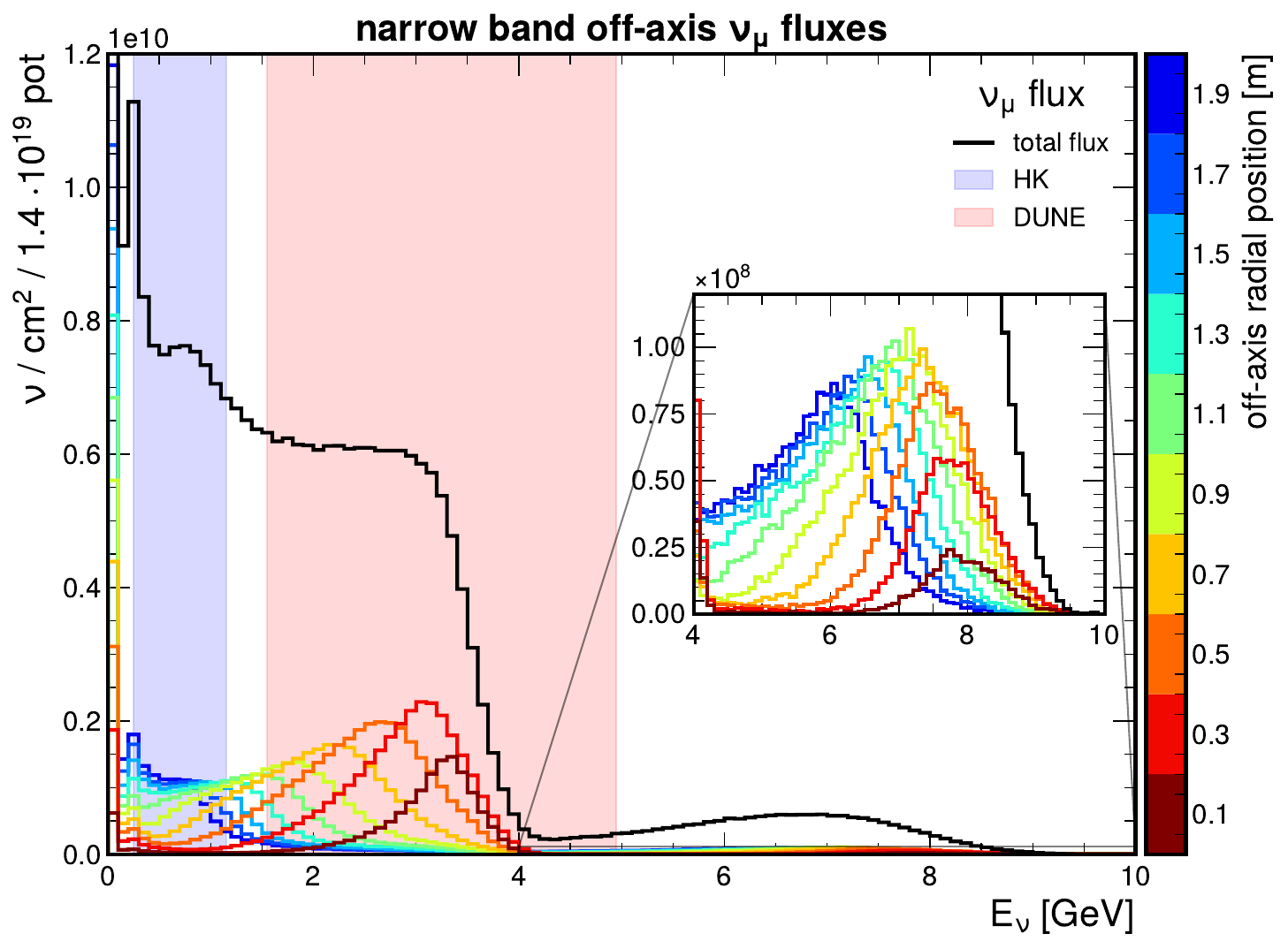}
        
		\caption{Total flux}
		\label{fig:nboa_flux_total}
	\end{subfigure}
    \begin{subfigure}{0.49\textwidth}
    \includegraphics[width=\textwidth]{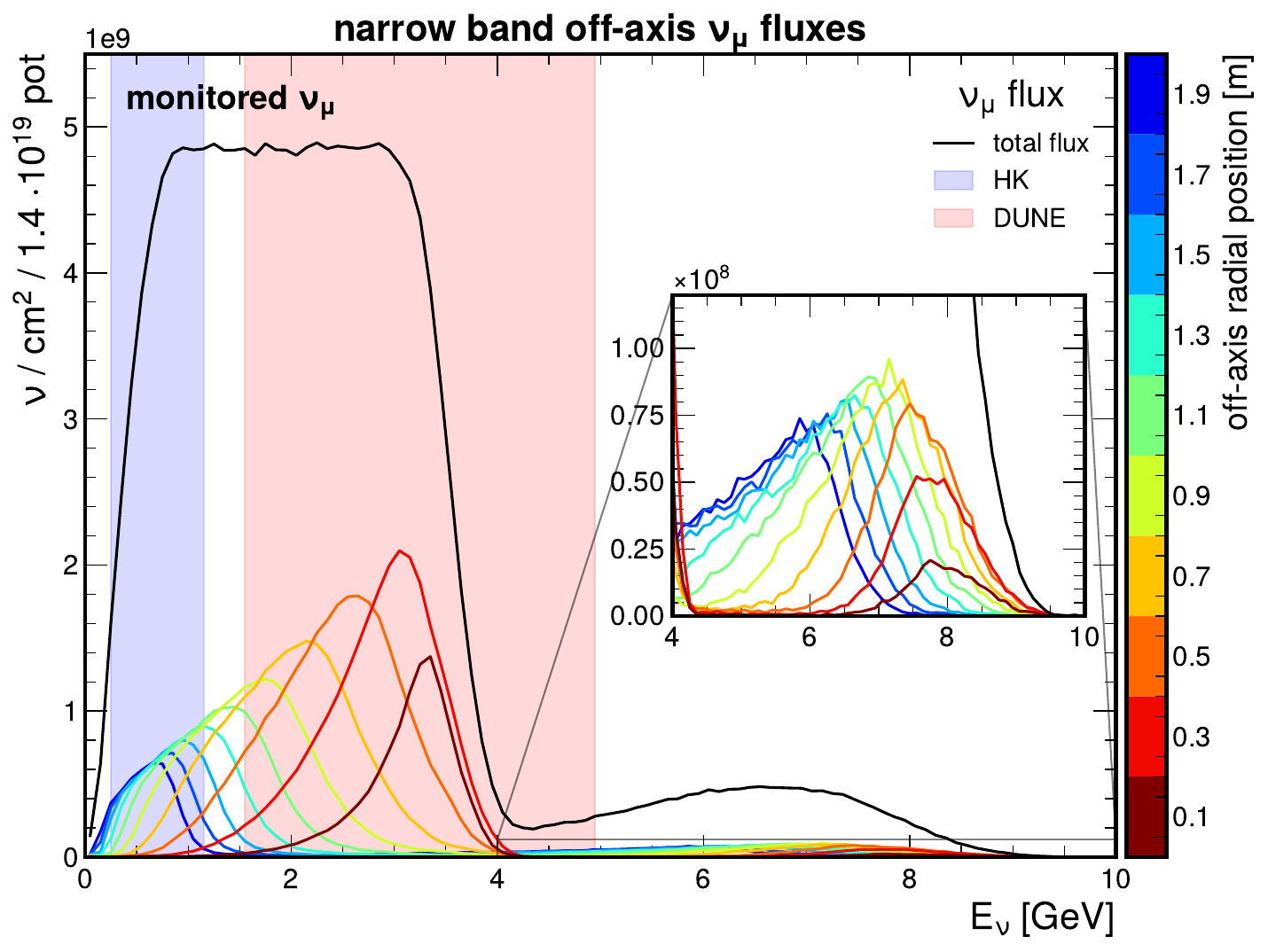}
    \caption{Monitored neutrinos}
    \label{fig:nboa_flux_monitored}
    \end{subfigure}
    \caption{\numu narrow band off-axis fluxes at different radial distances on the detector surface. The black line shows the total flux (i.e., considering the total squared surface of the exposed detector face) and the colored lines show the fluxes incident on filled discs of $\SI{20}{cm}$ at different radial positions on the detector surface. The radial position is given by the color scale alongside each plot. (a) shows the breakdown for all incoming neutrinos, and (b) shows the breakdown only for the monitored neutrinos, i.e., neutrinos produced from decays occurring along the instrumented decay tunnel. An inlay shows a zoom on the contribution of neutrinos from kaon parents in the $4-\SI{10}{GeV}$ energy range.}
    \label{fig:nboa_fluxes}
\end{figure}

In practice, the radial distances covered by the detector surface correspond to an angular range of $0-\ang{4.5}$. By selecting specific radial slices, a neutrino flux narrower than the total flux can be probed. This feature is leveraged in the experimental setup by defining 10 radial slices, each spanning a $\SI{20}{cm}$ window. The contribution of the total \numu flux within each radial slice is shown in Fig.~\ref{fig:nboa_flux_total}.
The total flux exhibits the characteristic dichromatic energy spectrum of a narrow-band beam, with a low-energy component from \pidecay and a high-energy component from \kdecay. At low energies, there is also a contamination of neutrinos from off-momentum mesons produced in the early stages of the beamline. Notably, this contamination is absent when considering only neutrinos produced in the decay tunnel (Fig.~\ref{fig:nboa_flux_monitored}). As expected, the neutrino energy decreases with increasing radial distances for both components.
Note that the sum of the NBOA fluxes will be slightly lower than the total flux, as only cylindrical sections inside the detector volume are considered, while the exposed face of the detector is square. Therefore, areas near the corners of the detector are not included in the analysis.

\begin{figure}[!htp]
	\centering
	\begin{subfigure}{0.49\textwidth}
		\centering
		\includegraphics[width=\textwidth]{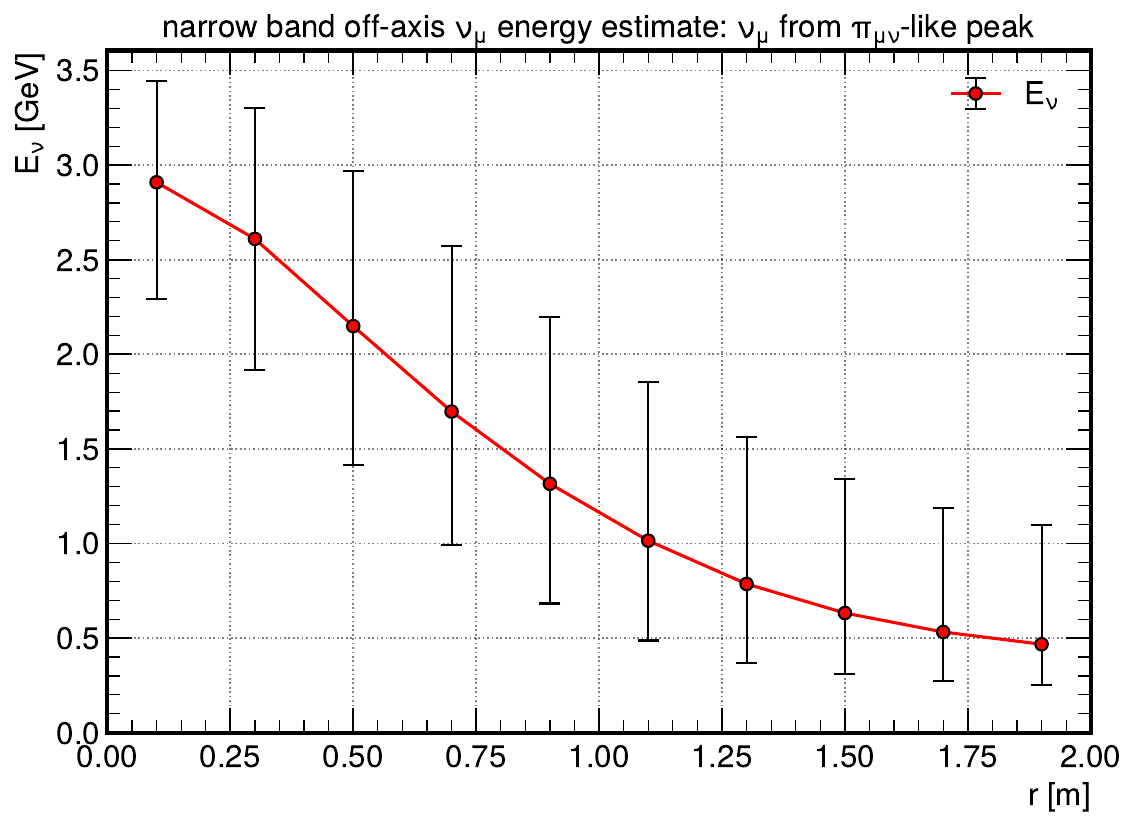}
		\caption{\pidecay peaks}
		\label{fig:nboa_resol_pidecay}
	\end{subfigure}
    \begin{subfigure}{0.49\textwidth}
    \includegraphics[width=\textwidth]{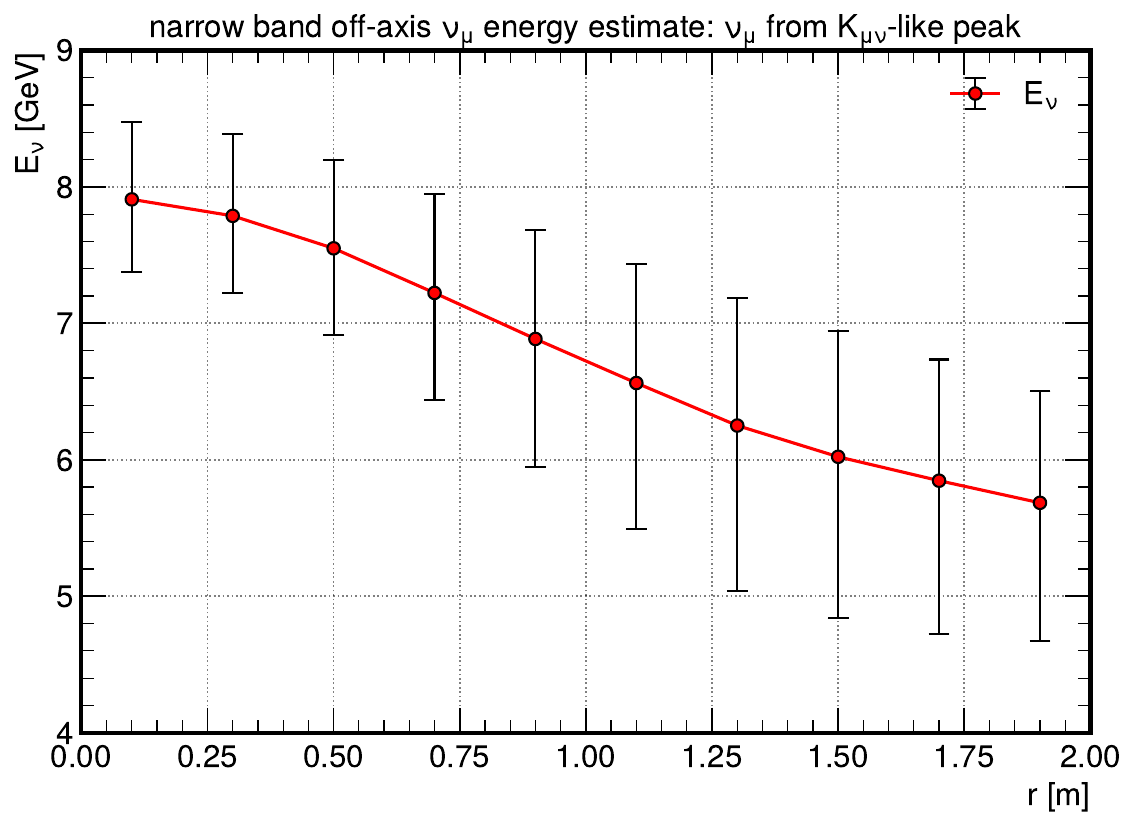}
    \caption{\kdecay peaks}
    \label{fig:nboa_resol_kdecay}
    \end{subfigure}
    \caption{Mean neutrino energy as a function of radial distance for NBOA interaction spectra obtained from \pidecay-like (a) and \kdecay-like (b) peaks. The asymmetric error bars correspond to regions of \enu containing 68\% of the flux integral on either side of the mean energy (above a threshold of $\SI{200}{MeV}$, see \autoref{sec:nu_fluxes}).
    }
    \label{fig:nboa_resolutions}
\end{figure}

The change in the probed neutrino energy profile with increasing off-axis angles is shown in Fig.~\ref{fig:nboa_resolutions}. The \pidecay and \kdecay contributions were separated using an energy cut placed at $\sim \SI{4}{GeV}$, depending on the off-axis configuration. Since their shape is not symmetrical, we report the width of NBOA spectra with asymmetric error bars representing the 68\% percentiles of their distributions with respect to the mean energy. Kaon peaks cover the range from $\sim$5 to $\sim$8~GeV, while pion peaks have energies particularly well-suited for cross-section measurements in the energy domain of next generation oscillation experiments (see Sec.\ref{sec:cross-section}).

It is worth noting that the facility could be operated focusing and transporting lower momentum mesons, by simply scaling the fields of the magnets: this way, a finer coverage of the region below 2 GeV could be attained. The performance of dedicated low-energy runs is currently under evaluation.

\subsection{Neutrino energy measurement from tagging}
\sisetup{parse-numbers=false}

\begin{figure}[h!]
    \centering
    \includegraphics[width=0.5\textwidth]{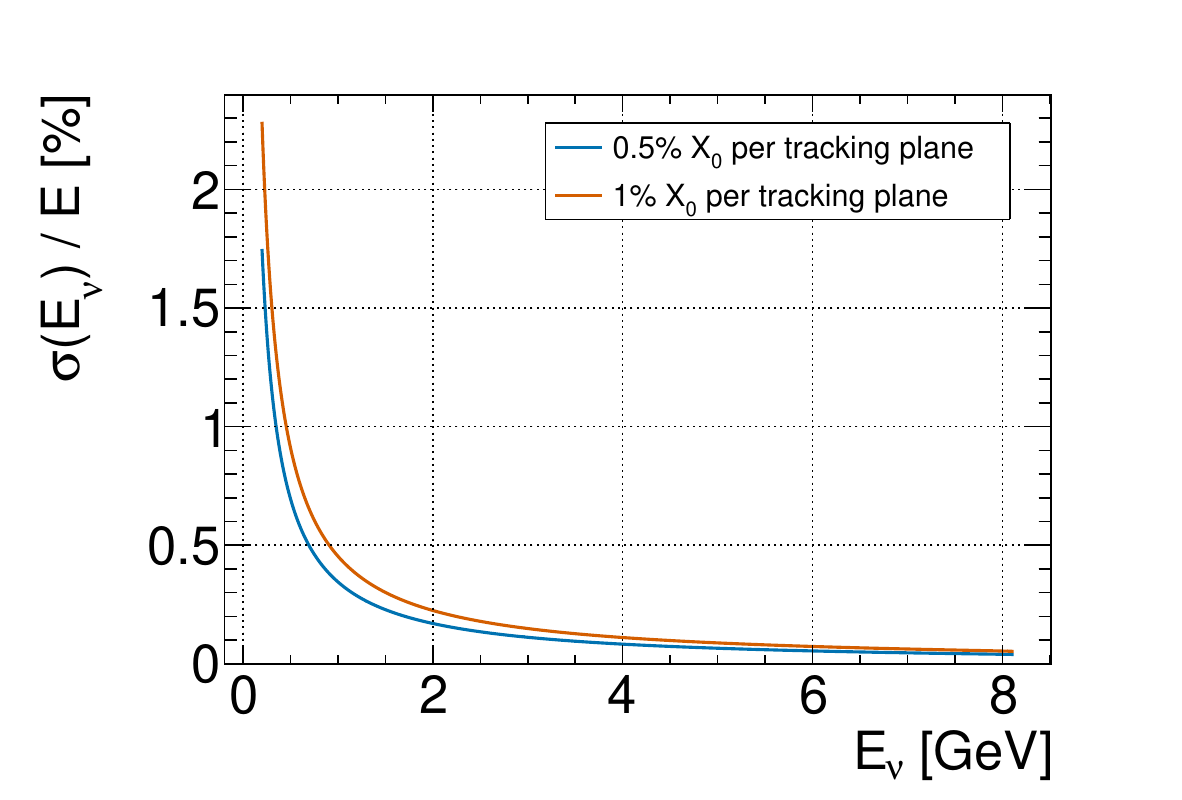}
    \caption{Expected energy resolution of tagged neutrino interactions from $\pi$ and $K$-decays, as a function of the true neutrino energy for a material budget per tracking plane of 0.5\% (blue) and 1\% (orange) of a radiation length .}
    \label{fig:tagging_resolution}
\end{figure}

Beam tagging allows to measure the energy of each tagged neutrino interaction (see Sec. \ref{sec:tagged_neutrino_interactions}) from the kinematics of the associated parent meson decay ($E_{\nu} = E_{\pi,K} - E_{\mu}$), independently of the neutrino detector response. The energy resolution is dominated by multiple scattering in the beam and muon spectrometers' tracking planes. Fig.~\ref{fig:tagging_resolution} shows the expected resolution for tracker planes of 0.5\% and 1\% radiation length (Sec. \ref{sec:silicon_detectors}). 

The neutrino energy resolution lies in the sub-\% range, below 0.4\% for $E_{\nu} > \SI{1}{GeV}$. This allows for measuring muon neutrino cross-sections as a function of energy with a fine-grained binning and negligible smearing between bins (see Sec. \ref{sec:cross-section}). It also gives the possibility to directly measure the energy smearing and bias in LAr, as needed by the DUNE experiment, by comparing the energy estimated in the LAr detector to that precisely measured by the taggers.

\FloatBarrier
\sisetup{parse-numbers=true}
\section{Neutrino cross-section measurements}
\label{sec:cross-section}

A monitored and tagged neutrino beam offers a wide range of unique and powerful cross-section measurements that can be tailored to characterise the physics responsible for the dominant systematic uncertainties in neutrino oscillation analyses. Past and present experiments such as MiniBooNE, T2K, MicroBooNE, NOvA and MINERvA have provided a wide range of measurements on different targets. This large body of measurements, primarily focused on quasi-elastic-like interactions and in the $\sim$1 GeV regime, has highlighted important shortcomings in the way in which we model neutrino-nucleus interactions. Furthermore, in the run-up to the DUNE experiment, there are no existing or planned high-statistics measurements using an argon target in the energy range relevant for DUNE beam oscillations (i.e., 2-3 GeV and above). In this section, we give a non-exhaustive overview of some of the measurements that could be made with nuSCOPE and how they can reduce uncertainties in neutrino oscillation experiments, with a focus on their application to supporting DUNE's physics program since we assume the use of the argon-based reference detector described in \autoref{sec:neutrino_detectors}. Whilst future work will report tailored studies assuming a water-based detector and will project cross-section measurements focused on supporting the physics program of Hyper-Kamiokande, in this section, we give qualitative expectations of what could be achieved. The physics goals listed in this section are part of a global strategy of cross-section measurements aimed at addressing the main sources of systematic uncertainties for future experiments. They complement past measurements made by the MiniBooNE, T2K, NOvA, MicroBooNE, and MINERvA experiments, and will build on the lessons we will learn from upcoming measurements at the Fermilab SBN program with ICARUS and SBND. They are also complementary to the measurements that will be performed in-situ with the DUNE and Hyper-Kamiokande near detectors. Note that this section focuses on measurements of neutrino cross-sections, but in principle, similar measurements could be made with an antineutrino beam. Exploration of this opportunity is also left for future work. 

We begin in \autoref{sec:nu_fluxes} with a description of the neutrino fluxes, and then in \autoref{sec:xsec_simulation} we provide an overview of the simulation to estimate the event rates that can be used for cross-section measurements.
We proceed to focus first on measurements that exploit the NBOA technique and the monitored neutrino beam to first make a \numu inclusive energy-dependent cross-section measurement in \autoref{sec:numu_ccinc} and then \numu and \nue differential cross-section measurements in \autoref{sec:diff_xsec}.
In \autoref{sec:prism_studies}, we show how the PRISM technique can be applied using the \numu NBOA fluxes to directly measure the \nue/\numu cross-section ratio. We further show how the NBOA technique can be used to constrain common backgrounds for neutrino oscillation analyses, focusing on the NC $\pi^0$ contribution in \autoref{sec:ncpizero}. We then proceed to consider measurements benefiting from neutrino tagging in \autoref{sec:nutag_xsec_measurements}, including direct measurements of neutrino energy reconstruction bias that can be used to calibrate DUNE alongside measurements of powerful kinematic quantities to characterise neutrino interactions that are usually only accessible to electron-scattering experiments. Finally, the cross-section measurement reach of nuSCOPE is summarised in \autoref{sec:xsec_summary}.

\subsection{Neutrino fluxes}
\label{sec:nu_fluxes}

\begin{figure}[!htp]
	\centering
	\begin{subfigure}{0.49\textwidth}
		\centering
		\includegraphics[width=\textwidth]{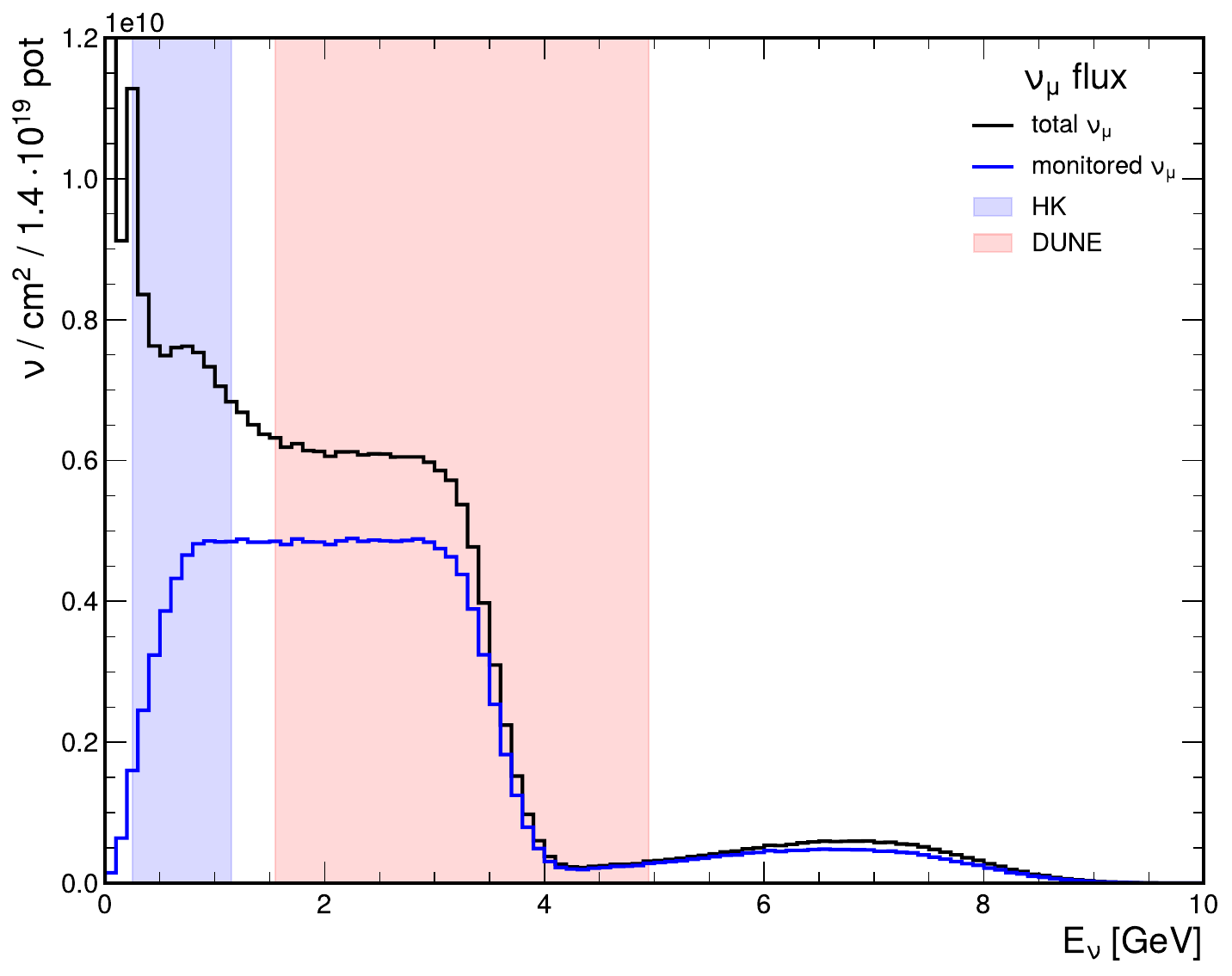}
		\caption{total \numu flux.}
		\label{}
	\end{subfigure}
    \begin{subfigure}{0.49\textwidth}
    \includegraphics[width=\textwidth]{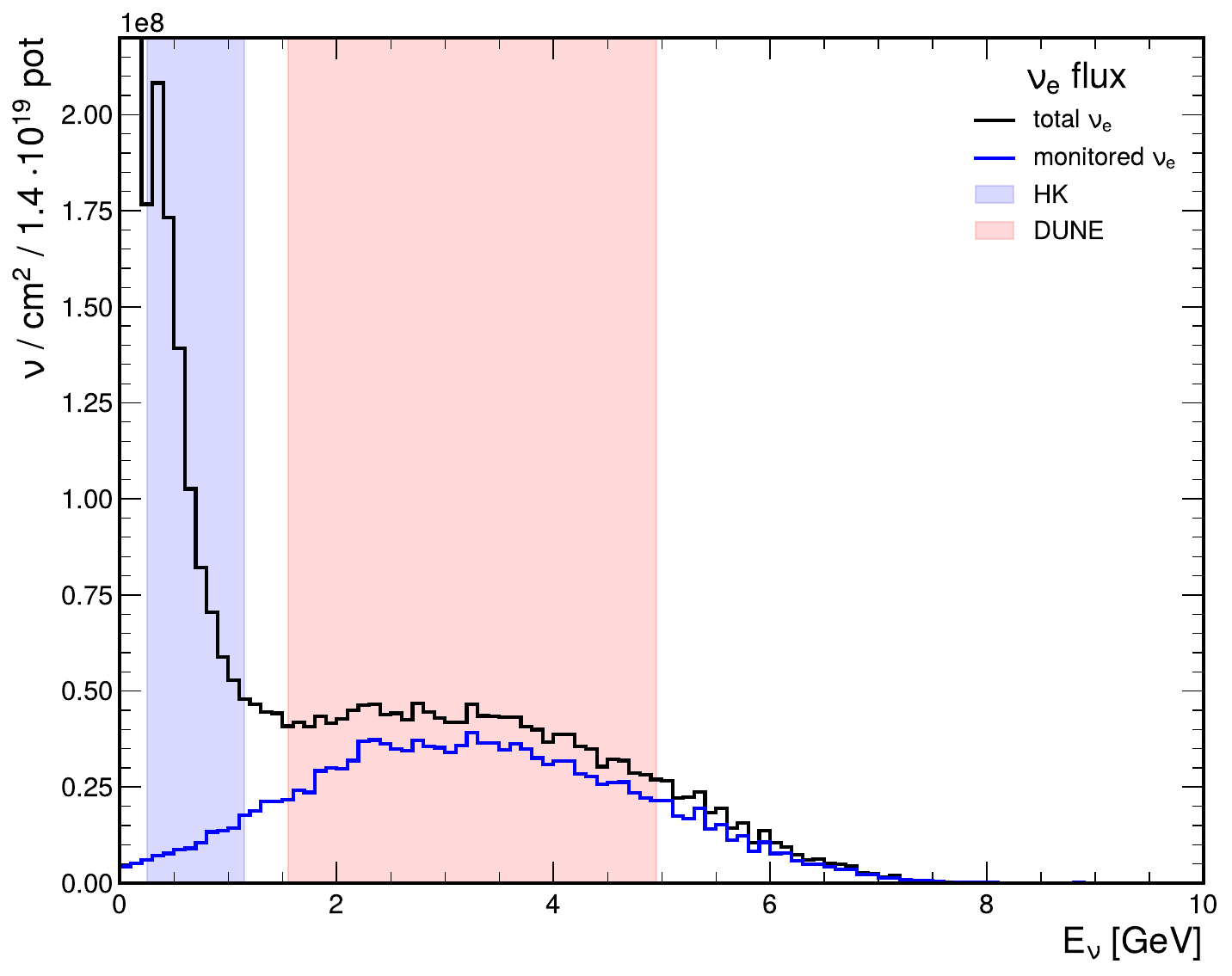}
    \caption{total $\nu_e$ flux.}
    \label{}
    \end{subfigure}
    \caption{Total $\nu_\mu$ (left) and $\nu_e$ (right) fluxes on the surface of the detector.
    The monitored neutrinos (blue) contributing to the total flux (black) are those produced from decays occurring along the decay volume, i.e. for which the associated charged lepton can be measured using either the instrumented decay tunnel (\numu from \kdecay and \nue from $K_{e3}$) or the instrumented hadron dump (\numu from \pidecay). Note that the $y$-axis range is limited to exclude the large spike at very low energies coming from neutrinos produced in the target region; the spike in the first bin extends to $\sim 8 \cdot 10^{10}$ and $\sim 7 \cdot 10^9$ $\nu \, / \, \si{cm}^2 \, / \, 1.4 \cdot 10^{19}$ PoT for the $\nu_\mu$ and $\nu_e$ fluxes respectively.}
    \label{fig:total_flux}
\end{figure}

The total neutrino fluxes are computed as the neutrinos in the detector acceptance and normalized to the neutrino detector $4 \times 4 ~\si{m}^2$ front face area and to the total protons on target exposure, which corresponds to $1.4 \cdot 10^{19}$ PoT.
The total obtained $\nu_\mu$ and $\nu_e$ fluxes are shown in \autoref{fig:total_flux}.

The beamline design is such to maximize the number of neutrinos produced from meson decays occurring along the instrumented decay tunnel, since these neutrinos can be monitored by measuring the associated charged leptons.
The large-angle positrons and muons produced from $K_{e3}$ and \kdecay kaon decays can be measured using the instrumented decay tunnel, whereas the forward-going muons produced in \pidecay -- outside the decay volume acceptance -- can be recorded using an instrumented hadron dump.
Nevertheless, a significant fraction of neutrinos in the detector acceptance do not come from meson decays occurring along the instrumented decay tunnel, but rather from other regions of the facility, and thus they cannot be monitored. 
These neutrinos come from interactions within the graphite target, along the beamline, and in the dumps.
In particular, the dominant contribution is below \SI{1}{GeV}, whereas for higher energies there are residual contributions due to neutrinos produced along the beamline after the second dipole (early decays before entering the tunnel) and in the hadron dump.
For the monitoring technique, these neutrinos represent a source of irreducible background in the energy spectra and, since they cannot be monitored, their contribution to the total flux will have a larger systematic flux uncertainty associated with them. For the tagging technique, their impact is reduced as, in most cases, they won't be matched to a mesonic decay.
A breakdown of the neutrino flux into monitored neutrinos and these additional background neutrinos is also shown in \autoref{fig:total_flux}.

In practice, although there is a large contribution from the low-energy unmonitored neutrinos to the total flux, their impact on the observable event rate will be minimal. At $E_\nu \lesssim
\textrm{200 MeV}$, the cross-section for charged current interactions is almost negligible, and it increases with neutrino energy. To improve the readability of the fluxes, we apply a 200 MeV cut to remove this contribution. This cut is applied to both the $\nu_\mu$ and $\nu_e$ fluxes.

As mentioned in Sec. \ref{sec:energy}, the monitored \numu flux exhibits the typical narrow-band beam dichromatic energy spectrum \cite{Charitonidis_2021} since two distinct peaks can be identified: a low energy peak at $E_\nu \lesssim \SI{4}{GeV}$ due to neutrinos produced by two-body decay in flight of pions ($\pi^+ \to \mu^+ \nu_\mu$) and a high energy peak at $E_\nu \gtrsim \SI{4}{GeV}$ coming from the two-body decay of kaons ($K^+ \to \mu^+ \nu_\mu$).
On the other hand, the \nue flux is due to neutrinos coming predominantly from the three-body kaon decays ($K^+\to e^+ \pi^0 \nu_e$), and a single large peak can be identified.

\subsection{Event rates and simulation}
\label{sec:xsec_simulation}

\subsubsection{Neutrino interaction models and simulation}
\label{sec:xsec_models}

We simulate neutrino interactions on an argon target using the GENIE event generator~\cite{Andreopoulos:2015wxa,GENIE:2021npt} and the total and narrow band off-axis fluxes shown in \autoref{fig:total_flux} and \autoref{fig:nboa_fluxes}, respectively.

The nominal configuration chosen for the simulation is the \argenie comprehensive model configuration (CMC), corresponding to the model chosen by the DUNE Collaboration for upcoming sensitivity studies~\cite{GENIE_30400tag}. The CMC uses a Local Fermi Gas model to describe quasi-elastic (CCQE) interactions~\cite{Nieves:2011pp}, with an additional contribution at high nucleon momenta mimicking the impact of short-range correlations (SRCs)~\cite{geniev304TN}. The nucleon axial form factor is described by the z-expansion parametrization, detailed in~\cite{Meyer:2016oeg}. Multi-nucleon interactions (mostly two-particle-two-hole interactions, or 2p2h) are simulated with the SuSAv2 model~\cite{RuizSimo:2016rtu,Megias:2016fjk}. Resonant interactions (referred to as RES in the remainder of the document) use the model by Rein and Sehgal~\cite{Rein:1980wg}. At higher energies, non-resonant hadronization processes are simulated using the custom AGKY model~\cite{Yang:2009zx} for interactions with an invariant hadronic mass $W$ below $\SI{1.7}{GeV}$, and using the PYTHIA 6 generator~\cite{Sjostrand:2006za} for $W > \SI{3}{GeV}$, with a linear interpolation between the two regions. Intra-nuclear transport processes (also known as final state interactions, or FSI) are described using the GENIE hA2018 model~\cite{Dytman:2021ohr}. This model uses tuned parameters according to the work in~\cite{GENIE:2021zuu}, in which the parameters have been tuned to neutrino-nucleon scattering measurements from bubble chamber experiments.

In some studies, we compare this model to an alternative CMC, \susagenie, whose implementation details can be found in~\cite{Dolan:2019bxf}. The model has two main differences with respect to the \argenie model. First, QE and 2p2h interactions are both simulated using the SuSAv2 model, but only using inclusive predictions (i.e., only as a function of lepton kinematics). One of the main consequences of this difference is that the models apply different treatments to the inclusion of long-range correlations inside the nucleus (also known as RPA, or random phase approximations), which will have a significant impact at low energy transfers. Second, the tuned parameters described in~\cite{GENIE:2021zuu} are not applied to this CMC, which results in a higher predicted cross-section for inelastic events. 

\subsubsection{Expected event rates}
\label{sec:xsec_eventrates}

We generate 1 million CC events using the total \numu flux, 1 million CC events using the total \nue flux, and 1 million CC events using each of the ten NBOA total fluxes shown in \autoref{fig:nboa_fluxes}. We then use the NUISANCE framework \cite{Stowell:2016jfr} to facilitate the analysis of these generated events.
The total number of expected events is obtained by multiplying the flux integral and the corresponding GENIE cross-section, and scaling this number according to the expected exposure, i.e. $1.4 \cdot 10^{19}$ PoT, and the number of argon nucleons in each corresponding detector volume: the whole detector volume for the total fluxes, and the volumes of hollow cylinders with $\SI{20}{cm}$ thick wall at each radial position. As mentioned above, we consider a total fiducial mass of 500 tons of argon in a $4\times4\times22.3~\si{m}^3$ volume located $\SI{25}{m}$ from the tunnel end.

Except for the study detailed in \autoref{sec:ncpizero}, we do not apply any detector effects, efficiencies, or smearing. For \numu and \nue CC inclusive interactions, we assume that the reconstruction efficiency for muons and electrons is 100\% and that the smearing on their momenta is much smaller compared to the size of the binning we choose in the studies. Note that this relies on the assumption that the detector setup can either fully contain muons, or is instrumented with a downstream magnetized spectrometer. The studies in this section aim to provide an overview of the types of measurements possible with this setup, notably in terms of expected statistics, and a complete study will need an estimation of detector effects in the projected measurements. 

\begin{figure}[!htp]
	\centering

    \begin{subfigure}{0.51\textwidth}
    \includegraphics[width=\textwidth]{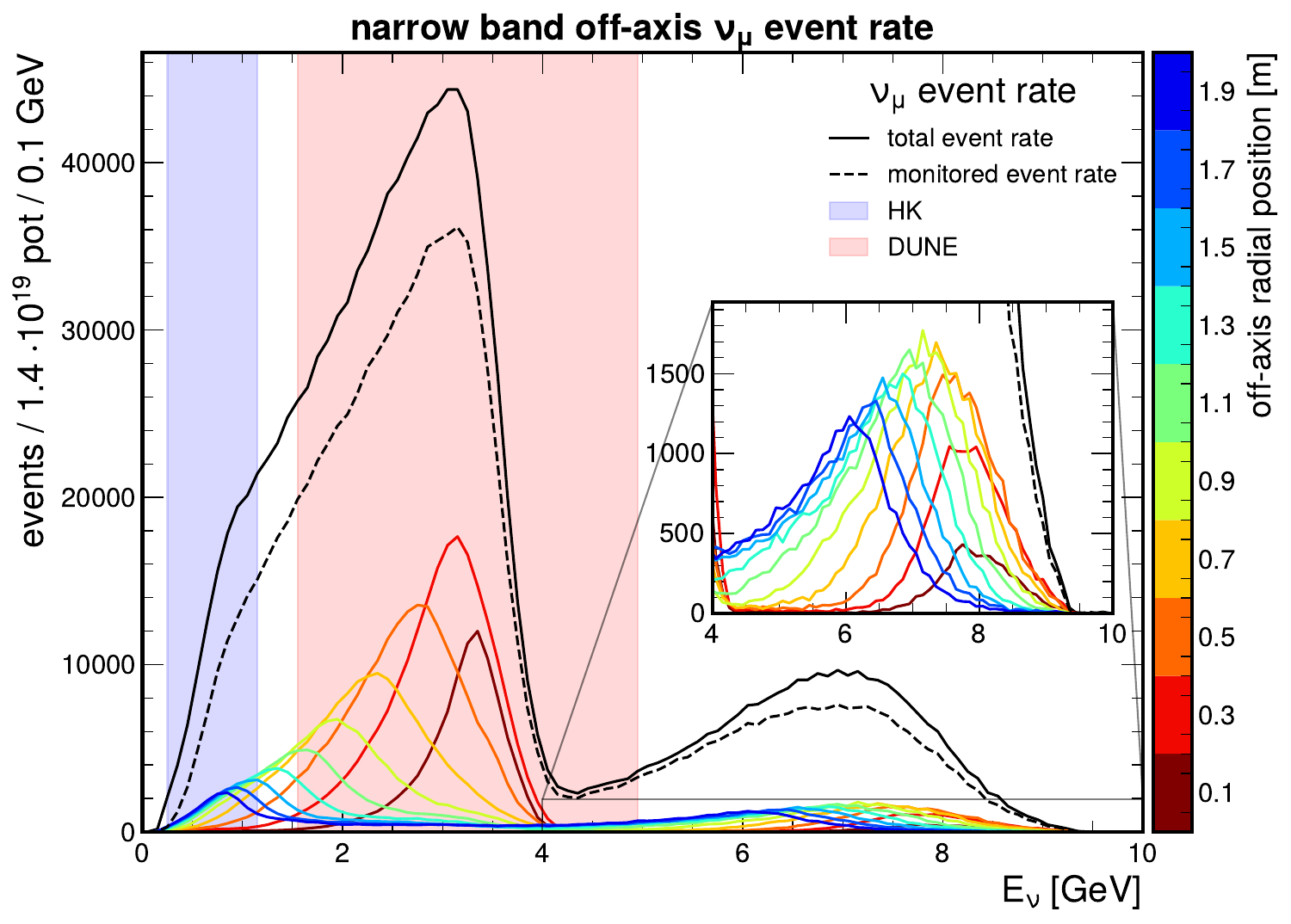}
    \caption{\numu}
    \label{fig:total_event_rate_numu}
    \end{subfigure}
        \begin{subfigure}{0.47\textwidth}
    \includegraphics[width=\textwidth]{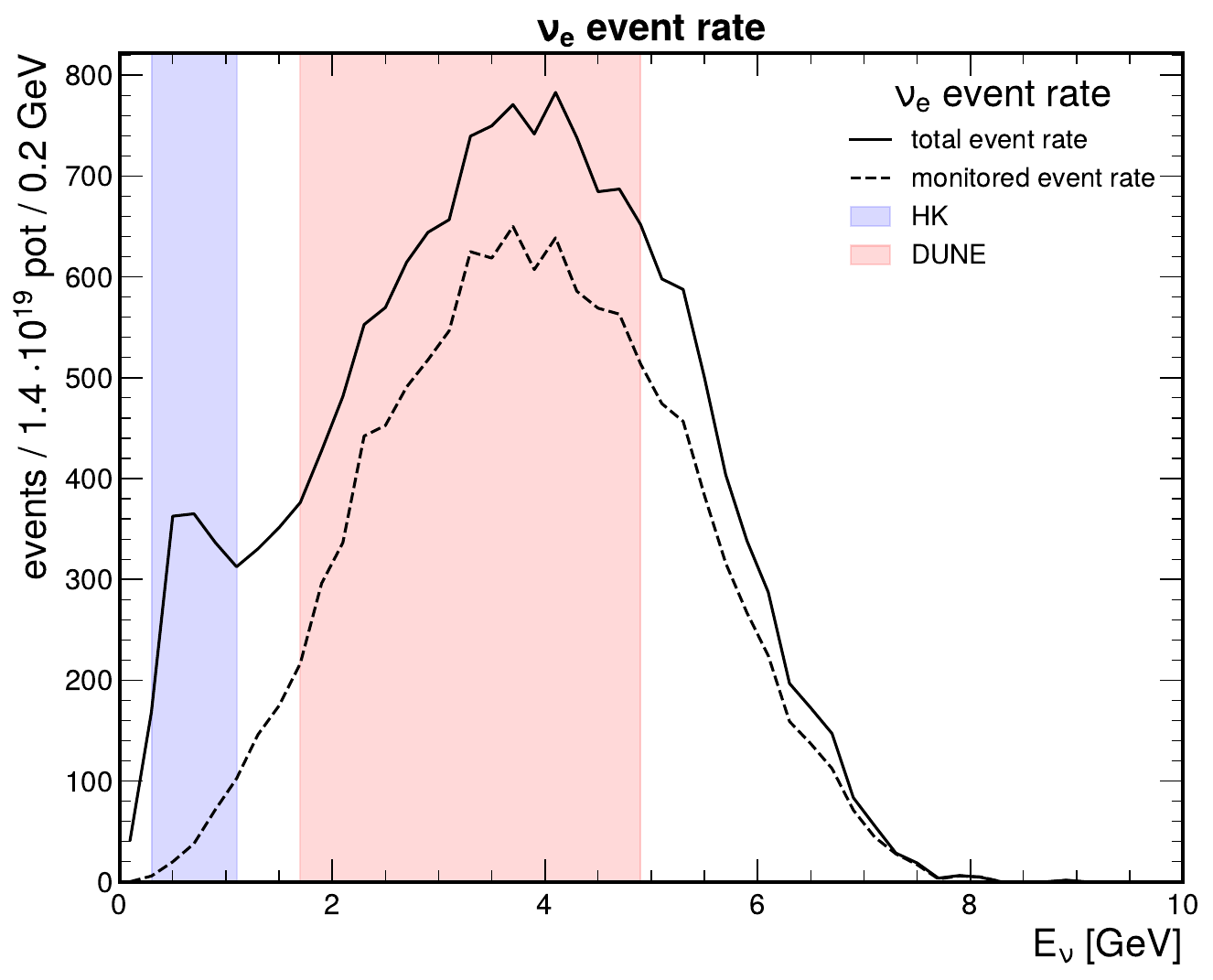}
    \caption{\nue}
    \label{fig:total_event_rate_nue}
    \end{subfigure}
    \caption{Total \numu (a) and \nue (b) charged current event rates. The black solid line shows the total event rate, and the dashed black line the monitored event rate. For the \numu case, the NBOA total fluxes are also shown with colored solid lines, where the colors correspond to different radial positions. The HK (blue) and DUNE (red) regions of interest are given by the shaded areas.}
    \label{fig:total_event_rate}
\end{figure}

The total CC event rates are shown in \autoref{fig:total_event_rate} and summarised in \autoref{tab:event_rate_integral}. As expected, the large contribution from the low-energy neutrinos outside of the monitored area in the flux is almost entirely suppressed by the low interaction cross-section. The event rates show that the two-peaked structure of the $\nu_\mu$ flux and the single peak of the $\nu_e$ flux are preserved, and the contributions are enhanced at higher energies due to the larger interaction cross-section. The clear separation between the \pidecay and \kdecay peaks seen in the simulated fluxes is preserved in the event rates. The event rates effectively cover the range of neutrino energies relevant to the DUNE experiment.

If a water-based target with a similar mass were used, the event rates would be expected to remain comparable, with nuclear effects and proton-neutron number differences causing a variation of less than 10\%.  A lower but still significant portion of the event rate is within the region of interest for Hyper-Kamiokande, and this can be well isolated using the most off-axis NBOA samples. Running with a lower meson beam energy may allow a higher statistics sample relevant to Hyper-Kamiokande.

\begin{table}
    \centering
    \begin{tabular}{c|c}
        channel & events / $1.4 \cdot 10^{19}$ PoT      \\ \hline
        off-axis $\nu_{\mu}$ at $r = \SI{0.1}{m}$   &   $1.0\times10^5$  \\ \hline
        off-axis $\nu_{\mu}$ at $r = \SI{0.3}{m}$   &    $2.2\times10^5$ \\ \hline
        off-axis $\nu_{\mu}$ at $r = \SI{0.5}{m}$   &   $2.2\times10^5$  \\ \hline
        off-axis $\nu_{\mu}$ at $r = \SI{0.7}{m}$   &   $1.8\times10^5$  \\ \hline
        off-axis $\nu_{\mu}$ at $r = \SI{0.9}{m}$   &   $1.3\times10^5$  \\ \hline
        off-axis $\nu_{\mu}$ at $r = \SI{1.1}{m}$   &   $1.0\times10^5$  \\ \hline
        off-axis $\nu_{\mu}$ at $r = \SI{1.3}{m}$   &   $8.3\times10^4$   \\ \hline
        off-axis $\nu_{\mu}$ at $r = \SI{1.5}{m}$   &   $6.8\times10^4$   \\ \hline
        off-axis $\nu_{\mu}$ at $r = \SI{1.7}{m}$   &   $5.9\times10^4$   \\ \hline
        off-axis $\nu_{\mu}$ at $r = \SI{1.9}{m}$   &   $5.2\times10^4$   \\ \hline
        total $\nu_{\mu}$                           &   $1.3\times10^6$ \\ \hline
        total $\nu_{e}$                             &   $1.7\times10^4$   \\ \hline
        total monitored $\nu_{\mu}$                 &   $1.0\times10^6$ \\ \hline
        total monitored $\nu_{e}$                   &   $1.2\times10^4$ \\ \hline
        total tagged $\nu_{\mu}$                    &   $7.6\times10^5$   \\
        
    \end{tabular}
    \caption{\numu and \nue event rate integral for \argenie . The event rate is given for the total number of events, as well as separated into different radial slices with respect to the center of the exposed detector face.}
    \label{tab:event_rate_integral}
\end{table}
\subsection{Flux-averaged \numu CC-inclusive cross-section measurement}
\label{sec:numu_ccinc}

One of the simplest yet most powerful measurements that can be made using a monitored neutrino beam is the measurement of flux-averaged CC cross-sections using different NBOA fluxes. Since the visible event rate for a neutrino experiment results from the convolution of the neutrino flux and the interaction cross-section, it is only possible to measure the total cross-section for a process \emph{averaged} over the range of available energies. The flux-averaged cross-section $\langle \sigma \rangle_{\Phi}$ (in $\si{cm}^2/\text{nucleon}$) is defined as:

\begin{equation}
    \langle \sigma \rangle_{\Phi} = \frac{N_{\text{events}}}{\Phi N_{\text{tgt}}N_{\text{PoT}}},
    \label{eq:flux_avg_xsec}
\end{equation}

\noindent where $N_{\text{events}}$ is the number of expected events in the detector (as predicted by GENIE), $\Phi$ is the neutrino flux integral (expressed in $\nu/\si{cm}^2/ \text{PoT}$), $N_{\text{tgt}}$ is the number of nucleons in the considered volume and $N_{\text{PoT}}$ is the total exposure in protons on target, i.e. $1.4 \cdot 10^{19}$ PoT.

This method does not rely on a fine reconstruction of all final-state particles. Indeed, it is equivalent to counting the number of neutrino interactions that have a charged lepton consistent with a muon in the final state, regardless of other products. For wide-band beams, the measurement of the CC inclusive cross-section is intrinsically limited by the range spanned by the neutrino flux, since the neutrino energy is not known on an event-by-event basis. However, by performing a flux-averaged measurement, we avoid building in model dependence into this analysis. To estimate the feasibility and precision of such a measurement, this method can be applied to all the NBOA fluxes. As shown in \autoref{fig:nboa_fluxes} and \autoref{fig:total_event_rate}, the \pidecay-like and \kdecay-like contributions to both the neutrino flux and the event rate are well separated at $\sim \SI{4}{GeV}$ in true neutrino energy. We therefore apply this $\sim \SI{4}{GeV}$ energy cut, depending on the off-axis configuration, to the obtained event rates\footnote{In fact, this cut in true neutrino energy introduces some model dependence in the analysis, as the cut would be made on events as a function of reconstructed neutrino energy. However, due to the small flux in this region and the natural separation between the fluxes, we anticipate a negligible contribution to the systematic budget of the experiment.} and extract two flux-averaged cross-sections for each NBOA flux -- once for \pidecay decays, and once for \kdecay decays.

\begin{figure}[!htp]
    \centering
    \includegraphics[width=0.7\linewidth]{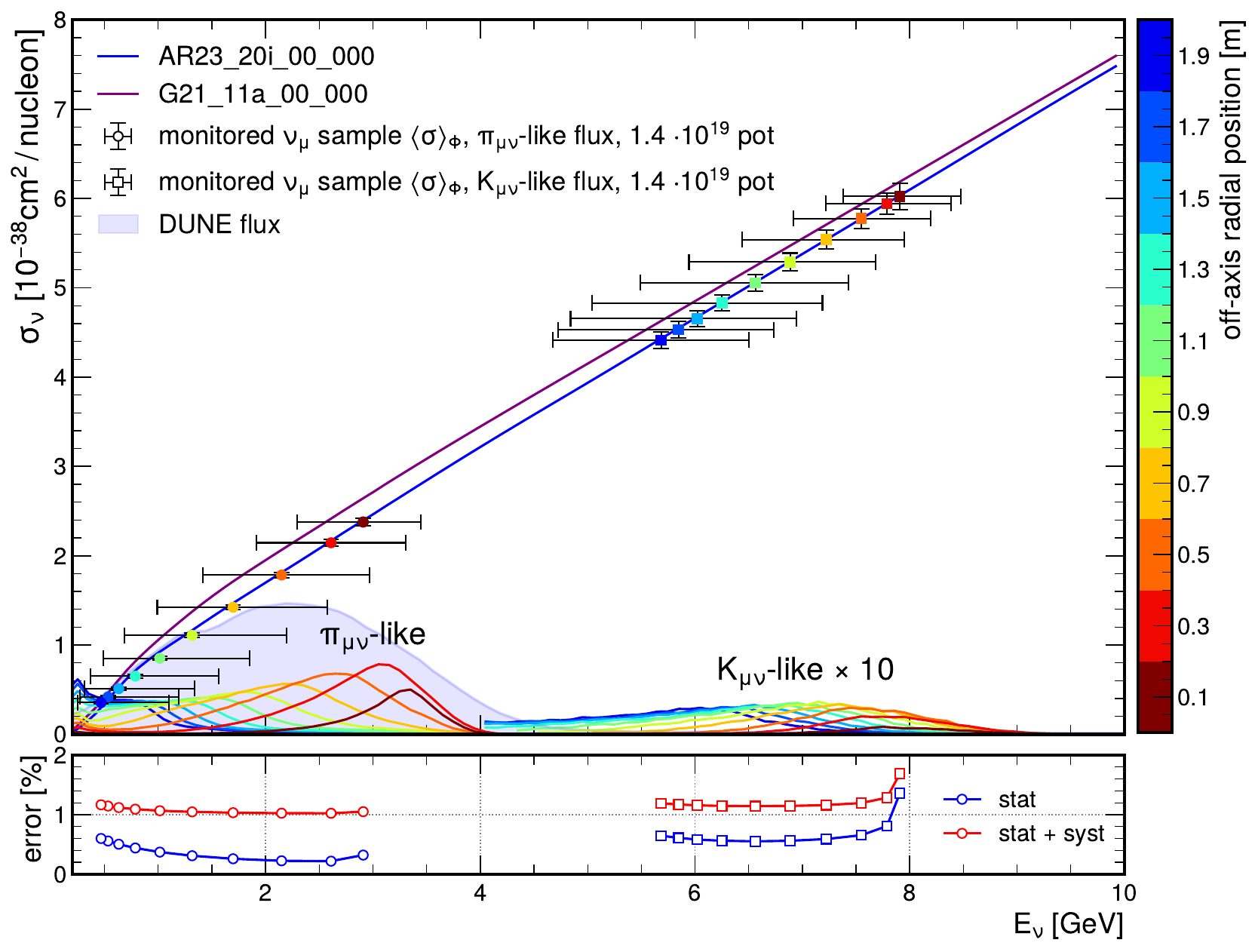}
    \caption{Flux averaged \numu CC inclusive cross-section as a function of neutrino energy using the NBOA technique. The colored lines correspond to the NBOA fluxes at different radial distances, given by the colored scale to the right of the figure. The \kdecay component of each flux has been artificially inflated by a factor of 10 for illustration purposes. Each NBOA flux has a corresponding predicted measurement point of the same color. Horizontal error bars encase the 68\% percentiles with respect to the mean energy for the NBOA fluxes. The underlying figure shows the size of uncertainties due to available statistics (blue) and considering also systematic uncertainties related to the monitored flux prediction, assumed to be $\sim$1\%, (red). The measurements are compared to the \argenie model (blue) and the \susagenie model (purple). The DUNE near-detector flux is shown for reference using an arbitrary normalization.}
    \label{fig:rainbow_plot_cc}
\end{figure}

The resulting flux-averaged cross-sections, along with statistical uncertainties, are shown in \autoref{fig:rainbow_plot_cc}. Several observations can be made:
\begin{itemize}
    \item Thanks to the large size of the detector assumed in this setup, the statistical errors associated with such a measurement are almost always below 1\%. The measurement would therefore be entirely limited by systematic uncertainties. In current high-statistics measurements, the largest source of systematic uncertainty usually stems from the knowledge of the neutrino flux~\cite{MicroBooNE:2024zwf,MINERvA:2022bno}. We have assumed a systematic uncertainty\footnote{Note that to complete this study, it is also essential to assign appropriate uncertainties to the unmonitored contributions to the neutrino flux, which will be investigated in future work.} on the normalization of the flux of 1\%, coming from the monitored beamline.
    \item The energy ranges spanned by the \pidecay and \kdecay peaks cover the vast majority of the DUNE on-axis flux. 
    \item The natural span of the \pidecay and \kdecay fluxes results in a ``gap'' in the inclusive cross-section measurements between 3--5 GeV. However, this region can be recovered using techniques such as obtaining a virtual flux from the real \numu fluxes (the PRISM technique), which we will describe in \autoref{sec:prism_studies}. 
\end{itemize}

\noindent
It is worth noting that, in principle, other experiments can perform such flux-averaged measurements using off-axis fluxes. This technique is proposed both for DUNE via the DUNE-PRISM~\cite{DUNE:2021tad} setup and by the SBND experiment~\cite{delTutto2021SBND}. We note, however, that in the case of DUNE-PRISM the off-axis fluxes are significantly wider than the NBOA fluxes considered in this work, making a fine-grained analysis of the evolution of the cross-section more difficult. They also cannot easily probe the tail of DUNE's neutrino energy distributions. Whilst the SBND-PRISM concept offers exciting possibilities for imminent measurements, it spans only a small range of neutrino energies compared to the NBOA fluxes in this analysis. More importantly, the measurement proposed in this section will benefit from the percent-level constraint on the integral of the neutrino flux from the monitoring instrumentation, whereas conventional beams are currently limited by flux systematic uncertainties of the order of 10\%, which may present a bottleneck to precision measurements.

Similar measurements could be made using a water-based target where analyses using the most off-axis NBOA fluxes would allow a measurement of the evolution of the cross-section in a region of interest for Hyper-Kamiokande with \%-level precision. 

\subsection{Flux-averaged \numu and \nue CC double-differential cross-section measurements}
\label{sec:diff_xsec}

Whilst the inclusive cross-section measurement presented in \autoref{sec:numu_ccinc} provides a very useful measurement, it only constrains the total CC cross-section, which gets contributions from several channels regulated by different dynamic processes. The relative contribution and the specific underlying physics of each process are fundamental information needed to ensure the success of future experiments. There are a variety of measurements that allow us to probe these individual mechanisms; in this section, we focus on only a few examples inspired by measurements made by current neutrino scattering experiments. 

\subsubsection{\numu \cczeropi differential cross-section}
\label{sec:numu_doublediff}

A first example is the measurement of differential cross-sections as a function of outgoing lepton kinematics. The simplest channel that we can target experimentally is the CC quasi-elastic (CCQE) process, also referred to as one-particle-one-hole (1p1h) process, containing a single lepton and nucleon in the final state. In fact, it is not possible to measure pure CCQE events, as the interaction takes place inside the nuclear medium and final state interactions result in the same interaction topology from different interaction channels. The closest visible final state in a detector is one with a final state lepton and no pions - known as the \cczeropi topology. This channel will have significant contributions from CCQE processes, but also from multi-nucleon interactions (2p2h), resonant pion production where the pion has been absorbed inside the nucleus (RES) or other processes with no pions in the final state. \cczeropi cross-sections have been extensively measured by neutrino scattering experiments (see, e.g., ~\cite{T2K:2023qjb, MINERvA:2018vjb,MicroBooNE:2024zkh}).

\begin{figure}
\subfloat{\includegraphics[width=0.34\textwidth]{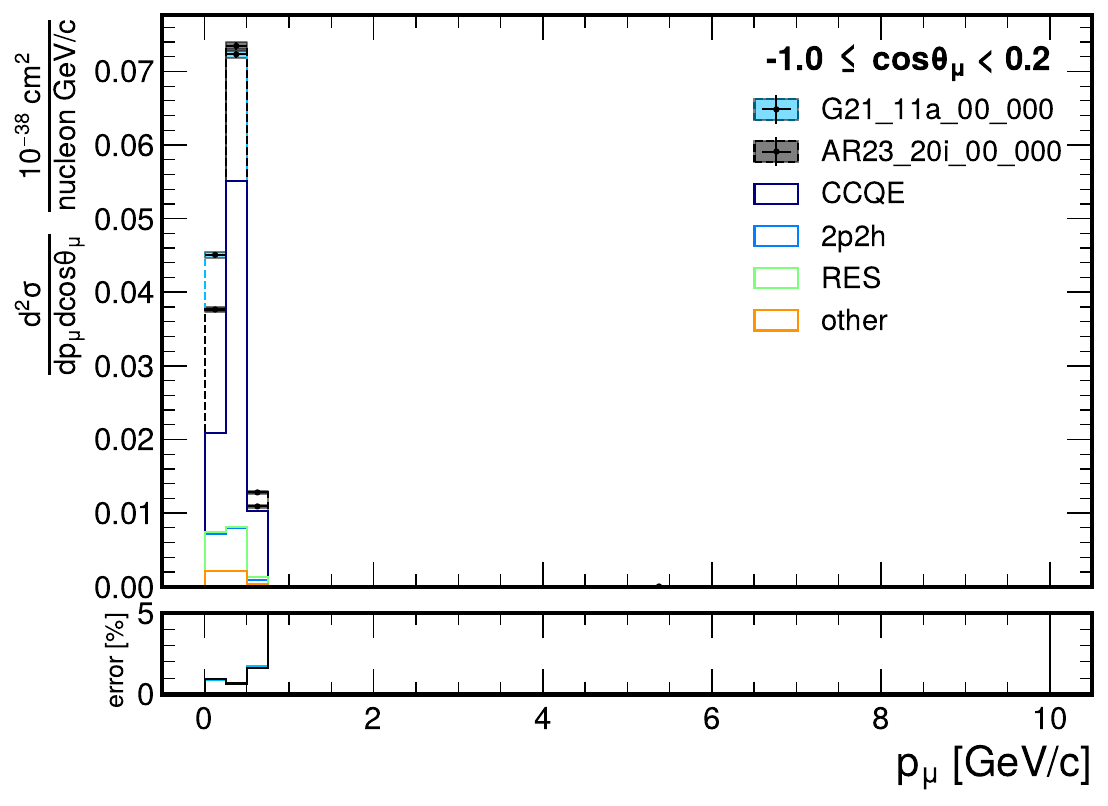}}
\subfloat{\includegraphics[width=0.34\textwidth]{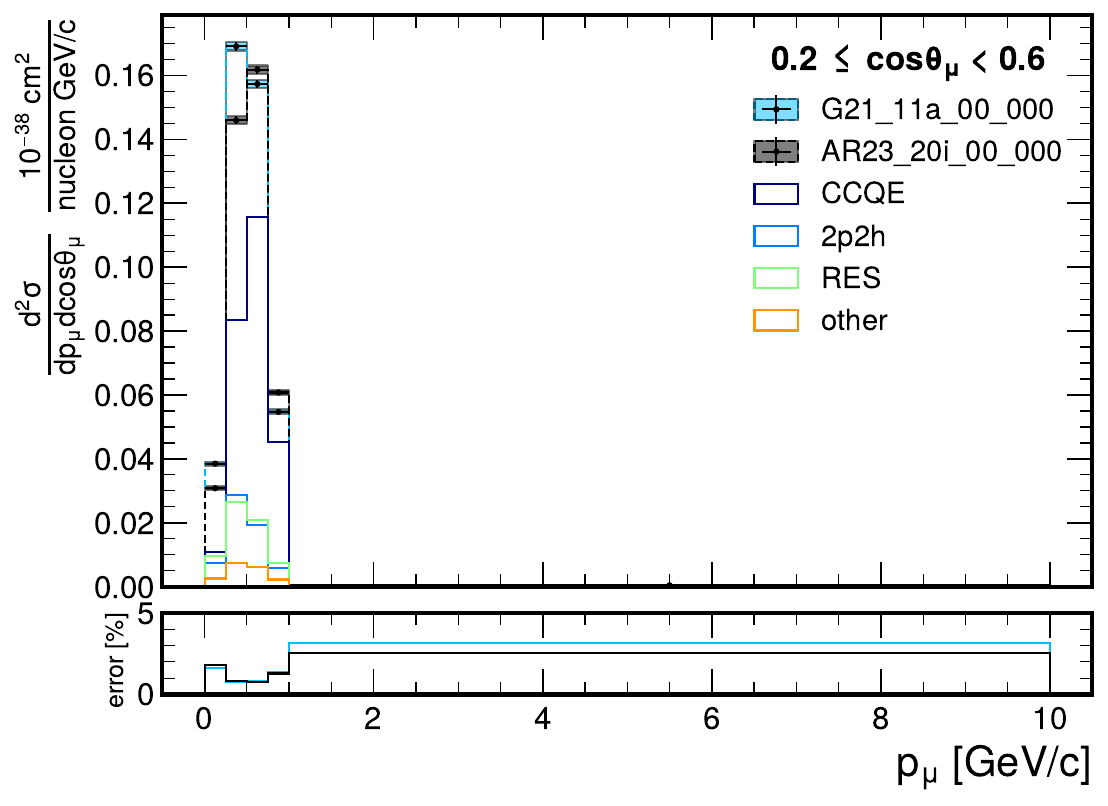}}
\subfloat{\includegraphics[width=0.34\textwidth]{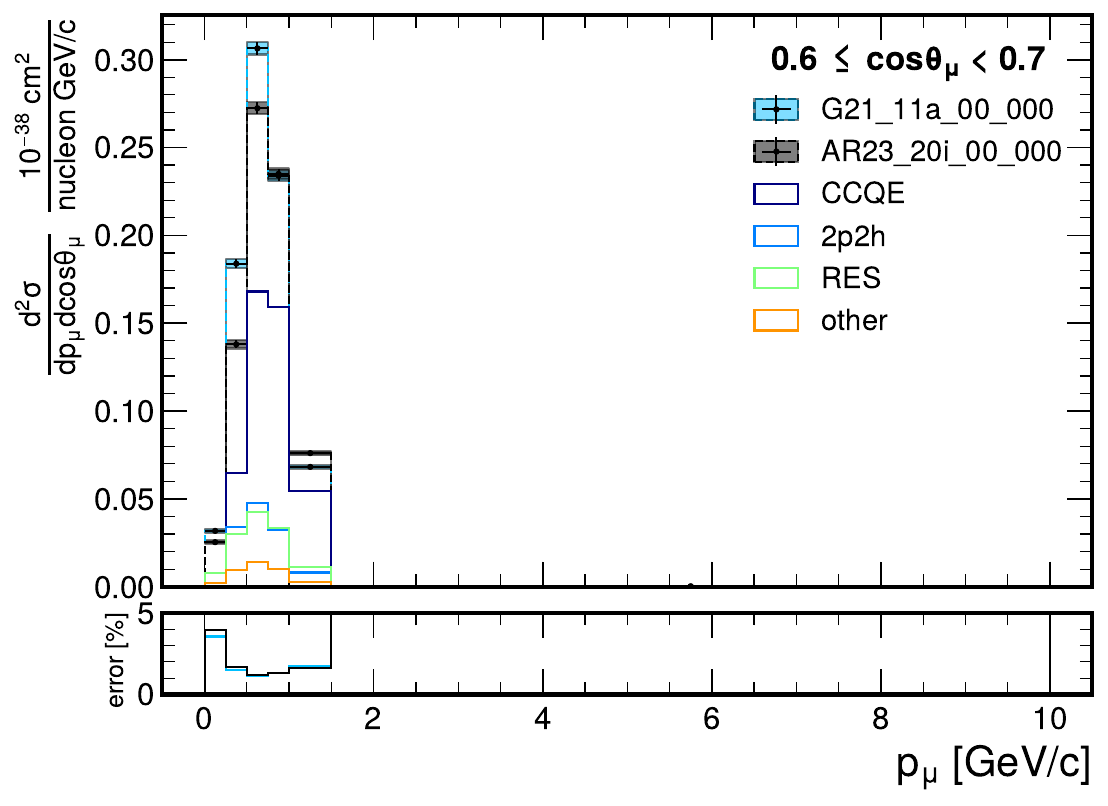}} \\
\subfloat{\includegraphics[width=0.34\textwidth]{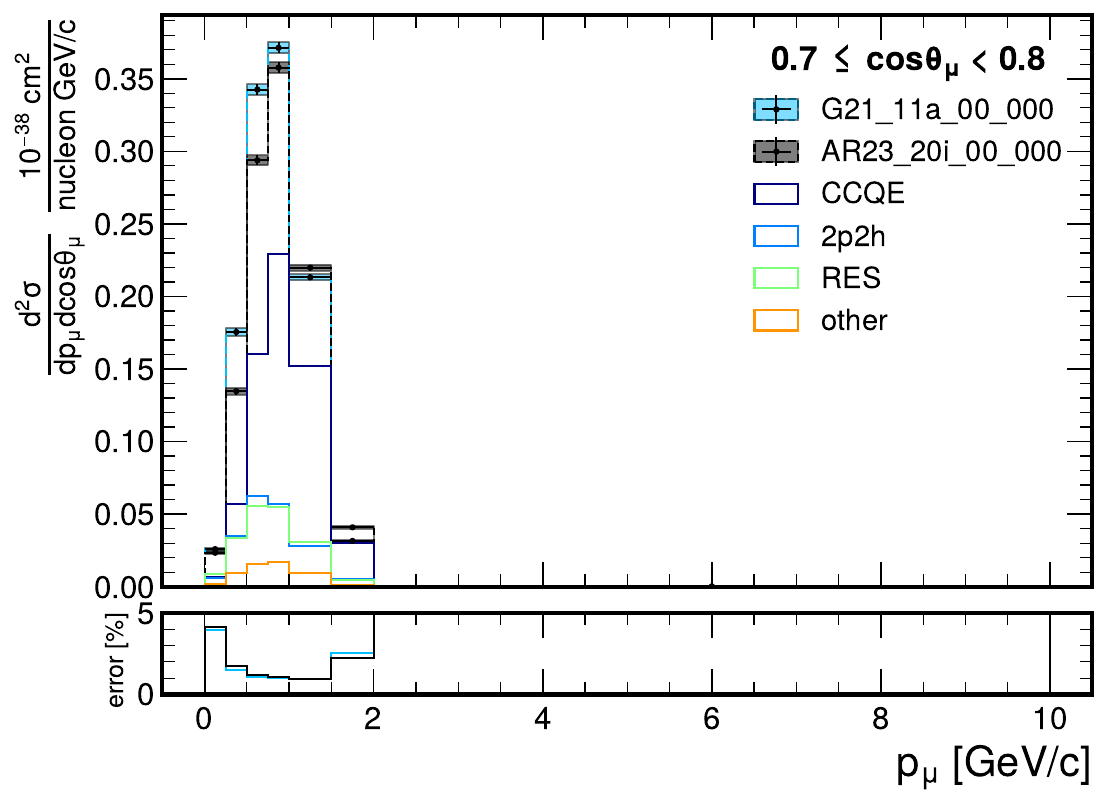}}
\subfloat{\includegraphics[width=0.34\textwidth]{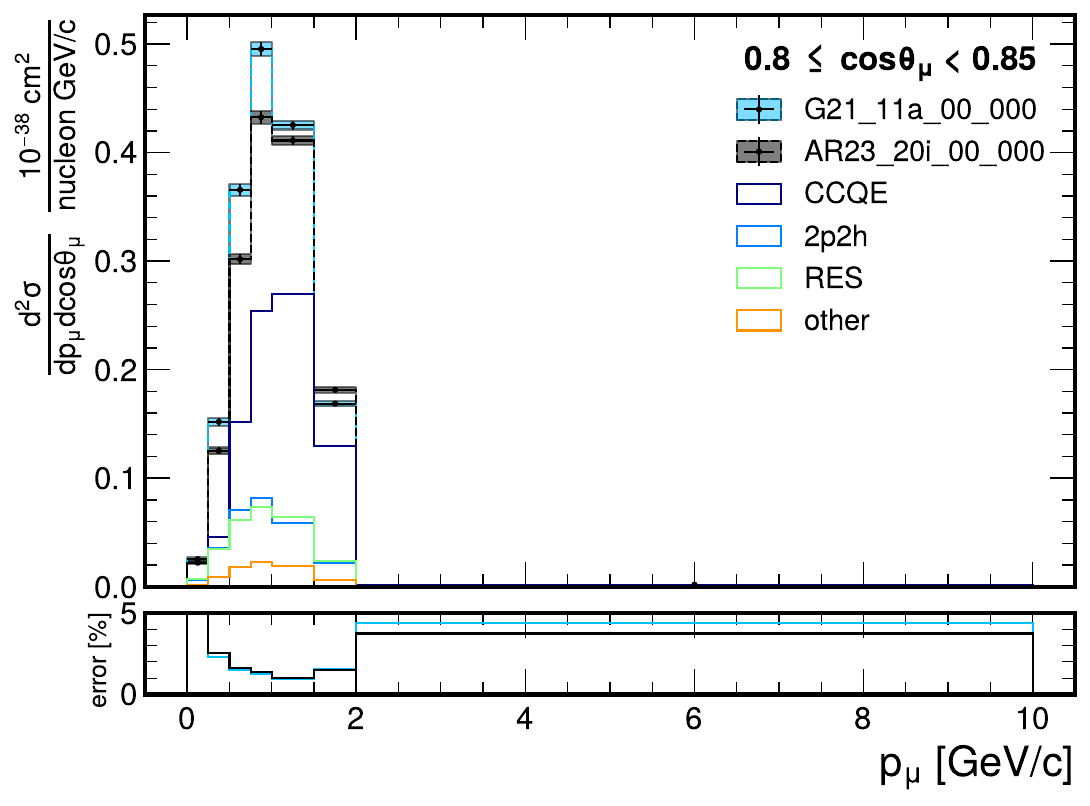}}
\subfloat{\includegraphics[width=0.34\textwidth]{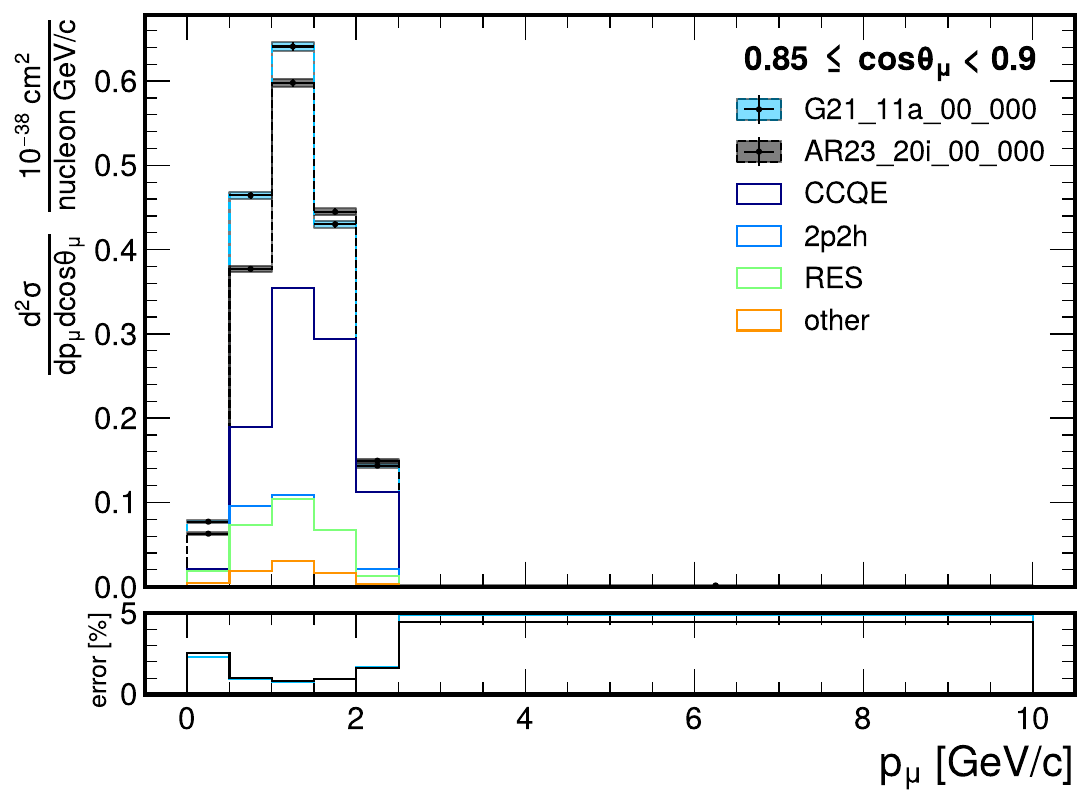}} \\
\subfloat{\includegraphics[width=0.34\textwidth]{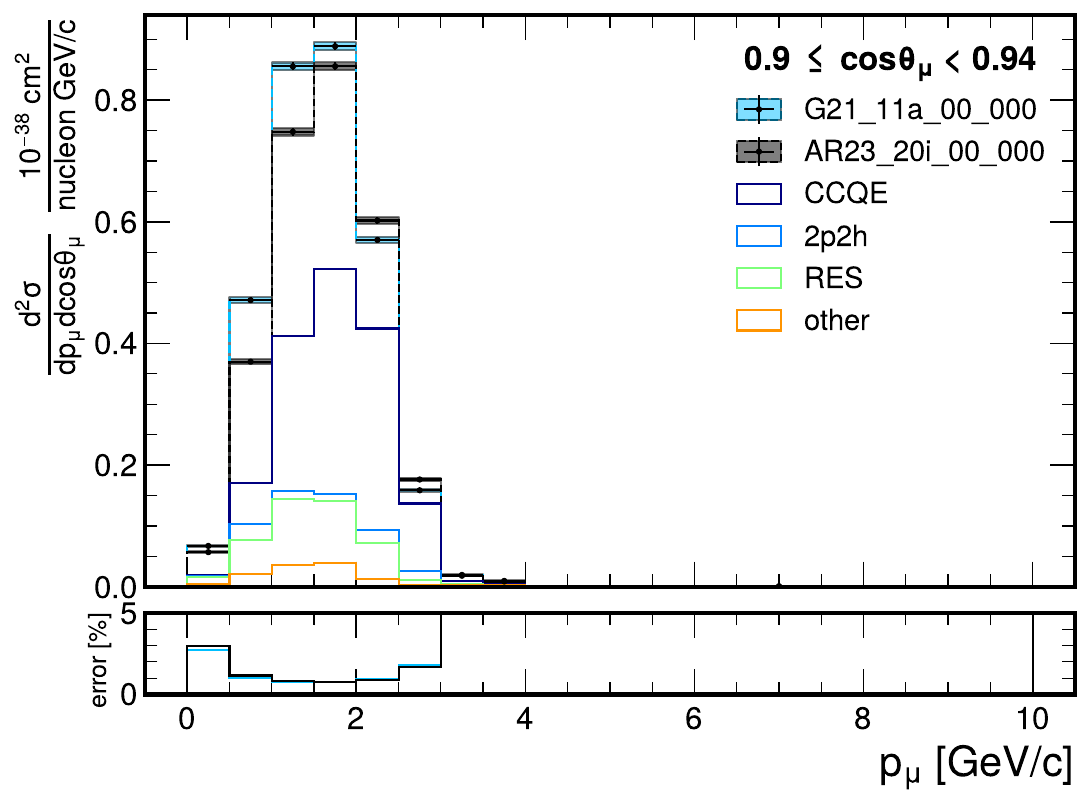}}
\subfloat{\includegraphics[width=0.34\textwidth]{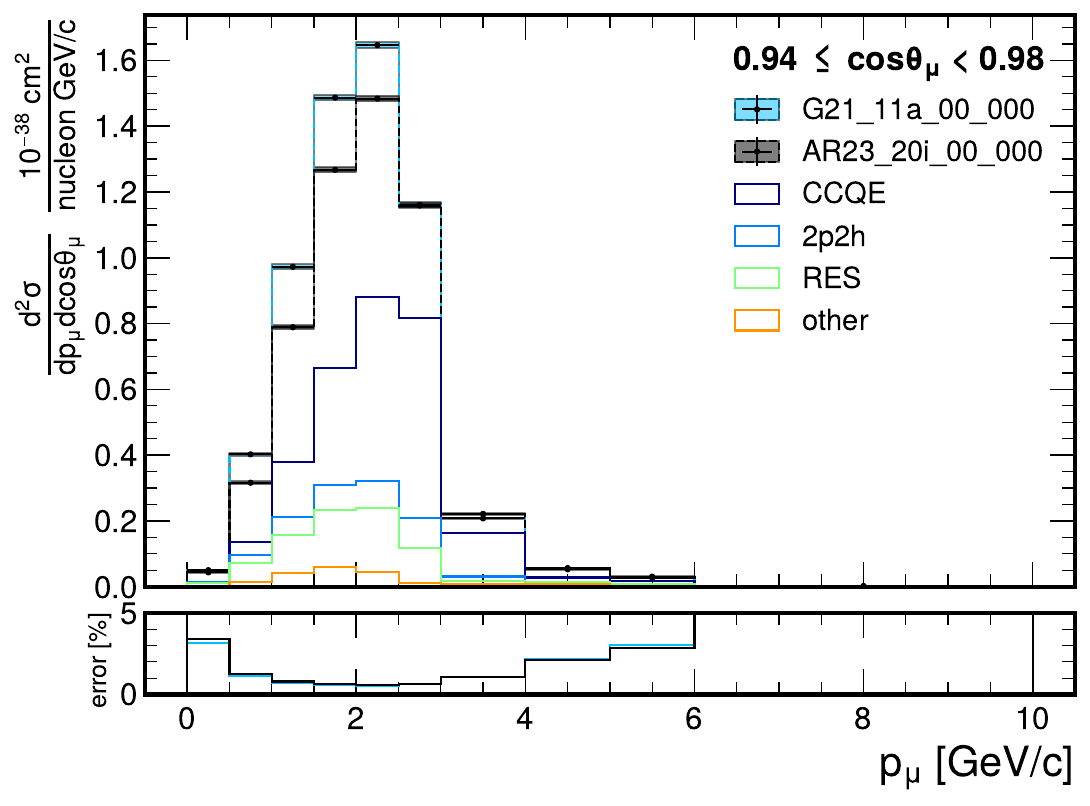}}
\subfloat{\includegraphics[width=0.34\textwidth]{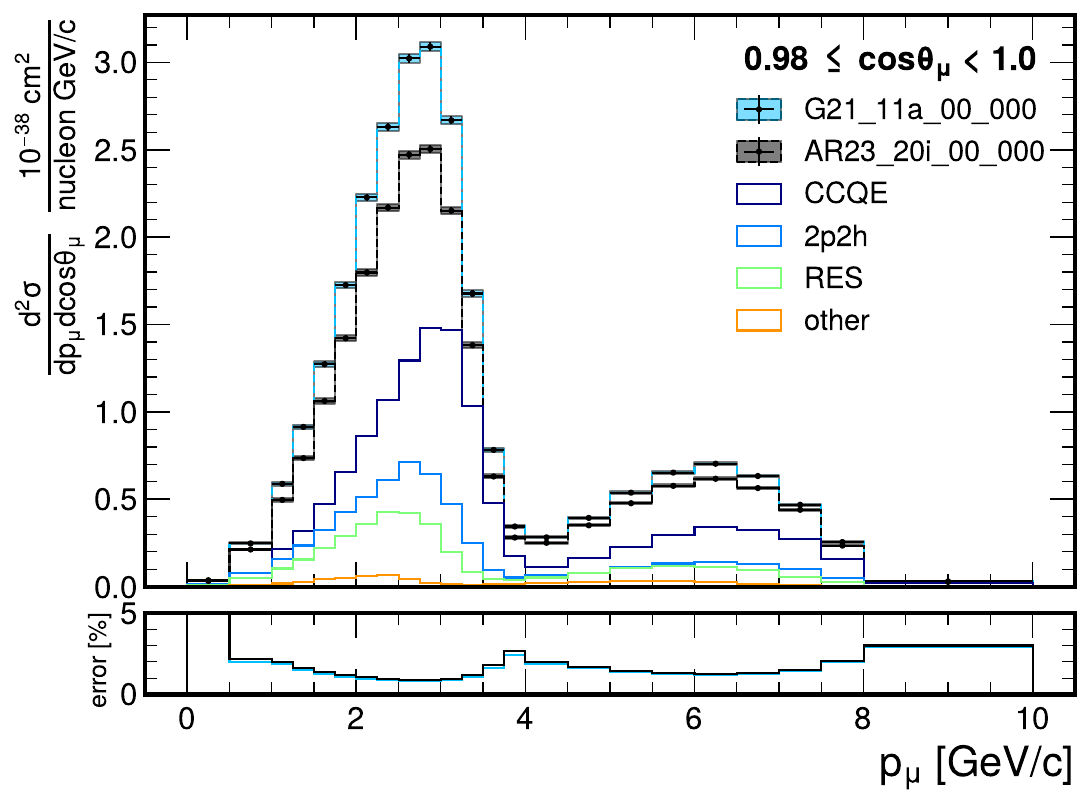}} \\

\caption{Projected measurement of \numu flux averaged cross-section for \cczeropi events as a function of muon momentum \pmu in bins of \cosmu, using the \argenie model (black line) and the \susagenie model (blue line) with error bars. The \argenie prediction is also broken down by true interaction mode, with each mode represented by solid colored lines, as indicated in the legend. The statistical error associated with each measurement is shown beneath each of the main figures}
\label{fig:d2sigma_dpmu_dcostheta_interaction_mode_CC0pi}
\end{figure}

The interest of measuring the differential cross-section as a function of outgoing lepton kinematics is that the latter maps to the momentum and energy transfer in a neutrino scatter, averaged over the range of available neutrino energies. In addition, inclusive predictions as a function of lepton kinematics are often the only type of predictions available inside neutrino generators from different models. A projected measurement of this double-differential cross-section, using the \argenie model, as a function of the muon momentum \pmu and direction with respect to the beam axis \cosmu is shown in \autoref{fig:d2sigma_dpmu_dcostheta_interaction_mode_CC0pi}.

The choice of \pmu and \cosmu binning was not optimized for the detector setup. Instead, we use the \cosmu binning employed by the T2K experiment in the measurement reported in~\cite{T2K:2023qjb}. The T2K \pmu binning was refined to have at least 50 expected events in each bin. From \autoref{fig:d2sigma_dpmu_dcostheta_interaction_mode_CC0pi}, it is apparent that the dominant channel across all \cosmu ranges is the CCQE process, but we also see contributions from 2p2h, RES and other processes, as expected for this topology. The muon angle is correlated with the energy transfer $\omega$ in the interaction -- it is clear that the range of muon momenta increases with increasing values of \cosmu. Importantly, we report the projected statistical error on the cross-section measurement, assuming two different neutrino interaction models. Across all bins, this error is below 10\%. In the peak regions, the error is below 1\%. It is important to note that this level of statistical uncertainty is comparable to that of conventional cross-section experiments, and at this level of precision, such measurements become systematically limited. However, the largest source of systematic uncertainty on conventional measurements comes from the modeling of flux uncertainties, at the level of 10\%, whereas the monitored neutrino beam allows this source to be reduced to the level of 1\%. 

The projected measurement is compared with one using a different model, \susagenie. The two models predict different values for the differential cross-section across all angular regions. The statistical power of the projected measurement is enough to discriminate between the two models, and the different kinematic regions are sensitive to different aspects of modeling differences. For example, at high scattering angles and intermediate momenta, the models differ primarily in the strength of the cross-section for resonant pion production processes. At forward scattering angles, the models apply different treatments for the description of collective nuclear effects (such as nuclear screening). The latter, in particular, is one of the dominant sources of systematic uncertainties for neutrino oscillation experiments, whether using beam or atmospheric neutrinos.. 

The same measurements presented here could be achieved using a water-based target, allowing similar precision measurements with compelling model discrimination potential for Hyper-Kamiokande. For a Hyper-Ka\-mio\-kan\-de-o\-rien\-ted analysis, it may be useful to perform this measurement using only the most off-axis NBOA fluxes, thereby tailoring the analysis to the most relevant neutrino energies. 

\subsubsection{\nue double differential cross-section}
\label{sec:nue_doublediff}

\begin{figure}[htbp]
\subfloat{\includegraphics[width=0.5\textwidth]{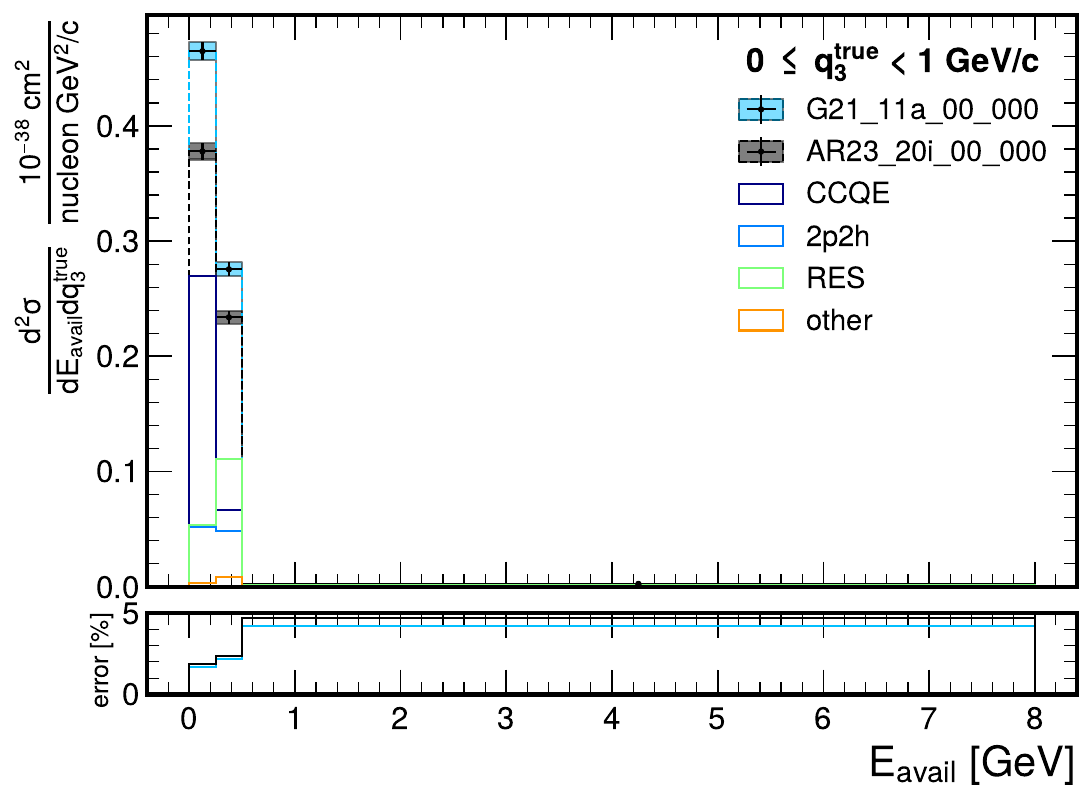}}
\subfloat{\includegraphics[width=0.5\textwidth]{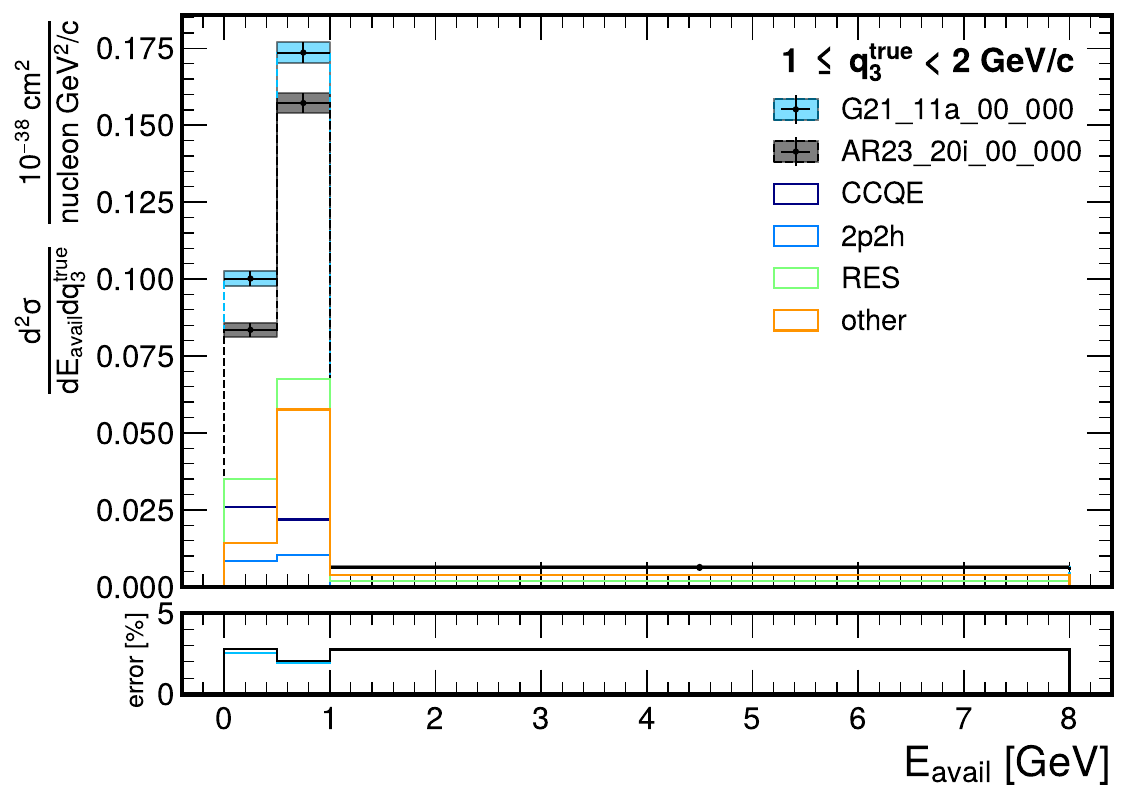}} \\
\subfloat{\includegraphics[width=0.5\textwidth]{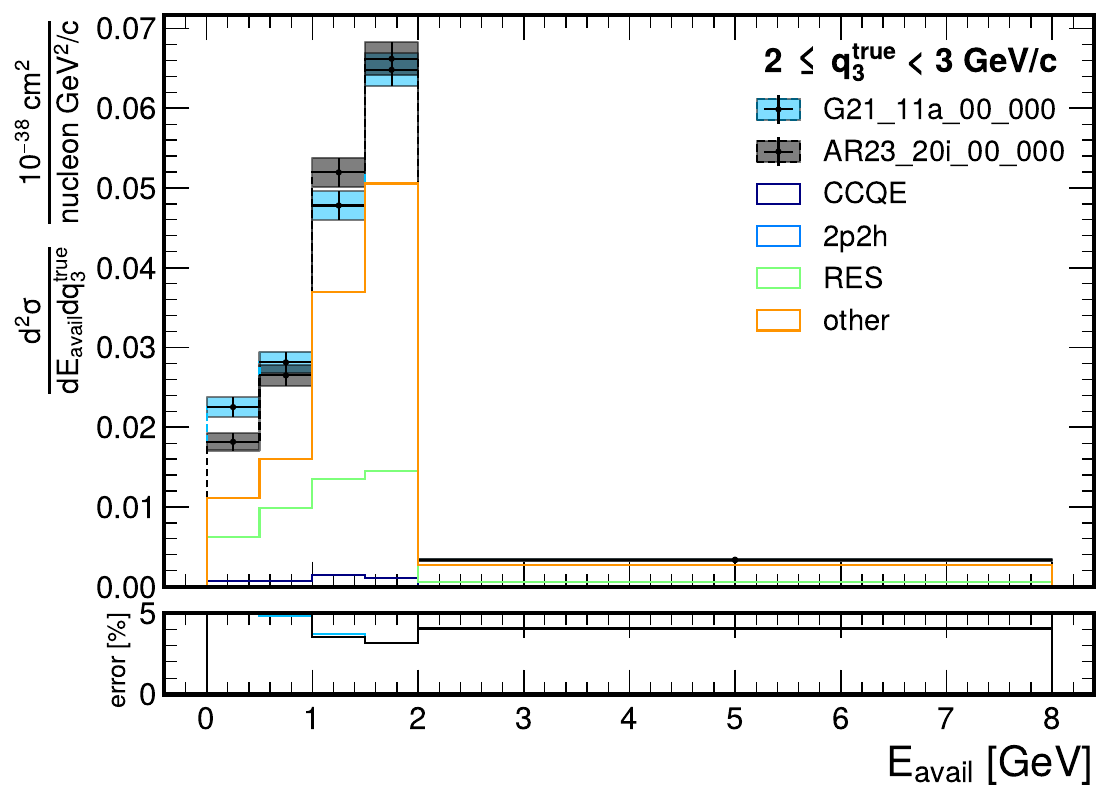}}
\subfloat{\includegraphics[width=0.5\textwidth]{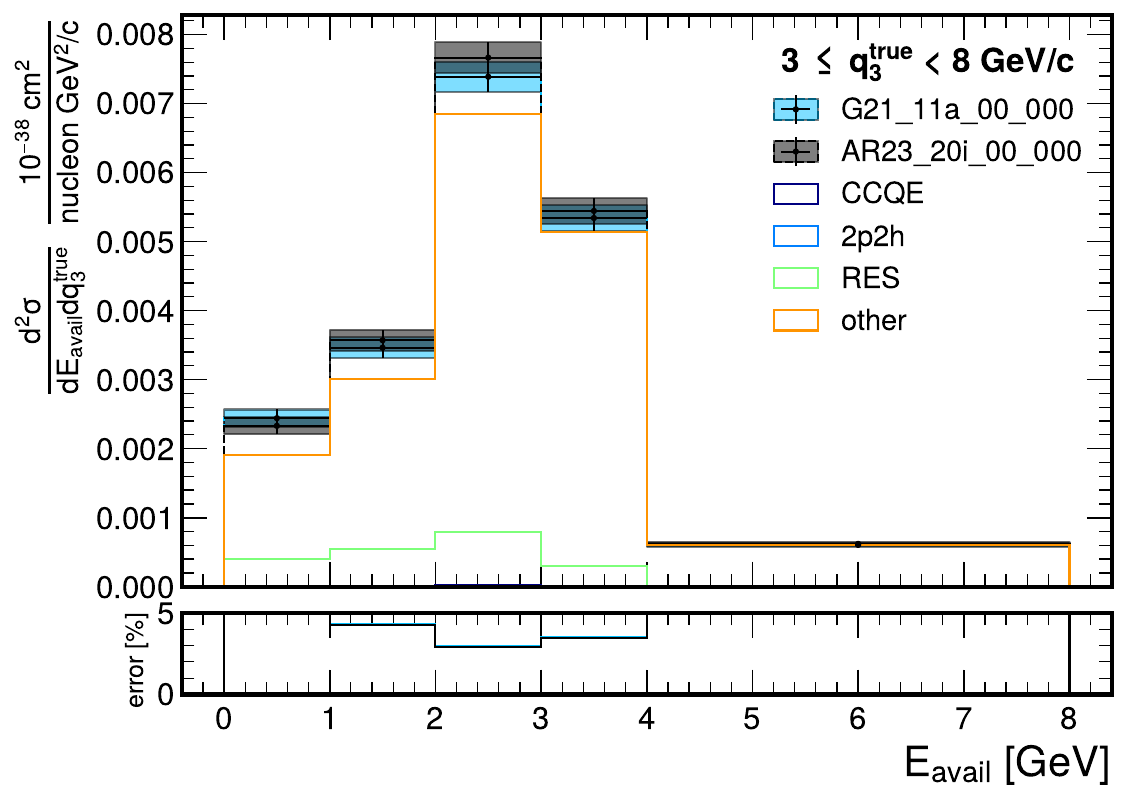}} \\

\caption{\nue flux averaged cross-section for CC inclusive events as a function of available energy \eavail in true three-momentum transfer $q_3^{\text{true}}$ bins, broken down by interaction mode.}
\label{fig:nuedifxsec}
\end{figure}

Similarly to the projected measurement for muon neutrinos, there is also interest in measuring the differential cross-section for electron neutrinos. We report a projected measurement in \autoref{fig:nuedifxsec}, as a function of calorimetric observables, \eavail and \qthree. The available, or recoil, energy \eavail is the calorimetric sum of the outgoing hadronic state. It is defined as:
\begin{equation}
\eavail = \sum_{i=\pi^{\pm}, p} T_i + \sum_{i=\pi^0,\gamma} E_i,
\end{equation}
where $\sum_{i=\pi^{\pm}}T_i$ is the sum of proton and pion kinetic energies, and $\sum_{i=\pi^0,\gamma} E_i$ is the sum of total energies deposited by neutral pions and photons. This is equivalent to MINERvA’s definition of \eavail~\cite{MINERvA:2023ner}. This variable is particularly interesting as it provides a proxy for the energy transfer in a neutrino interaction. \qthree is the projection of the momentum transfer vector onto the incoming neutrino direction. It is calculated as
\begin{align*}
    \qthree &= \sqrt{Q^2+q_{0}^{2}}\\
    Q^2 &= 2(E_{lep}+q_0)(E_{lep}-|\vec{p}_{lep}|\cos\theta_{lep})-m_{lep}^2
\end{align*}
where $Q^2=-q^2$ is the square of the four momentum transfer and $q_0=\omega$ is the energy transfer. $E_{lep}, \vec{p}_{lep}, \cos\theta_{lep}$ and $ m_{lep}$ are the outgoing lepton energy, momentum vector, direction with respect to the neutrino beam, and lepton mass, respectively.
For this illustration, we assume that the reconstructed \qthree from particle kinematics has been unfolded to the true value of \qthree \cite{MINERvA:2015ydy}. We note that this is a model-dependent procedure when using a wide neutrino beam, but the model dependence is mitigated at nuSCOPE by using tagged neutrinos, whose energy is well-known. We will describe the prospects for such measurements in \autoref{fig:nutag_escat_omega}.

As in the case of the \numu differential cross-section measurement, the projected statistical errors are below 5\% for the vast majority of analysis bins. As in the \numu CC double-differential measurement, we compare the reference model to the alternative \susagenie model. Large differences between the two models can be seen, in particular, at low values of \eavail and low \qthree. This is a region corresponding to low energy and momentum transfer, where differences between \numu and \nue cross-sections become significant due to the relative contributions of the lepton mass differences. This proof of principle illustrates that this measurement can be achieved with very good statistical precision, and, as in the case of the \numu CC double differential measurement, with low associated flux uncertainties of $\sim 1\%$.

Whilst such a measurement could be repeated on a water target, the $\nu_e$ flux is farther from Hyper-Kamiokande's region of interest than is the case for DUNE. However, running with a lower energy meson beam or making further kinematic constraints on the analyzed events may be able to provide a more tailored $\nu_e$ cross-section measurement for Hyper-Kamiokande.  

\subsection{Measurement of the \nue/\numu cross-section ratio}
\label{sec:prism_studies}

One of the largest sources of systematic uncertainty for future oscillation experiments is related to the differences between electron and muon neutrino cross-sections. Oscillation experiments measure \nue appearance in a \numu beam, but have few direct constraints on the \nue cross-section  -- the latter often represent small intrinsic contamination in the neutrino flux probed with the near detector, and they span energies often higher than those of appearance events. The strategy of oscillation experiments is to extrapolate the constraints obtained at the near detector using primarily muon neutrinos, by relying on models to translate the constraints on the electron neutrino cross-section. If lepton universality is assumed, the only differences between \numu and \nue cross-sections are related to terms in which the lepton mass becomes significant, which become important at relatively low energy transfers~\cite{Nikolakopoulos:2019qcr,Dieminger:2023oin}. In these regions, nuclear models predict differences of the order of 3\%~\cite{Dieminger:2023oin} on the \nuemuxsec ratio. To be competitive with the differences predicted by theoretical models, a direct measurement of \nuemuxsec would require an associated error of less than 3\%. 

Such a direct measurement is challenging to perform, as current artificial beams do not produce similar fluxes of \numu and \nue, meaning that cross-section measurements of each are integrated over different fluxes and cannot be directly compared.. In this section, we show the prospects of measuring the \nuemuxsec ratio with a monitored neutrino beam using the NBOA technique to obtain a virtual flux of muon neutrinos that matches the shape of the flux of electron neutrinos from three-body decays. This measurement employs the PRISM technique, which consists of performing linear combinations of real fluxes in order to obtain an arbitrary target flux shape. These fluxes will be used to extract the total flux-averaged cross-section for \nue and \numu interactions, and their associated ratio. More information about the PRISM technique can be found in~\cite{nuPRISM:2014mzw}.

\begin{figure} 
    \centering
    \includegraphics[width=0.51\textwidth]{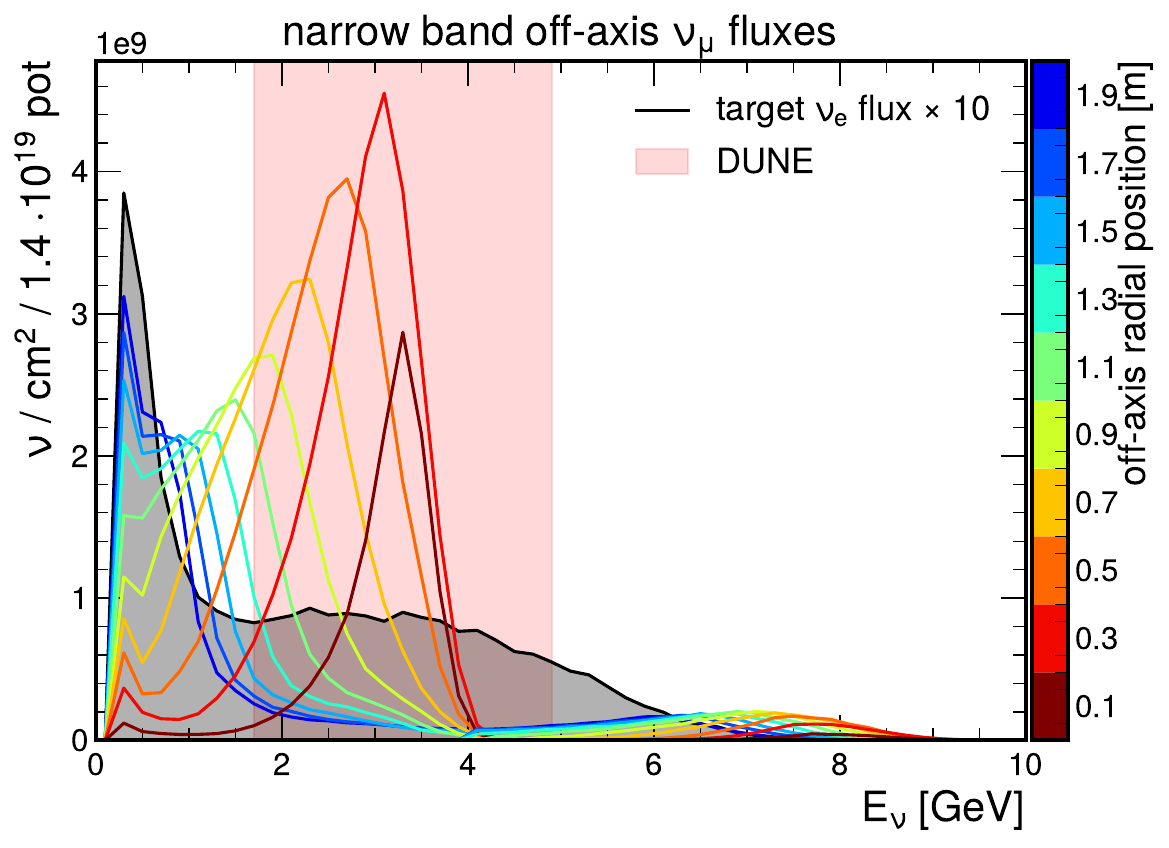}
    \includegraphics[width=0.47\textwidth]{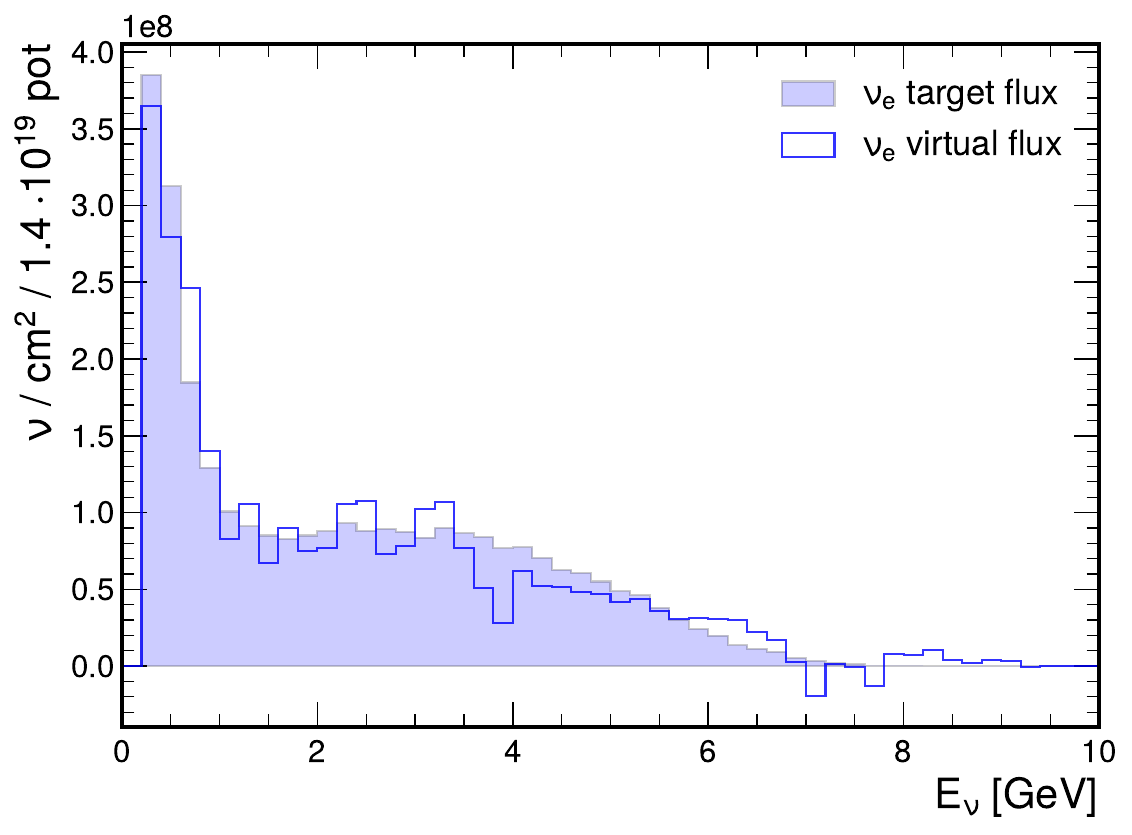}
    \caption{Left: \nue target flux and \numu narrow band off-axis fluxes. The \nue target flux is shown in grey and inflated by a factor of 10 to make it comparable with the \numu fluxes. The colored lines represent different NBOA fluxes, and the line color corresponds to the color scale showing the radial position with respect to the center of the exposed detector face. Right: the virtual \numu flux (blue lines) obtained by applying the PRISM technique, compared with the target \nue flux (filled blue area). }
    \label{fig:prism_nboa}
\end{figure}

The \numu NBOA fluxes obtained at different radial positions with respect to the center of the neutrino detector can be linearly combined to create a \emph{virtual flux}, $\varphi$, to reproduce the shape of a given \emph{target flux}, $\phi$. We build a virtual flux to match an arbitrary target flux $\phi (\enu)$  from a set of linear combinations of off-axis fluxes $\Phi_j (\enu)$ measured at different off-axis (i.e., radial) positions $j$ according to:
\begin{equation} \label{eq:prism_id}
    \phi (\enu) = \sum_j c_j \Phi_j (\enu).
\end{equation}
where $c_j$ are the coefficients for the linear combination. In this analysis, we set the \nue flux as the target flux and the NBOA off-axis fluxes as the input fluxes (split into the pion and kaon parent contributions with a cut at $\sim \SI{4}{GeV}$ neutrino energy as described in \ref{sec:numu_ccinc}). The different input and target fluxes are shown in the left plot of \autoref{fig:prism_nboa}.

It is important to stress that the set of linear equations encoded does not have a unique solution, since it is an ill-posed linear algebra problem.
If the set of equations were solved with a naive least-squares analysis, statistical fluctuations in the target flux would lead to large variations in a highly degenerate solution~\cite{Hasnip:2023ygr}. Any subsequent analysis using the virtual flux would have very large statistical uncertainties stemming from the implicit subtraction of large numbers in \autoref{eq:prism_id}. To choose a solution with less variance, we employ Tikhonov regularization~\cite{10.1145/321105.321114, tikhonov1977solutions} to find a stable approximated solution, where the variations between adjacent elements of $\vec{c}$ are reduced. This introduces a bias to reduce the variance, which can be adjusted via a regularisation strength. We set the strength to ensure \%-level precision on the resultant projected virtual-flux-integrated cross-section measurement. Details about the application of this regularization method can be found in~\cite{Hasnip:2023ygr}. To facilitate a stable solution, we use a coarse bin width for the analysis ($\SI{0.2}{GeV}$) and remove the first bin from the analysis (where the cross-section is negligible). The resulting virtual flux, compared to the target flux, is shown in the right figure of \autoref{fig:prism_nboa}. Even if there is room for further optimisation of the regularisation and binning, the fluxes match quite well. The remaining discrepancy could be accounted for using small model-dependent corrections, where the model dependence is further mitigated by neutrino tagging. 

Using the \argenie model, we obtain a virtual-flux-integrated \numu cross-section measurement with a 2\% statistical uncertainty. A \nue flux-integrated measurement can be made using the \nue flux directly with an estimated statistical precision of $\sim$1\%. Therefore, assuming the residual differences in the virtual flux can be accounted for, we project a measurement of the \numu/\nue cross-section ratio averaged over the \nue flux with a statistical precision of $\sim$2\%. The monitored beam is expected to facilitate flux systematic uncertainties at or below this level, and so a direct measurement of the \numu/\nue cross-section ratio at a level of precision that is comparable to theory calculations is achievable. As shown in \autoref{fig:prism_nboa}, this would be a measurement within the DUNE experiment's region of interest, providing a unique constraint on a source of systematic uncertainty that is currently projected to limit the experiment's sensitivity to constraining CP-violation. A measurement using a water-based target and a lower energy meson beam may be able to achieve the same result for Hyper-Kamiokande.

\subsection{Flux-averaged \numu NC $\pi^0$ cross-section measurement}
\label{sec:ncpizero}

In addition to providing measurements of the CC-inclusive cross-section, nuSCOPE is also well suited to provide a measurement of neutral current (NC) processes. NC interactions constitute a source of background for neutrino oscillation measurements. Although no charged lepton is produced at the primary vertex, hadrons or photons produced in NC interactions can be mis-reconstructed as electrons following photon conversion processes. This causes some NC events to be misattributed to CC events with a final state electron. This is notably one of the main sources of background for neutrino oscillation sources, in which the flavor of the lepton is used to measure electron (anti-)neutrino appearance. The main NC channel that contributes to this background is the production of neutral pions in NC interactions, which we will refer to as the \ncpizero topology. The neutral pions can be produced via a resonance, coherent scattering off the nucleus, or via neutral pion production during reinteractions inside the nucleus. This process has been measured by several neutrino scattering experiments (see e.g. a measurement performed by the MicroBooNE collaboration in ~\cite{MicroBooNE:2022zhr}, which we use as a reference for this study). Still, the measurements are limited to energies below $\sim$1 GeV. In this section, we present the prospects of measuring this process using a monitored neutrino beam and using the NBOA technique.

We follow the same approach as in \autoref{sec:numu_ccinc}. To reduce backgrounds, MicroBooNE separates the \ncpizero sample into two contributions from events with two photon showers ($2\gamma$) and either no tracked protons ($0p$) or one tracked proton ($1p$). We select events with a single $\pi^0$ in the final state and either zero or one proton with momentum above $\SI{300}{MeV/c}$ (corresponding approximately to the tracking threshold for a liquid argon detector such as MicroBooNE, quoted as a kinetic energy of 50 MeV).
We apply an efficiency correction factor of 10\% to the total number of \ncpizero events to match the efficiencies obtained by MicroBooNE in~\cite{MicroBooNE:2022zhr}.

\begin{figure}[!htp]
    \centering
	\begin{subfigure}{0.49\textwidth}
		\centering
        \includegraphics[width=\textwidth]{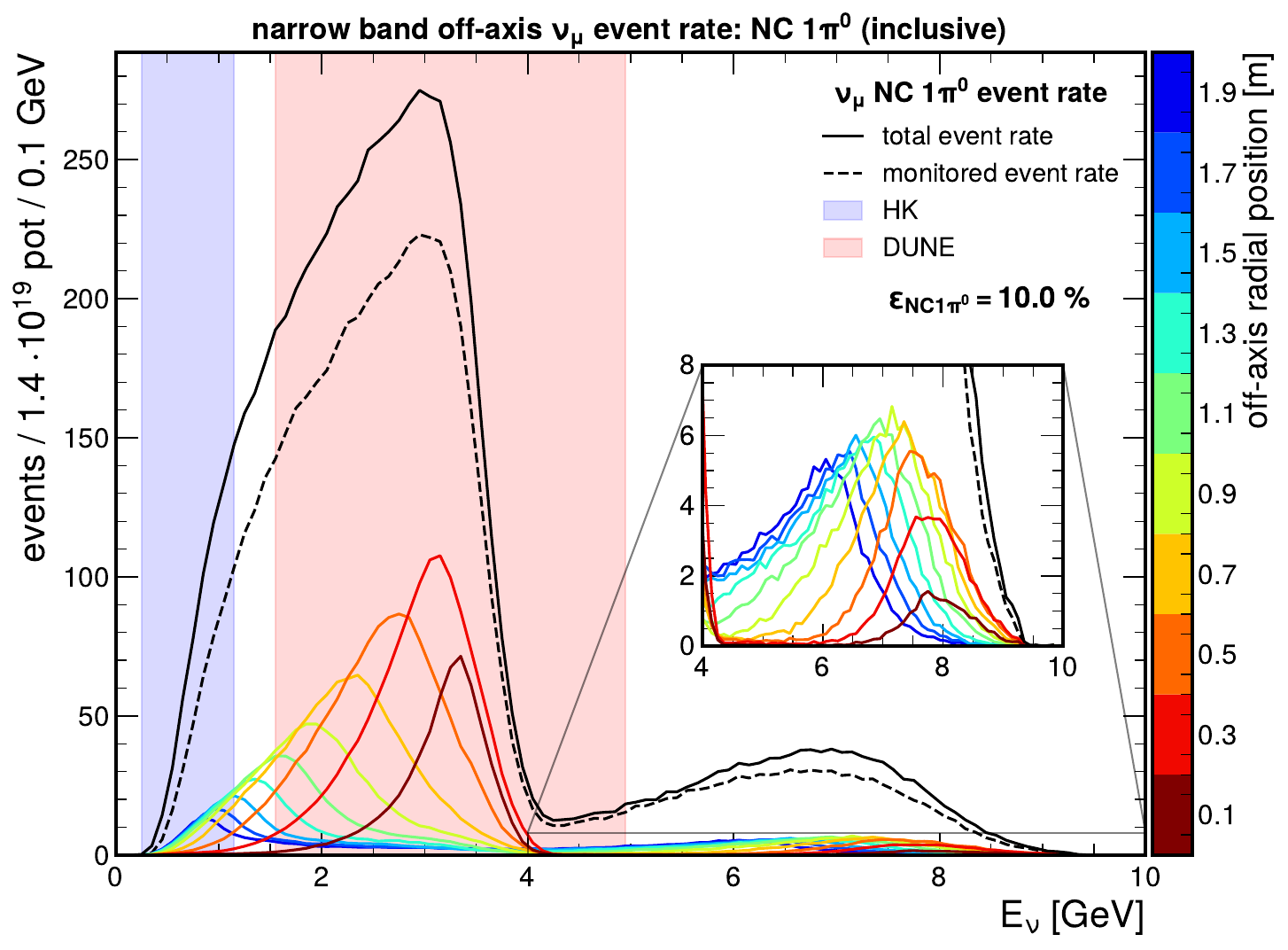}
        \caption{\ncpizero event rate}
        \label{fig:ncpi0_eventrate_nboa}
        \end{subfigure}
	\begin{subfigure}{0.49\textwidth}
		\centering
		\includegraphics[width=\textwidth]{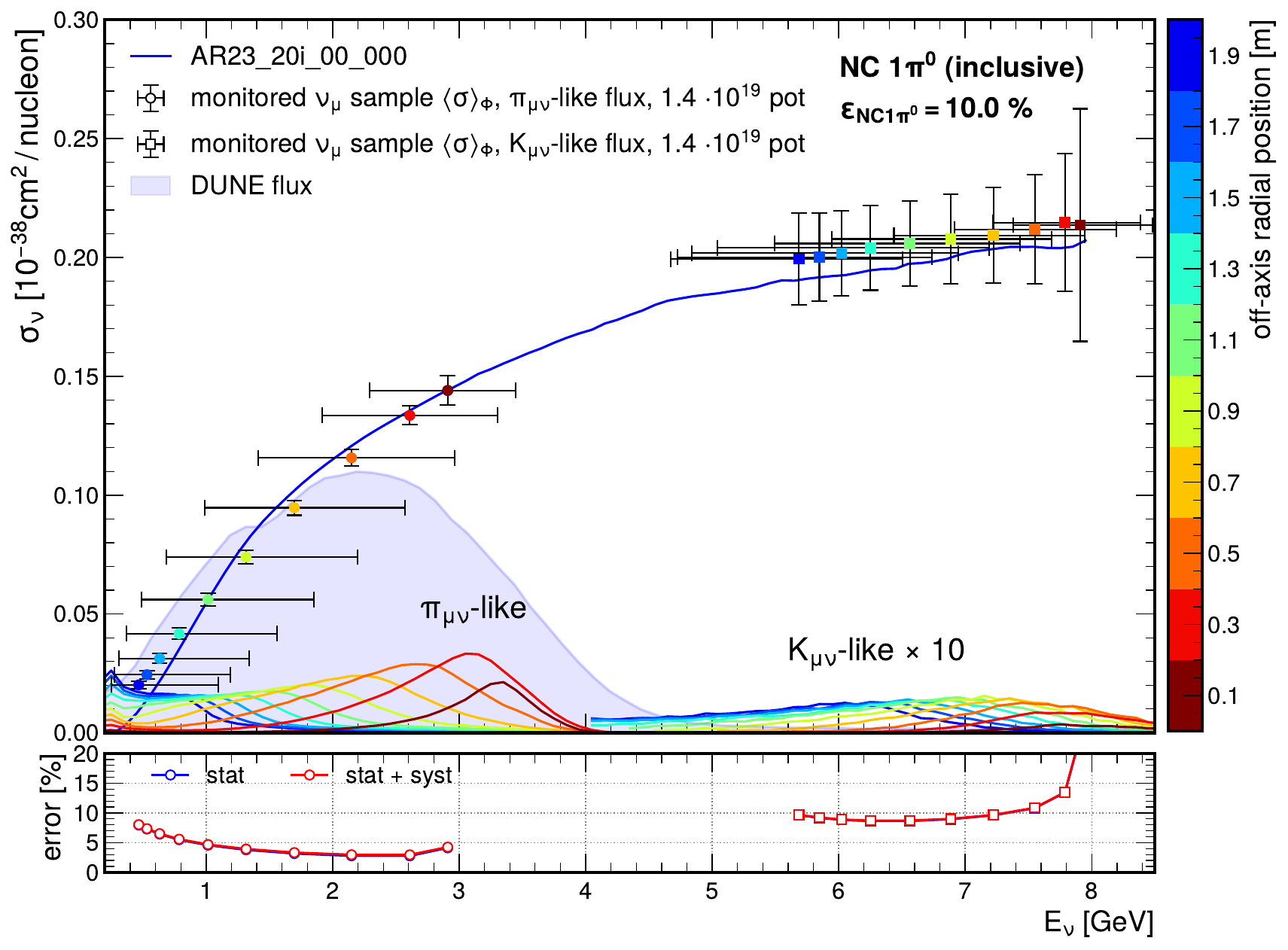}
    \caption{\ncpizero flux-averaged cross-section}
    \label{fig:ncpi0_inc}
    \end{subfigure}
    \caption{Expected performance of the setup for an \ncpizero measurement using the NBOA technique. The radial position is given by the associated color scale. The red and blue shaded areas highlight the DUNE and Hyper-Kamiokande regions of interest, respectively. The assumed selection efficiency ($\varepsilon_{\text{NC1}\pi^0}$) is of 10\%. (\textbf{Left}) Expected \ncpizero event rates using the GENIE \argenie model. The black line shows the expected event rate using the total flux, and the colored lines show the expected contributions at different radial positions in the detector. (\textbf{Right}) Flux averaged \numu \ncpizero inclusive cross-section as a function of neutrino energy. The \kdecay component of each flux has been artificially inflated by a factor of 10 for illustration purposes. Each NBOA flux has a corresponding predicted measurement point of the same color. Horizontal error bars encase the 68\% percentiles with respect to the mean energy for the NBOA fluxes. The underlying figure shows the size of the statistic (blue) and the statistic + systematic (red) errors corresponding to each measurement. The measurements are compared to the reference \argenie simulation (blue). The DUNE near-detector flux is shown for reference using an arbitrary normalization.}
\label{fig:ncpi0_results}
\end{figure}

The obtained \numu \ncpizero event rate using the NBOA technique is shown in \autoref{fig:ncpi0_eventrate_nboa}. Despite the reduction in statistics due to the efficiency correction and the lower cross-section for NC processes, the NBOA fluxes are still visible and well separated. 

Using each NBOA flux, we extract the flux-averaged inclusive \ncpizero cross-section\footnote{In this measurement, we report the ``total'' \ncpizero cross-section as the ratio of the event rates as predicted by GENIE using the \argenie model divided by the incoming monitored neutrino flux. Since \ncpizero is a final state topology rather than an elementary interaction channel, GENIE does not provide a single cross-section for these processes. In these results, any discrepancies between cross-section curves and measurement points are due to statistical fluctuations due to reduced statistics of this topology.} as a function of the incoming neutrino energy, shown in \autoref{fig:ncpi0_inc}. We note that the expected size of statistical errors is below 10\% across the majority of DUNE energies and below 5\% in the peak region. For reference, the statistical error associated with the MicroBooNE measurement is of $\sim 6\%$. However, it is important to note that the latter has an associated systematic error of $\sim 16\%$, of which the dominant contribution comes from neutrino flux uncertainties at the order of $\sim 12\%$. With the NBOA technique, the contribution of the flux uncertainty (assumed to be 1\%) becomes subdominant compared to the statistical error. Moreover, the MicroBooNE measurement provides a single flux-averaged cross-section measurement for a mean energy of $\sim \SI{0.8}{GeV}$. Using the NBOA technique, the cross-section for this process can be measured for a multitude of mean neutrino energies, covering the relevant energies for DUNE and Hyper-Kamiokande.

\subsection{Cross-section measurements with the tagged neutrino sample}
\label{sec:nutag_xsec_measurements}

In addition to the measurements using a monitored neutrino beam and the NBOA technique, nuSCOPE provides the unique opportunity to perform measurements using tagged neutrinos. This provides prospects for performing measurements where the neutrino energy is known on an event-by-event basis. In this section, we present several measurements that can be made by exploiting this new capability and discuss their impact for future generation experiments. 

\begin{figure}[!htbp]
	\centering
		\includegraphics[width=0.48\textwidth]{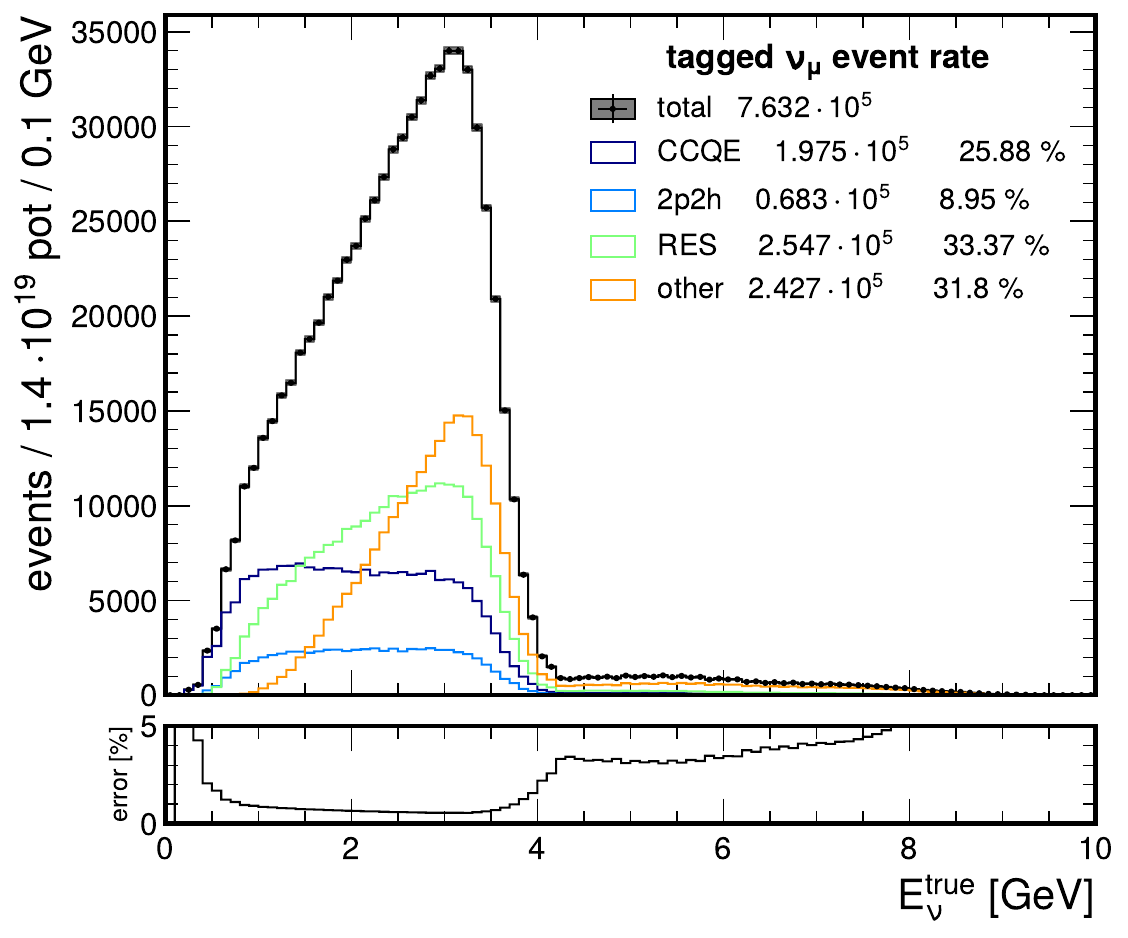}
        \includegraphics[width=0.5\textwidth]{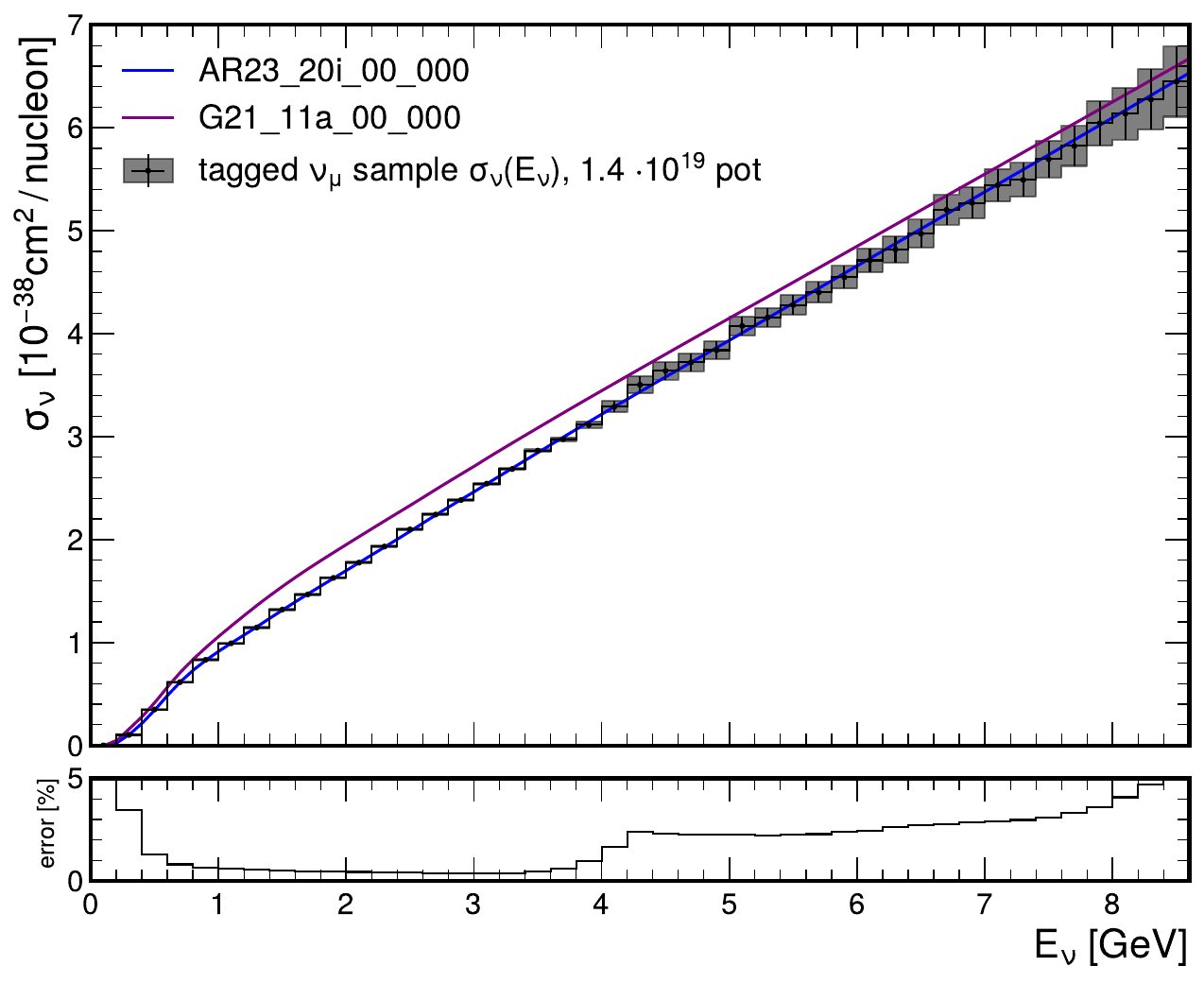}
        
        \caption{Left: Projected measurements of the event rate using tagged neutrinos as a function of the true neutrino energy derived from the tagging procedure. The total event rate (black line) is also broken down by the expected event rates from each interaction mode (colored lines). The total event rates and their percentages with respect to the total are given in the legend. Right: Projected measurement of the \numu CC inclusive integrated cross-section as a function of neutrino energy using tagged neutrinos. The reference model (\argenie, in blue) is compared to the alternative model \susagenie, shown in purple. The error bars represent the statistical error expected on the measurement, also shown below the main figure.}
	\label{fig:nutag_eventrates}
\end{figure}

First, the expected event rate from tagged neutrinos using the \argenie model is shown in \autoref{fig:nutag_eventrates}. Despite the reduction in statistics because tagged neutrinos can only be found in the decay tunnel, and taking into account the tagging efficiency, the total event rate is significant and comparable to that obtained using the NBOA technique discussed in \autoref{sec:xsec_eventrates}. As can be seen from the fluxes in \autoref{fig:tagging_fluxes}, the \kdecay contribution is smaller than the NBOA case due to the lower tagging efficiency in this region. The \pidecay contribution is similar to the one obtained with the monitored beam.

In the following sections, we present a few examples of measurements that can be achieved by exploiting the tagging technique.

\subsubsection{\numu CC-inclusive cross-section}
\label{sec:nutag_ccinc}

As a natural extension to the study shown in \autoref{sec:numu_ccinc}, the total \numu CC inclusive cross-section can also be measured with tagged neutrinos. The advantage of the tagged neutrino beam allows us to perform the measurement directly as a function of the neutrino energy, instead of extracting a flux-averaged cross-section for each off-axis position. The projected measurement is shown in the right plot of  \autoref{fig:nutag_eventrates}.

The neutrino energy binning employed for this measurement is uniform in steps of $\SI{200}{MeV}$, which is coarser than the neutrino energy resolution shown in  \autoref{fig:tagging_resolution}. The binning scheme can further be optimized in future studies.
The total statistics at each neutrino energy allow us to extract a cross-section with associated statistical uncertainties of less than 5\% across the entire considered energy range. The statistical error on the measurement is below $\sim 1\%$ between 1-$\SI{4}{GeV}$, corresponding to the higher statistics coming from \pidecay decays. The small increase in statistical error around $\SI{4.5}{GeV}$ is due to the transition region between the \pidecay and \kdecay populations, and the statistical error stays at the level of 2-3\% for the \kdecay region. It is important to note that thanks to the tagging, this measurement has negligible flux systematic uncertainties. 

\subsubsection{Measuring the neutrino energy bias}
\label{sec:nutag_energybias}

Due to their requirements for high-intensity neutrino beams, future neutrino experiments such as DUNE and Hyper-Kamiokande cannot employ the tagging and monitoring technique in their experimental setups. As a result, in order to infer the true neutrino energy from the interaction products, they rely on a neutrino interaction model that relates the amount of visible energy in the detector to the true neutrino energy. There are large uncertainties associated with neutrino interaction models, and this constitutes a large source of systematic uncertainty for oscillation measurements.

Using tagged neutrinos, whose true energy is known on an event-by-event basis, it is possible to measure directly the relationship between the true and reconstructed neutrino energies, as seen by future generation neutrino experiments. 
We examine the prospects for such a measurement for a liquid argon detector employing calorimetric energy reconstruction methods\footnote{We note that a different energy reconstruction method should be used for a water Cherenkov detector such as Hyper-Kamiokande, and the sources of bias will be different. Future studies will examine the prospect of neutrino energy bias measurements using kinematic reconstruction methods such as the one which will be used by Hyper-Kamiokande.}.

\begin{figure}[!htp]
	\centering
	\begin{subfigure}{0.49\textwidth}
		\centering
		\includegraphics[width=\textwidth]{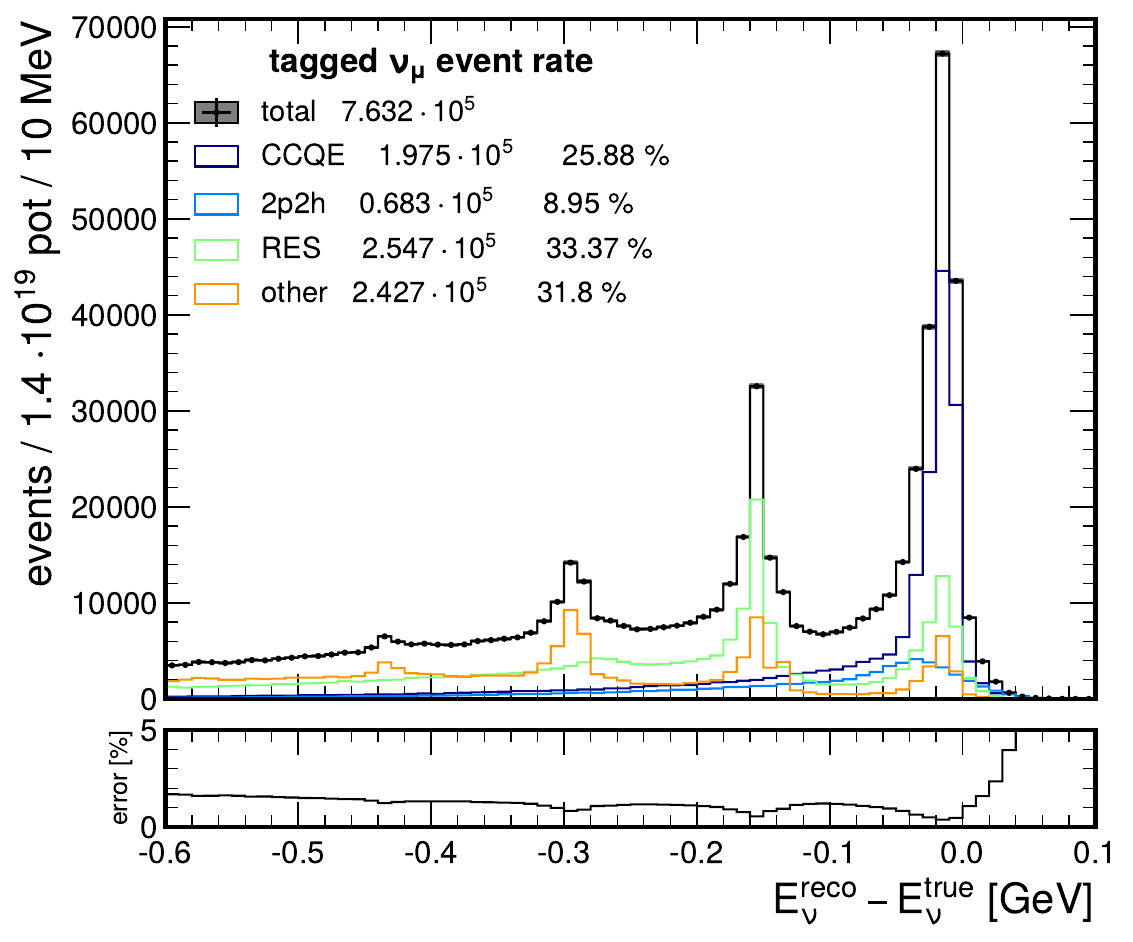}
		\caption{Absolute bias}
		\label{fig:nutag_absbias_bymode}
	\end{subfigure}
    \begin{subfigure}{0.49\textwidth}
    \includegraphics[width=\textwidth]{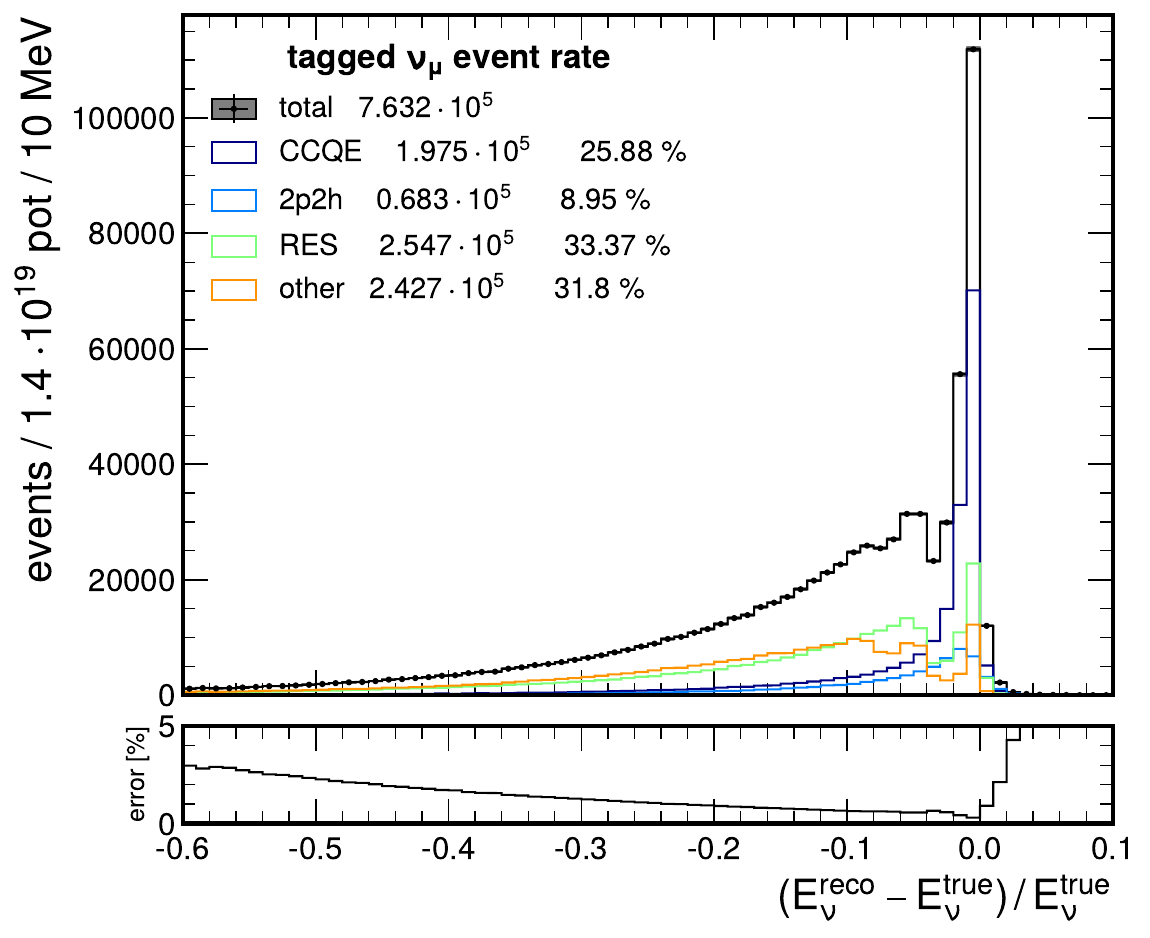}
    \caption{Relative bias.}
    \label{fig:nutag_relbias_bymode}
    \end{subfigure}
    \caption{Expected event rates using the tagging technique as a function of the absolute and relative neutrino energy bias, using the \argenie model. The distributions are split by true interaction mode, indicated in the legend. \enureco is defined as in \autoref{eq:enurec}, whilst \enutrue is the true neutrino energy derived from the tagging procedure. Error bars are given for the total predictions from the model. The colored lines show the breakdown by interaction channel as predicted by the model. Each figure is accompanied by an underlying plot showing the evolution of the statistical error on the cross-section measurement as a function of each quantity.}
    \label{fig:nutag_bias_evrate}
\end{figure}

The absolute bias in neutrino energy measurements is given by the difference between the reconstructed and true neutrino energy. For a liquid argon detector, we assume the reconstructed neutrino energy to be the sum of the visible energy deposited in the detector:
\begin{equation}
\enureco=E_\mu+\sum_{i=\pi^{\pm}, p} T_i + \sum_{i=\pi^0,\gamma} E_i,
\label{eq:enurec}
\end{equation}
where $E_\mu$ is the muon energy, $\sum_{i=\pi^{\pm}, p} T_i$ is the sum of kinetic energies of protons and charged pions and $\sum_{i=\pi^0,\gamma} E_i$ is the sum of total energies of neutral pions and photons. The difference between the true and reconstructed neutrino energy under this assumption will thus have three main contributions:
\begin{itemize}
    \item The missing energy due to the removal energy of nucleons inside the nucleus.
    \item For each charged pion, a contribution equal to one pion mass (i.e. $\SI{139.6}{MeV/c^2}$), as we assume that the number of pions is not determined in the analysis, as they might not always leave an identifiable Michel electron tag and a fraction of their energy is carried away by neutrinos.

    \item Invisible energy carried by undetected particles -- this mainly corresponds to neutrons produced primarily through final state interactions inside the nucleus (but some neutrons also come from the elementary interaction processes).
\end{itemize}
The expected event rates as a function of the absolute and relative neutrino energy bias are shown in \autoref{fig:nutag_absbias_bymode} and \autoref{fig:nutag_relbias_bymode}, respectively. From \autoref{fig:nutag_absbias_bymode}, it is visible that the distribution exhibits sharp peaks spaced by integer multiples of pion masses. The majority of single pions come from RES interactions. At higher energies, the pion peaks are populated by ``other'' interactions, which mainly correspond to deep or shallow inelastic scattering processes. Each peak has a width of the order of $\SI{50}{MeV}$, which corresponds to the span in nucleon removal energies associated with the ground state of the nucleus. The absolute bias distributions from pion production and CCQE events sit on top of a slowly decaying contribution, coming from all interaction modes, corresponding to events in which a fraction of the energy is carried away by neutrons. As can be seen in \autoref{fig:nutag_relbias_bymode}, the fraction of neutrino energy carried away by events producing charged pions and neutrons (primarily in the tail of the distribution) is responsible for a fraction of invisible energy ranging between 10\%-50\%.

\begin{figure}[!htp]
	\centering
	\begin{subfigure}{0.5\textwidth}
		\centering
		\includegraphics[width=\textwidth]{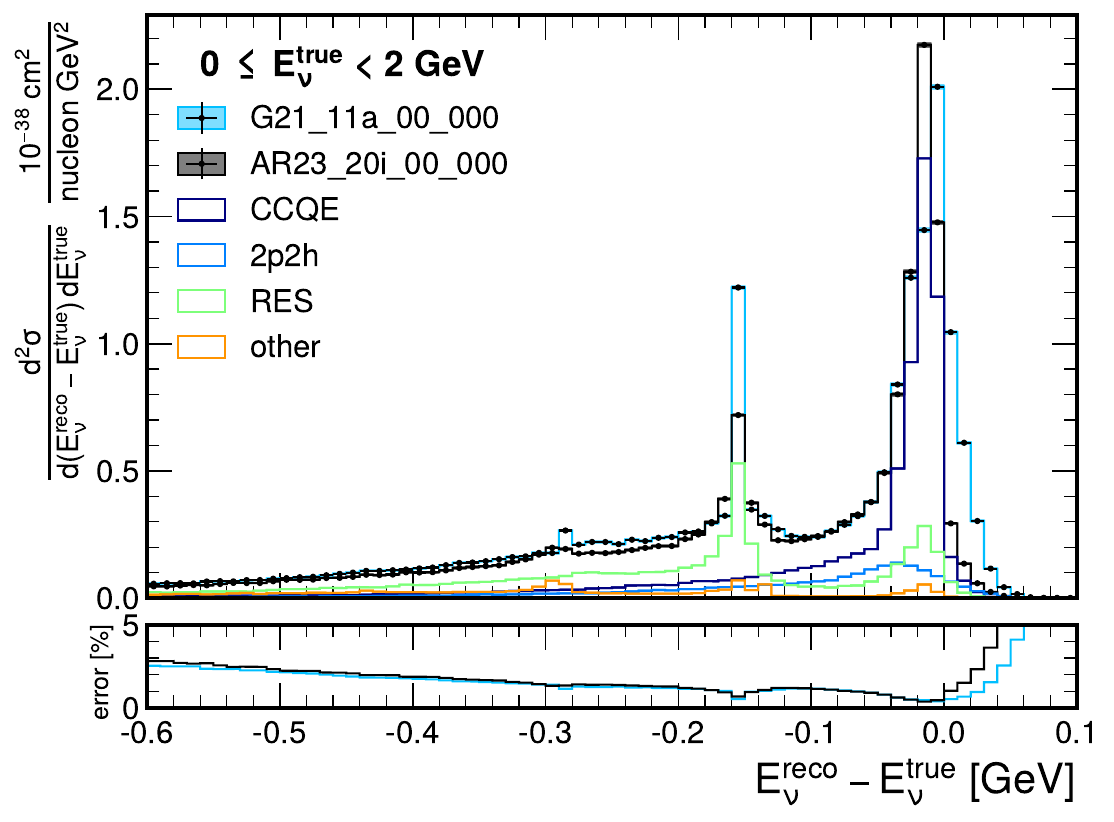}
		\label{fig:nutag_enubias_abs_xsec_low}
	\end{subfigure}
    \begin{subfigure}{0.49\textwidth}
    \includegraphics[width=\textwidth]{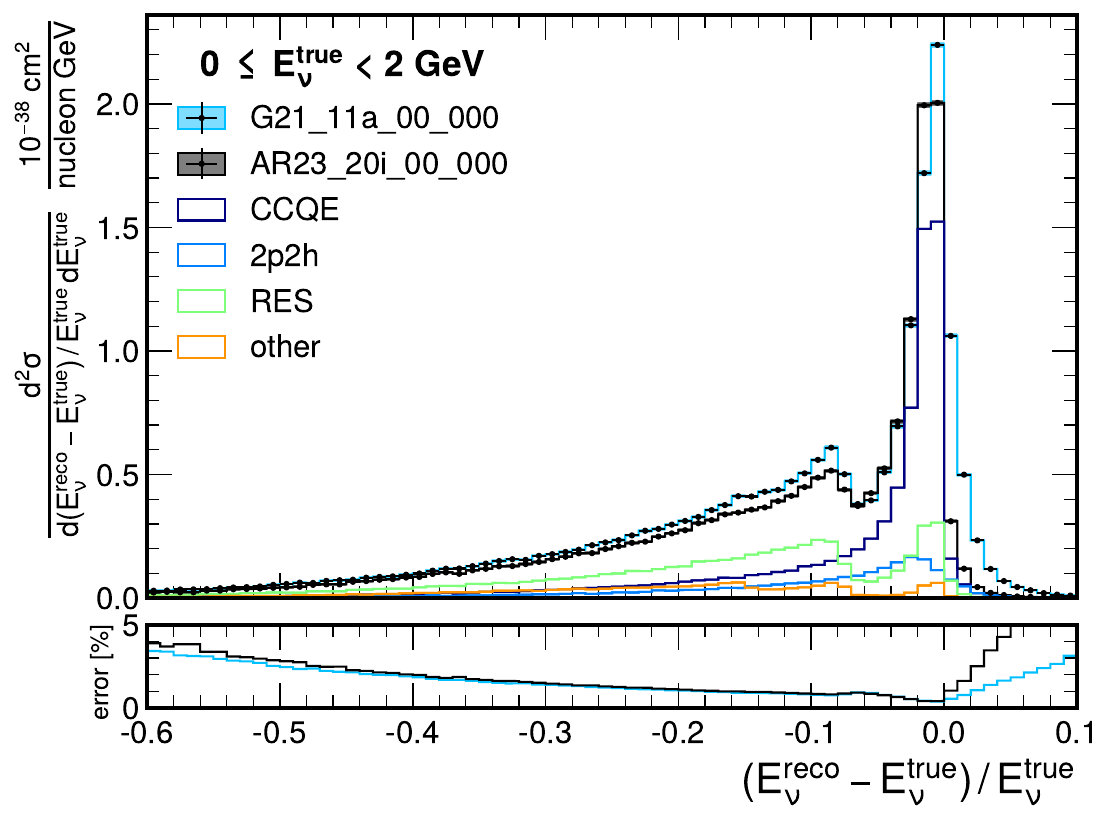}
    \label{fig:nutag_enubias_rel_xsec_low}
    \end{subfigure}

	\begin{subfigure}{0.5\textwidth}
		\centering
		\includegraphics[width=\textwidth]{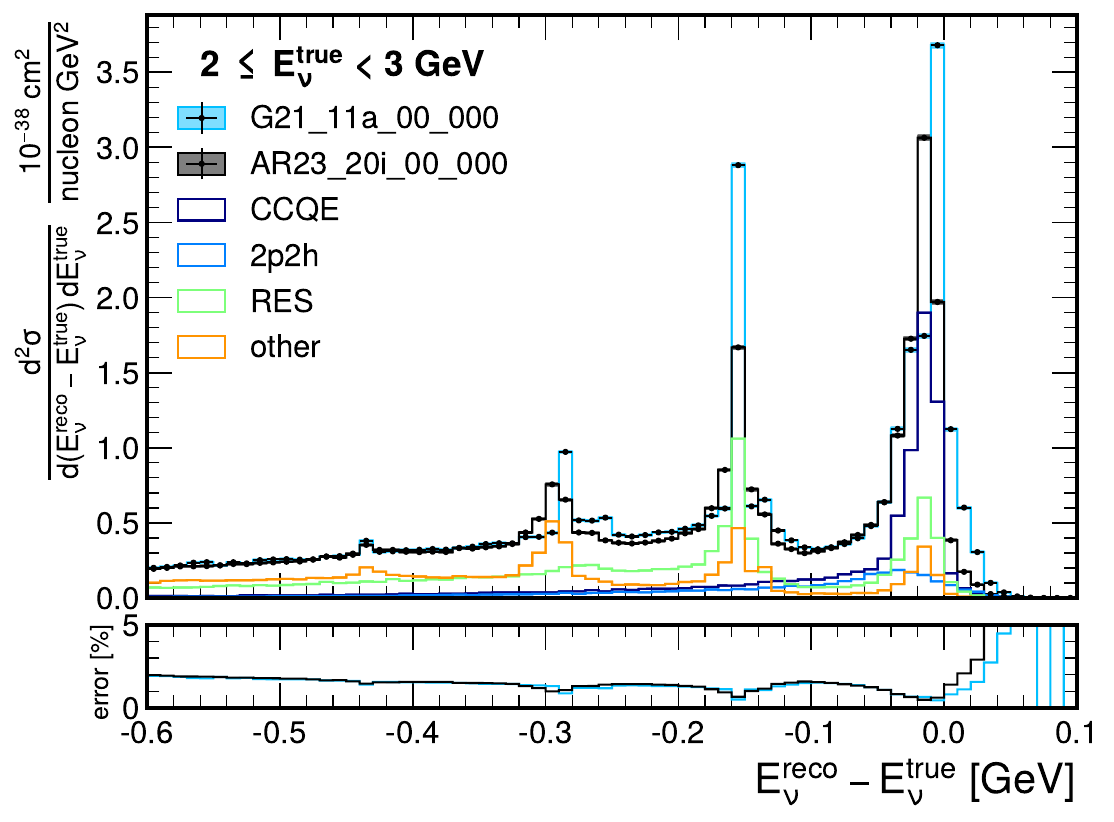}
		\label{fig:nutag_enubias_abs_xsec_mid}
	\end{subfigure}
    \begin{subfigure}{0.48\textwidth}
    \includegraphics[width=\textwidth]{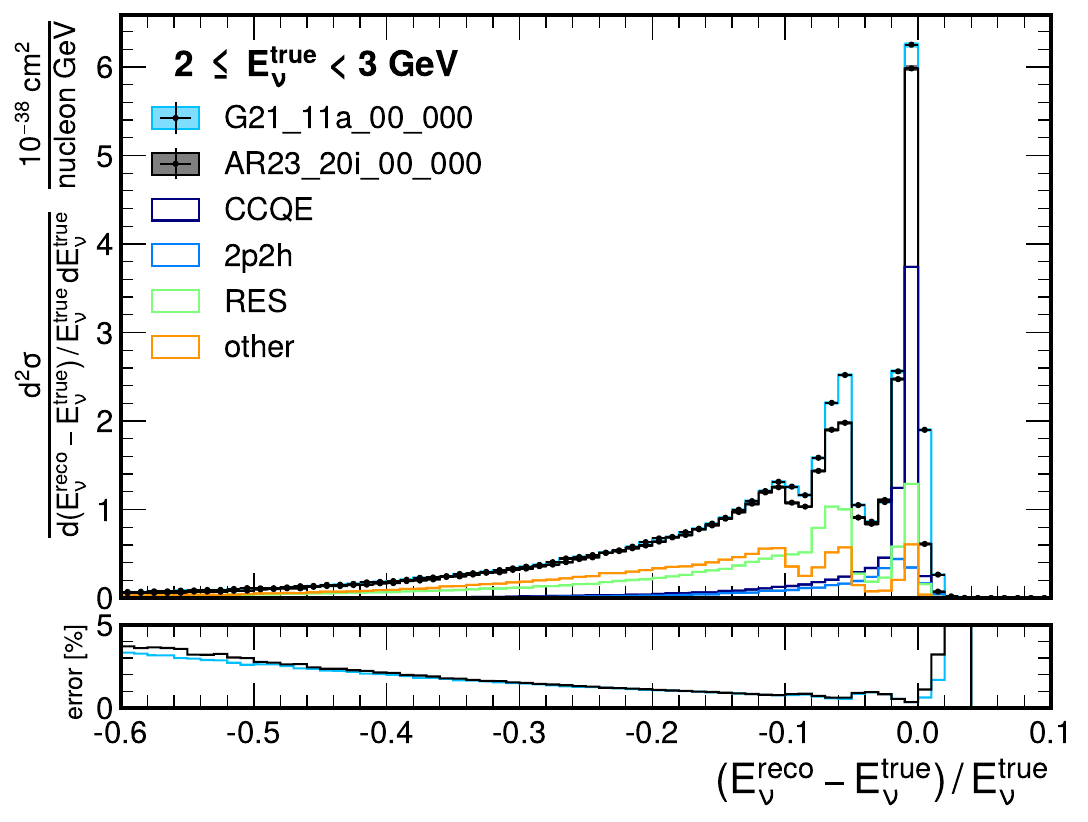}
    \label{fig:nutag_enubias_rel_xsec_mid}
    \end{subfigure}

	\begin{subfigure}{0.5\textwidth}
		\centering
		\includegraphics[width=\textwidth]{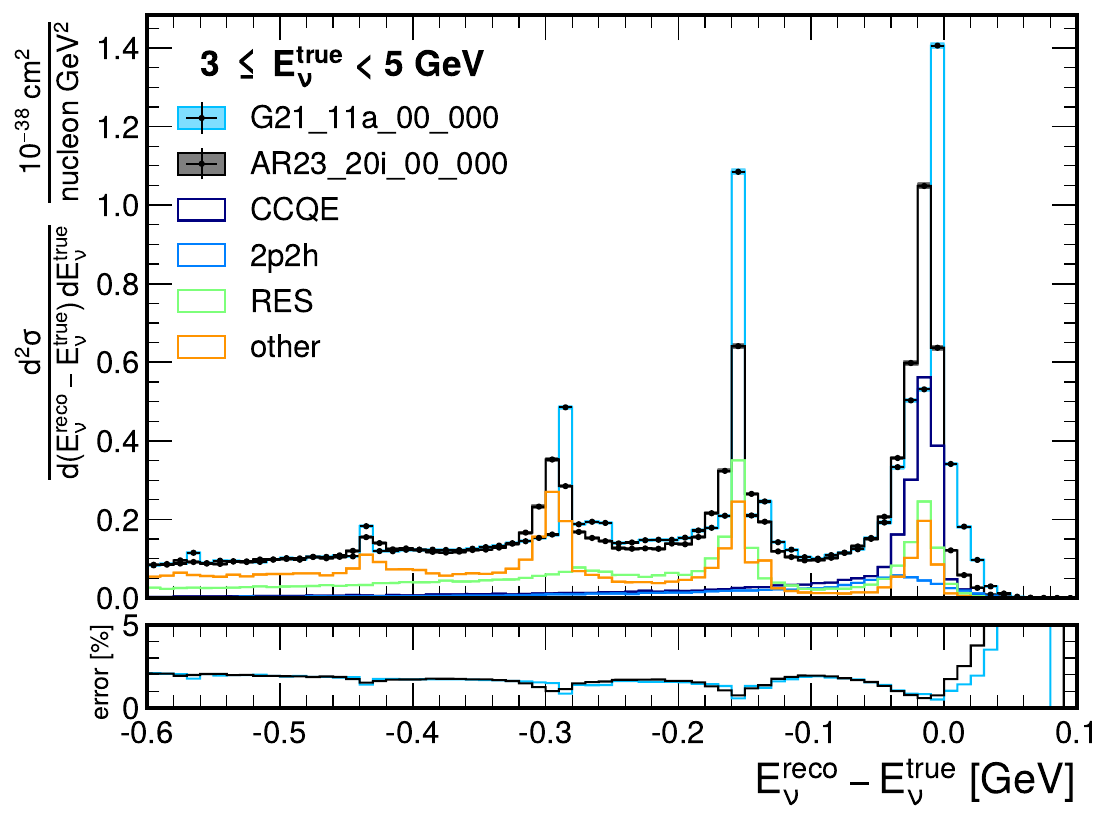}
	\label{fig:nutag_enubias_abs_xsec_high}
	\end{subfigure}
    \begin{subfigure}{0.49\textwidth}
    \includegraphics[width=\textwidth]{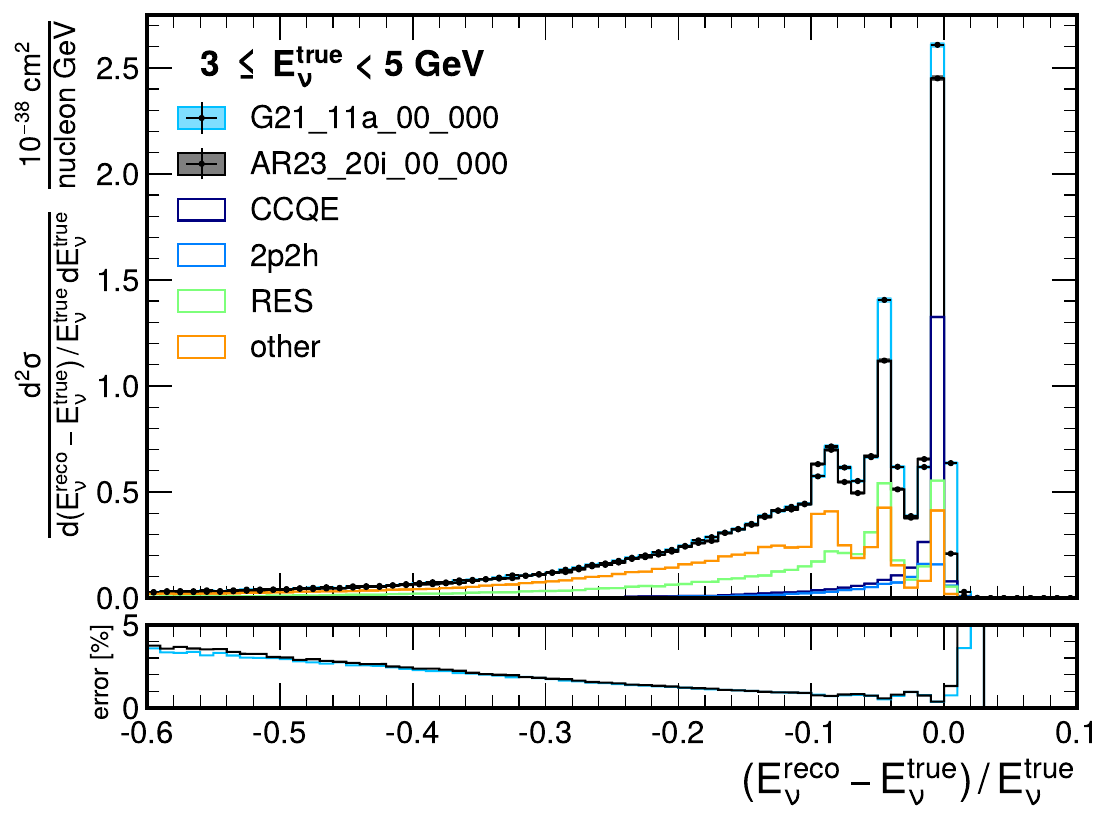}
    \label{fig:nutag_enubias_rel_xsec_high}
    \end{subfigure}
    \caption{Projected measurements of the CC cross-section as a function of the absolute ($\enureco-\enutrue$, left) and relative ($(\enureco-\enutrue)/\enutrue$, right) neutrino energy bias using tagged neutrinos, for different neutrino energy regions. \enureco is defined as in \autoref{eq:enurec}, whilst \enutrue is the true neutrino energy derived from the tagging procedure. Error bars are given for the total predictions from the \argenie (black) and the \susagenie (light blue) models. The colored lines show the breakdown by interaction channel as predicted by the \argenie model. Each figure is accompanied by an underlying plot showing the evolution of the statistical error on the cross-section measurement as a function of each quantity.}
    \label{fig:nutag_enubias_xsec_all}
\end{figure}

Using the a priori knowledge of the energy of tagged neutrinos, we show the projected measurement of the absolute and relative neutrino energy for different regions of neutrino energy in \autoref{fig:nutag_enubias_xsec_all}. The energy ranges considered are between $[0, 2] ~\si{GeV}$ (low), $[2, 3]~\si{GeV}$ (mid) and $[3, 5]~\si{GeV}$ (high). The choice of energy intervals can be optimized, and was chosen to illustrate that the contributions to the neutrino energy bias change as a function of neutrino energy. This is particularly important for oscillation experiments, as the constraints obtained with the near detector must be propagated to the far detector, where oscillations change the shape of the neutrino flux.

\autoref{fig:nutag_enubias_xsec_all} highlights that this measurement can be performed using the setup described in \autoref{sec:detector_parameters} with a statistical precision much below 5\% across the vast majority of the spectrum. A similar measurement has been performed by the e4nu collaboration by using electron scattering measurements~\cite{CLAS:2021neh}, providing crucial benchmarks for neutrino interaction models. However, this measurement is only able to probe the vector part of the interaction. We note that, with the tagged neutrino beam and within the considered energy ranges, it is possible not only to measure the vector and axial contributions from neutrino interaction models, but also to use these measurements to effectively calibrate the neutrino energy bias of the DUNE experiment's far detectors as a function of neutrino energy across the full range of relevant energies.

\subsubsection{Electron scattering-like measurements with tagged neutrinos}
\label{sec:nutag_escat_measurements}

The measurements described in \autoref{sec:nutag_ccinc} and \autoref{sec:nutag_energybias} provide direct insight into the quantities that matter for future generation neutrino oscillation analyses, but do not attempt to place constraints on individual interaction channels or the dynamics of the underlying processes. Electron-scattering experiments have historically sought to characterize the nuclear dynamics in these processes by measuring differential cross-sections as a function of quantities such as the energy and momentum transfer (\etransfer and \qthree), the invariant mass of the hadronic system (\wexp), or the Bjorken $x$ and $y$ variables.

However, electron scattering measurements are only sensitive to the vector part of the interaction process, as electrons interact electromagnetically, and are not well suited to probe the axial part of nuclear dynamics. With a tagged neutrino beam, such measurements become accessible using neutrinos as a probe, as their energy is known on an event-by-event basis, and allows us to directly measure the aforementioned quantities. Such measurements can give targeted insight into the exact processes that govern nuclear effects, which currently dominate neutrino oscillation analyses. In this section, we present a few selected examples of measurements that can be made with the tagging technique to refine nuclear models.

\begin{figure}[!htp]
	\centering
	\begin{subfigure}{0.49\textwidth}
		\centering
        \includegraphics[width=\textwidth]{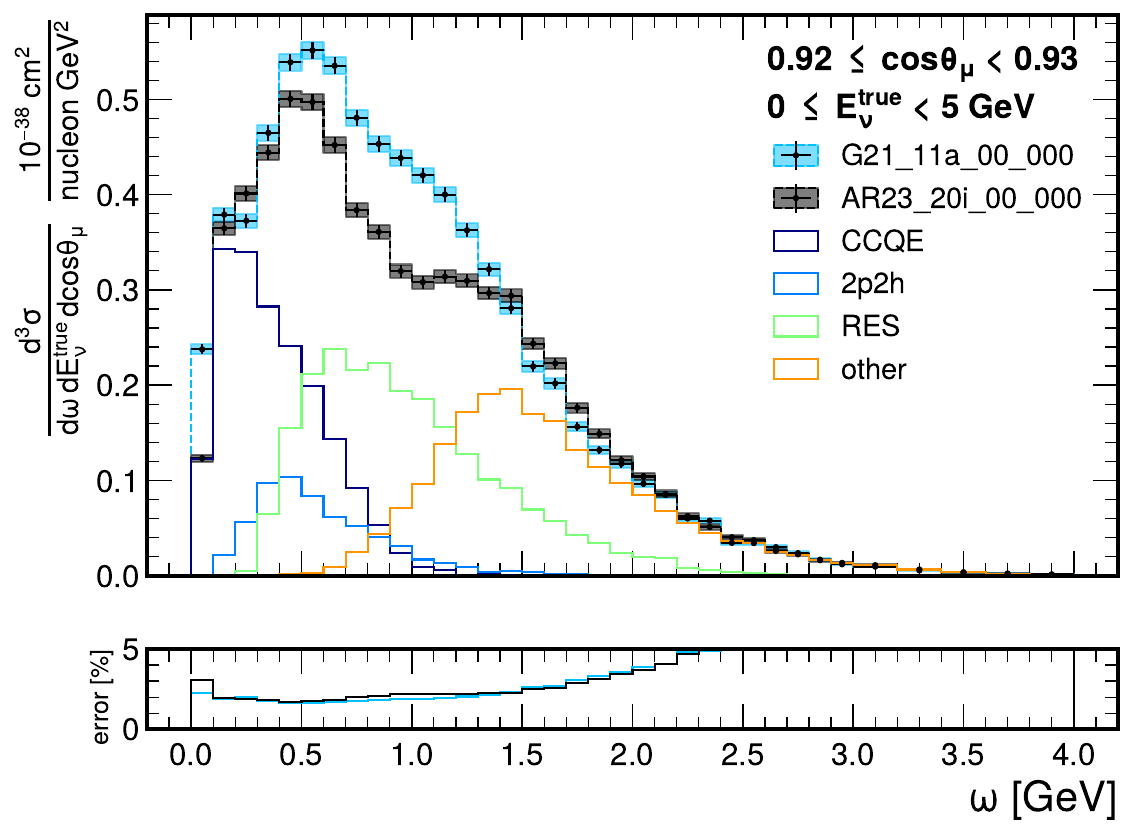}
	\end{subfigure}
    \begin{subfigure}{0.49\textwidth}
    \includegraphics[width=\textwidth]{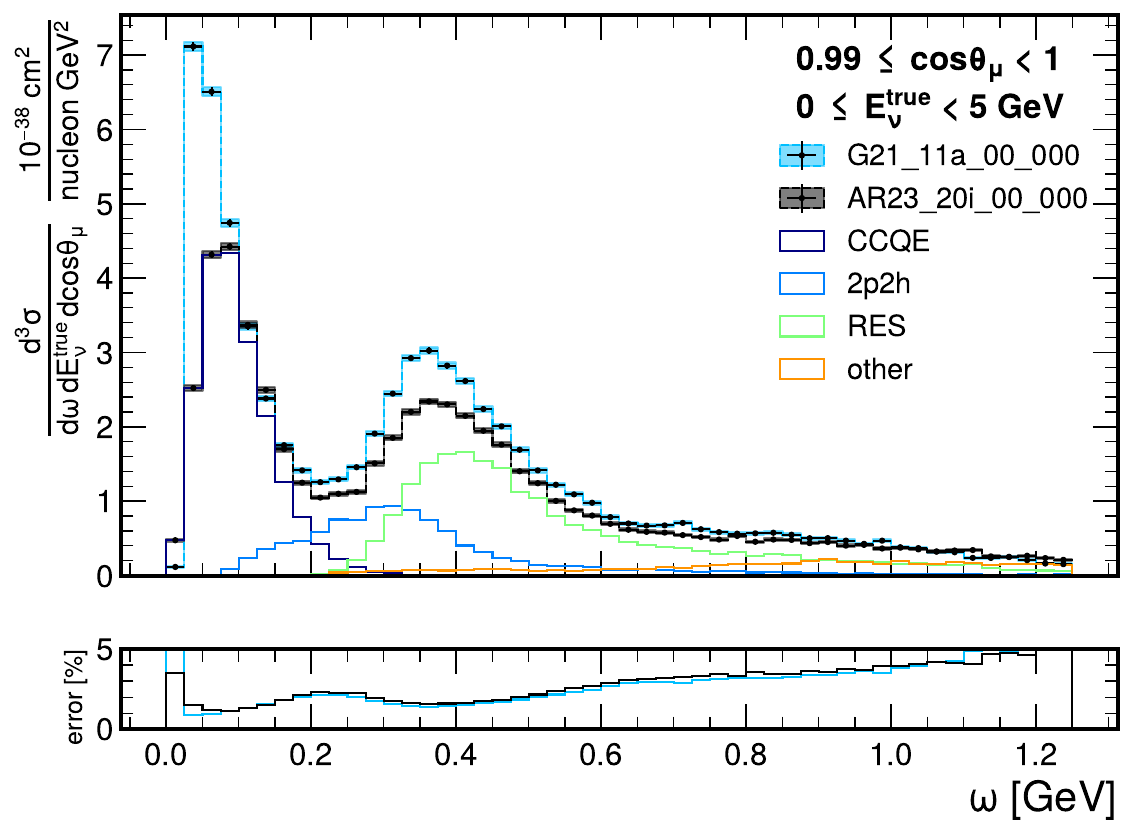}
    \end{subfigure}
    \caption{CC \numu triple differential cross-section as a function of energy transfer \etransfer, for different muon scattering angle \cosmu and neutrino energy \enutrue regions. The filled regions show the associated statistical uncertainty for the \argenie (black) and \susagenie (light blue) models. The breakdown by interaction mode is given for the \argenie model. Each figure is accompanied by the evolution of the associated statistical uncertainty on the measurement, shown underneath.}
    \label{fig:nutag_escat_omega}
\end{figure}

A first example is shown in \autoref{fig:nutag_escat_omega}. The figure compares the expected measurement using muon neutrinos coming primarily from \pidecay decays. The resulting triple differential cross-section is reported as a function of the energy transfer, \etransfer, the muon scattering direction with respect to the neutrino direction, \cosmu and for neutrino energies in the $[0,5]\textrm{ }\si{GeV}$ range. The energy transfer is defined as 
\begin{equation}
    \etransfer=E_\mu - \enu,
\end{equation}
where $E_\mu$ is the muon energy. This observable is only accessible thanks to the a priori knowledge of the neutrino energy with the tagging technique and cannot be accessed using conventional broad-band beams. For brevity, we only show two angular regions that highlight differences between the \argenie and \susagenie models. The first of the two ($\cosmu\in[0.92,0.93]$) highlights differences at relatively high energy transfer values, which stems from the different parameters related to nucleon-level form factors within the GENIE event generator. The \argenie model applies parameters tuned to bubble chamber measurements of neutrino-nucleon scatters~\cite{GENIE:2021zuu}, which notably reduces the cross-section of RES interactions and deep and shallow inelastic scatters. The second region ($\cosmu\in[0.99,1.00]$) is dominated by very forward scatters and showcases primarily the difference in the treatment of collective nuclear effects (such as the nuclear screening) applied by the different models in the QE region. In both cases, the associated statistical uncertainty is below 5\% across the entire spectrum.

\begin{figure}[!htp]
\subfloat{\includegraphics[width=0.34\textwidth]{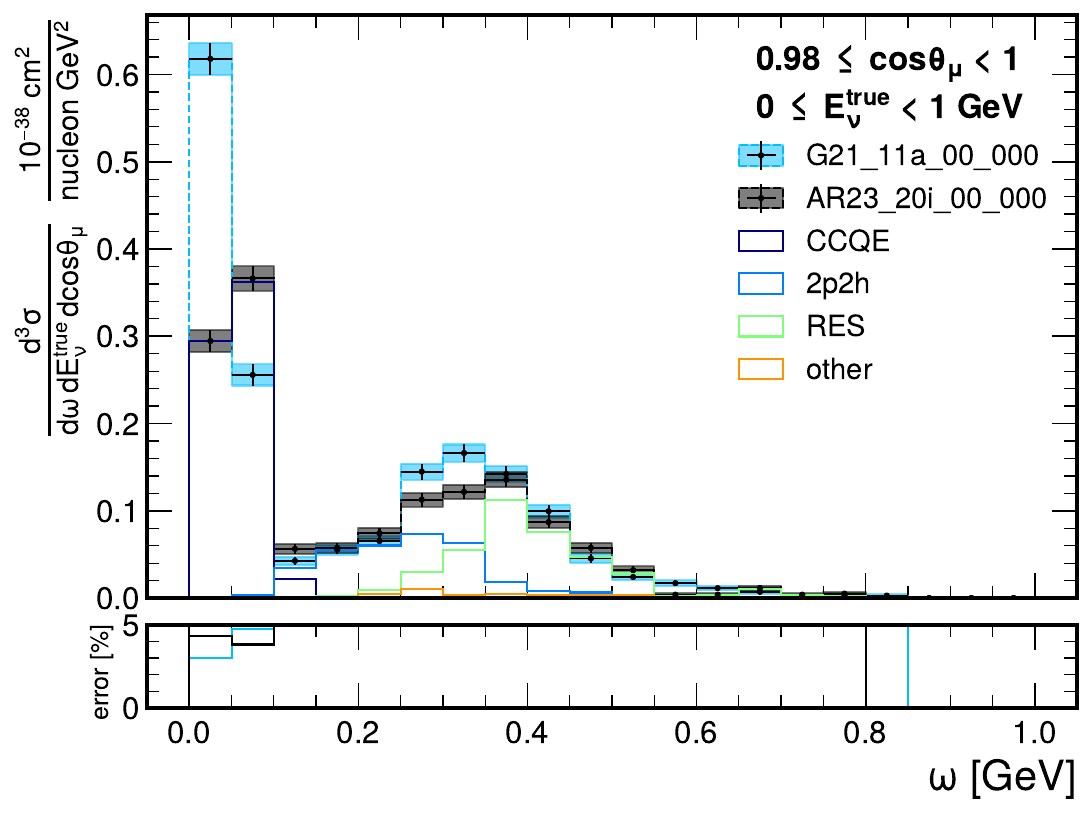}}
\subfloat{\includegraphics[width=0.34\textwidth]{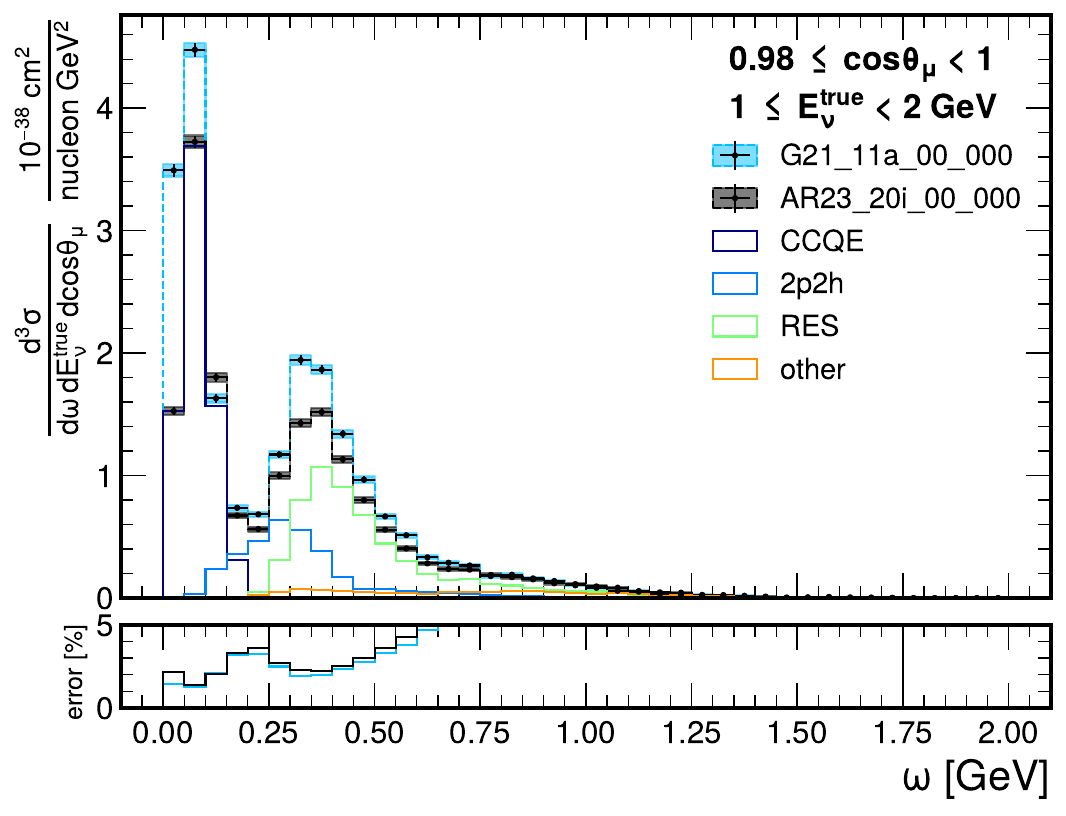}}
\subfloat{\includegraphics[width=0.34\textwidth]{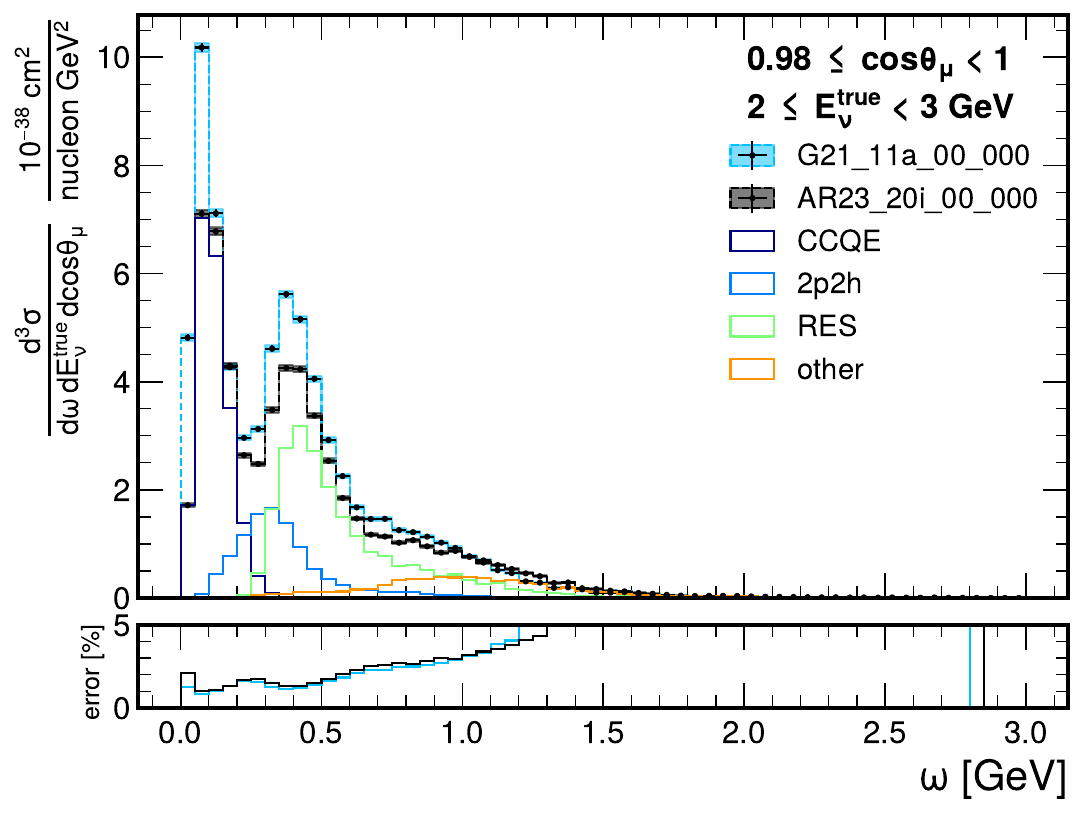}} \\
\subfloat{\includegraphics[width=0.34\textwidth]{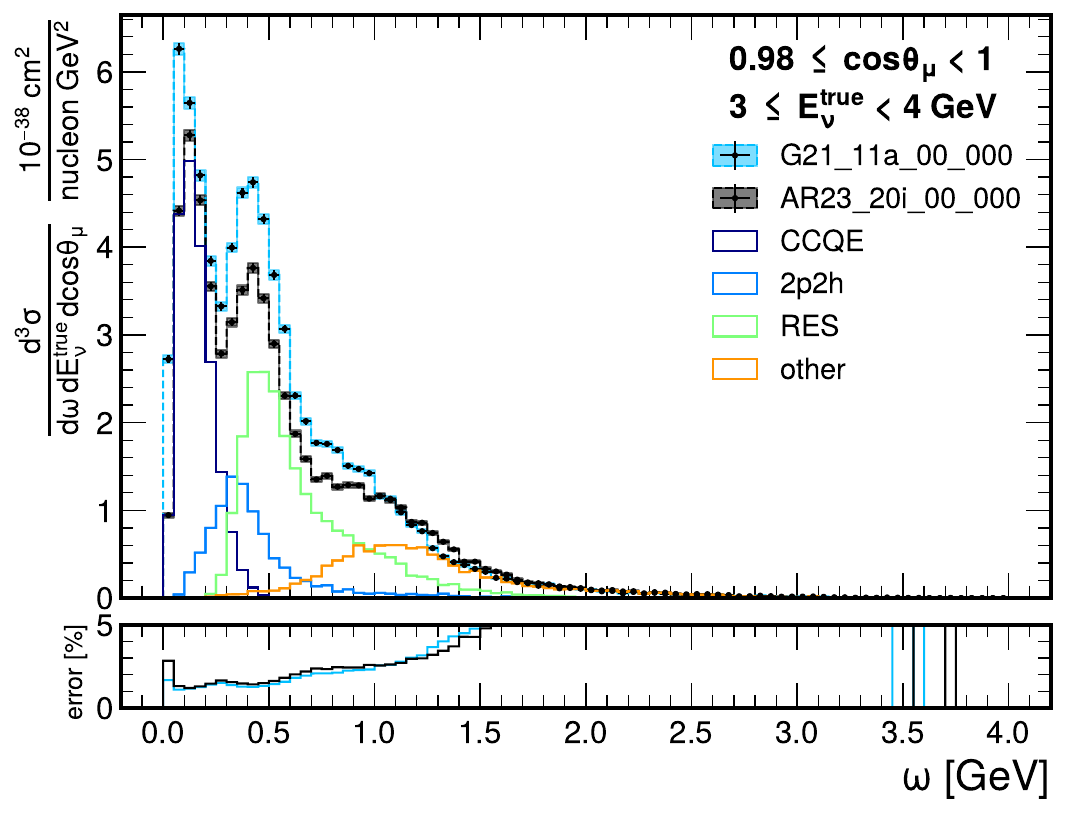}}
\subfloat{\includegraphics[width=0.34\textwidth]{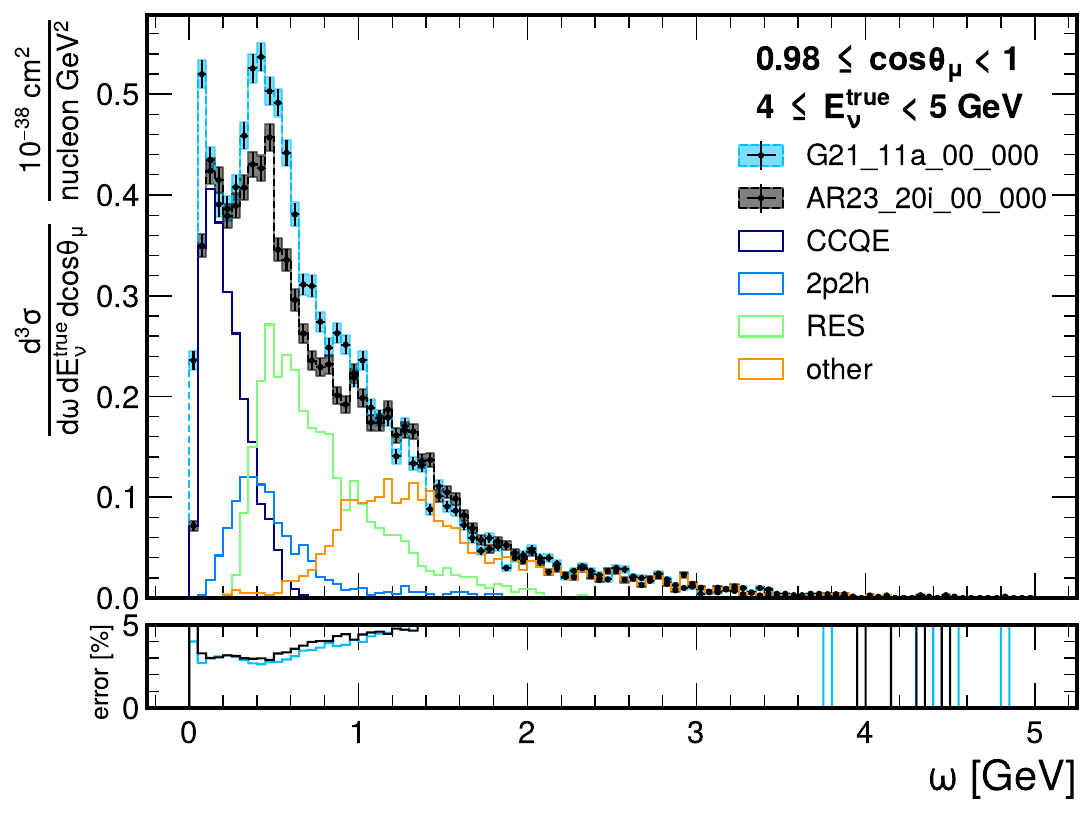}}

\caption{CC \numu triple differential cross-section as a function of energy transfer \etransfer, for very forward muon scattering directions, $\cosmu\in[0.98,1]$ and different neutrino energy \enutrue regions. The filled regions show the associated statistical uncertainty for the \argenie (black) and \susagenie (light blue) models. The breakdown by interaction mode is given for the \argenie model. Each figure is accompanied by the evolution of the associated statistical uncertainty on the measurement, shown underneath.}
\label{fig:nutag_escat_omega_fwd}
\end{figure}

The forward-angle, low energy transfer region allows us to probe the physics responsible for the main systematic uncertainties for beam and atmospheric oscillation analyses~\cite{T2K:2024wfn}. As previously discussed, it is essential to have an accurate description of such processes as a function of neutrino energy, as neutrino oscillations modify the fluxes probed with the near and the far detectors. The energy dependence of nuclear effects is also particularly important for oscillation analyses involving atmospheric neutrinos, where there is no prior constraint like that obtained with a near detector. This type of measurement can also be achieved with the tagging technique, as illustrated in \autoref{fig:nutag_escat_omega_fwd}. The size and shape of the differences predicted by the two models change as a function of the neutrino energy range. Across the majority of the spectra, the statistical errors are below 5\%, except for low energies (i.e. $\enu < \SI{1}{GeV}$) where the measurement in the current binning scheme becomes statistically limited. We note that there is potential for further optimization of the binning for such a measurement. 

Finally -- the second example of an electron scattering-like measurement -- we investigate is the double-differential cross-section as a function of neutrino energy and the invariant mass of the hadronic system, \wexp. In this example, \wexp is defined as: 
\begin{equation}
\label{eq_Wreco}
    \wexp= \sqrt{M_N^2 + 2M_N\etransfer - Q^2},
\end{equation}
where $M_N$ is a nucleon mass (i.e. a neutron or a proton, depending on the interaction channel), and $Q^2$ is the square of the four-momentum transfer ($Q^2= -q^2$, where $ q = (\etransfer,\vec{\qthree})$).

\begin{figure}[!htp]
\subfloat{\includegraphics[width=0.34\textwidth]{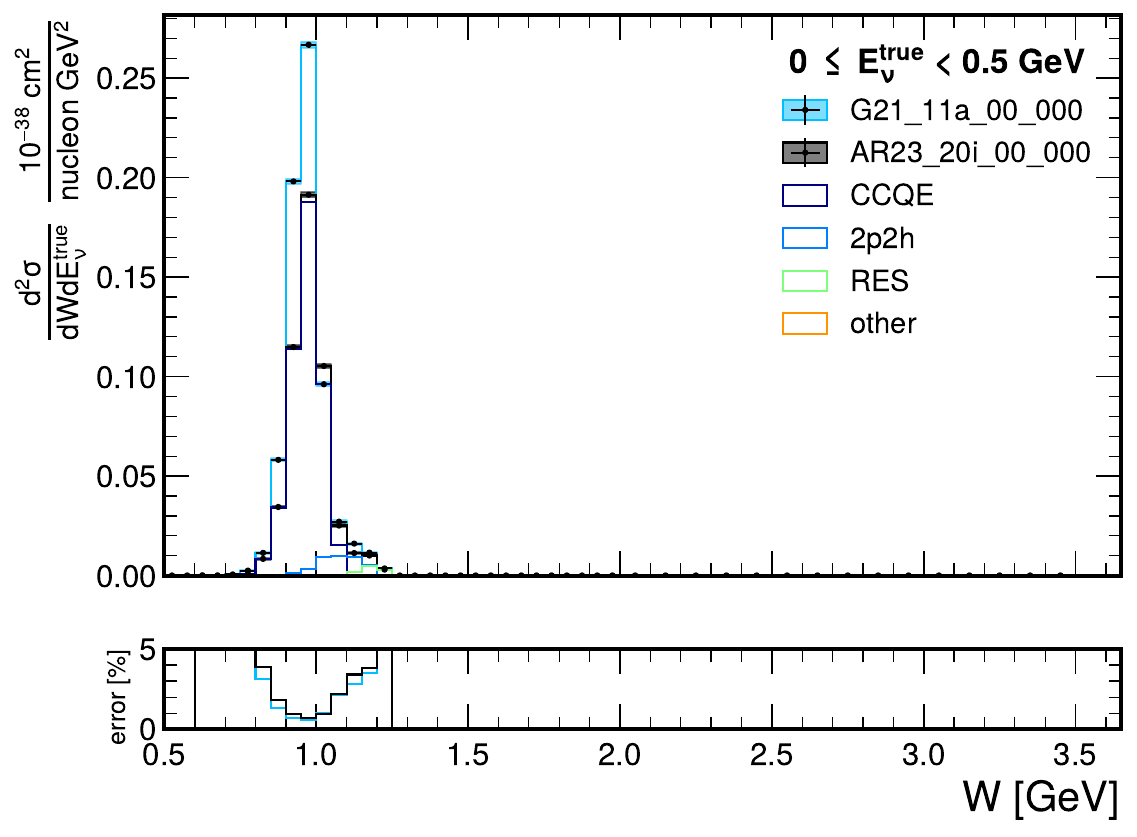}}
\subfloat{\includegraphics[width=0.34\textwidth]{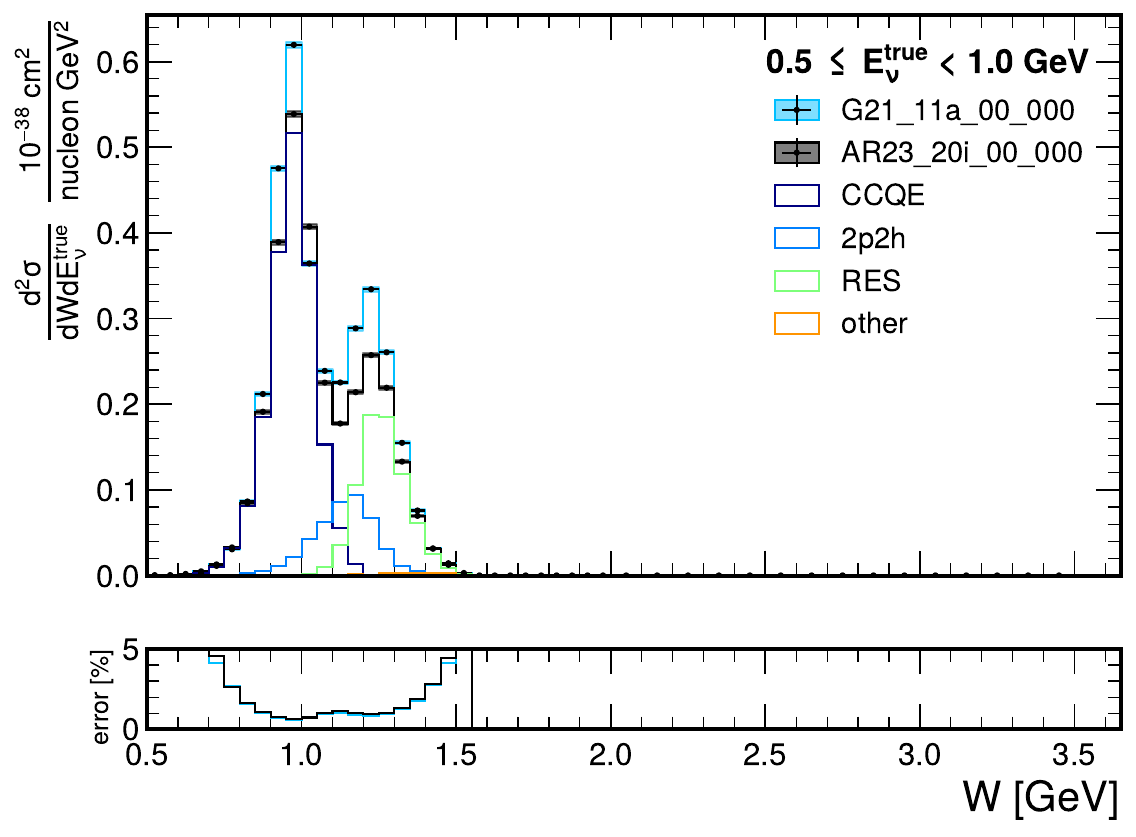}}
\subfloat{\includegraphics[width=0.34\textwidth]{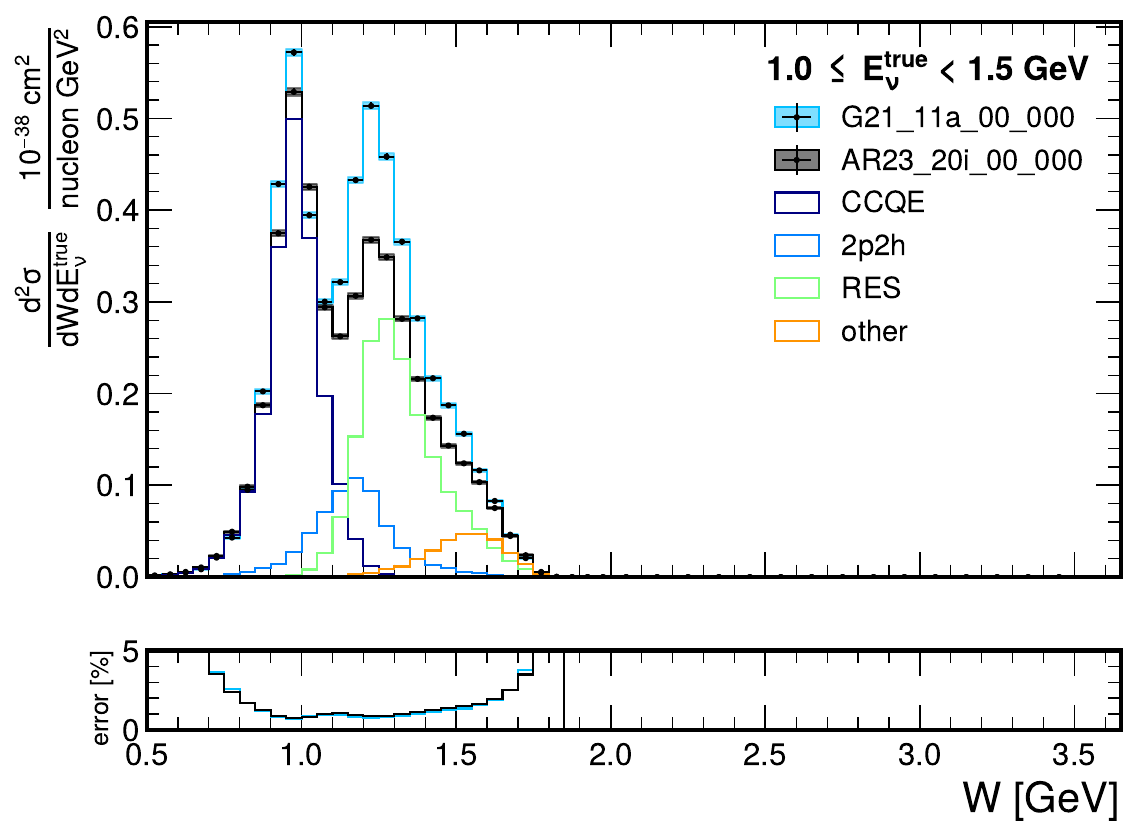}} \\
\subfloat{\includegraphics[width=0.34\textwidth]{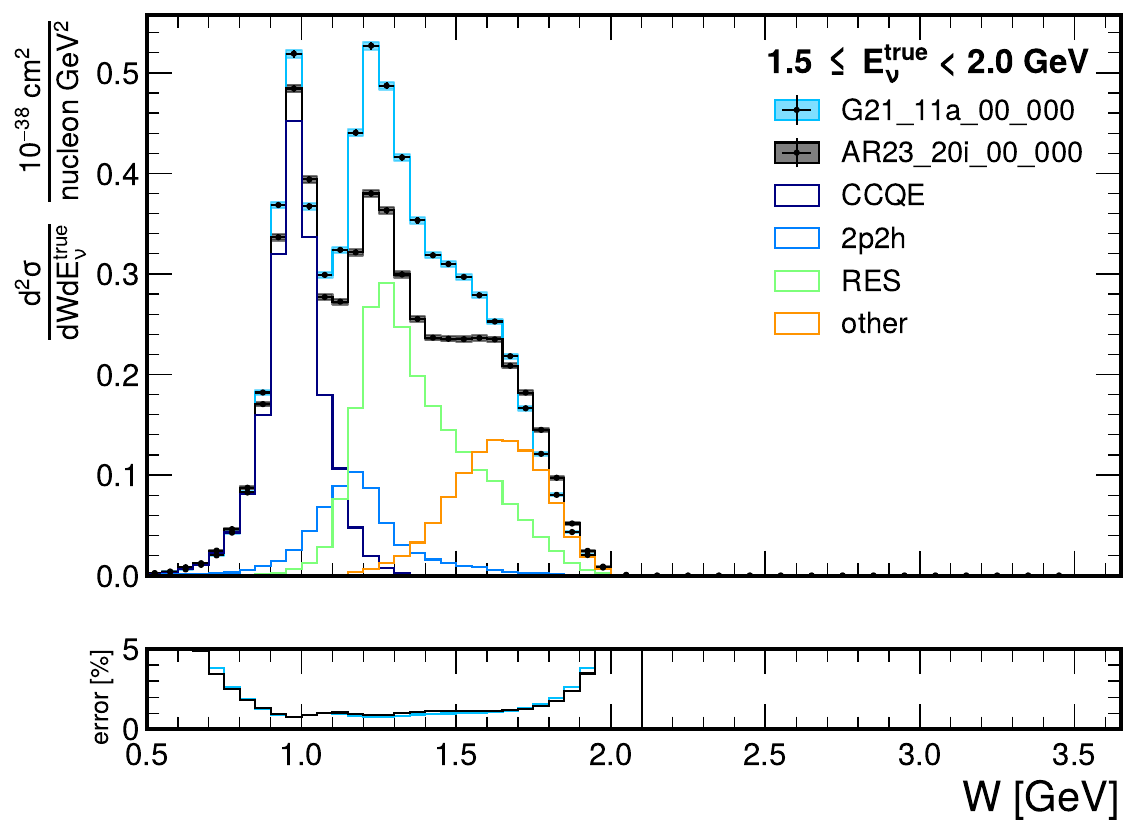}}
\subfloat{\includegraphics[width=0.34\textwidth]{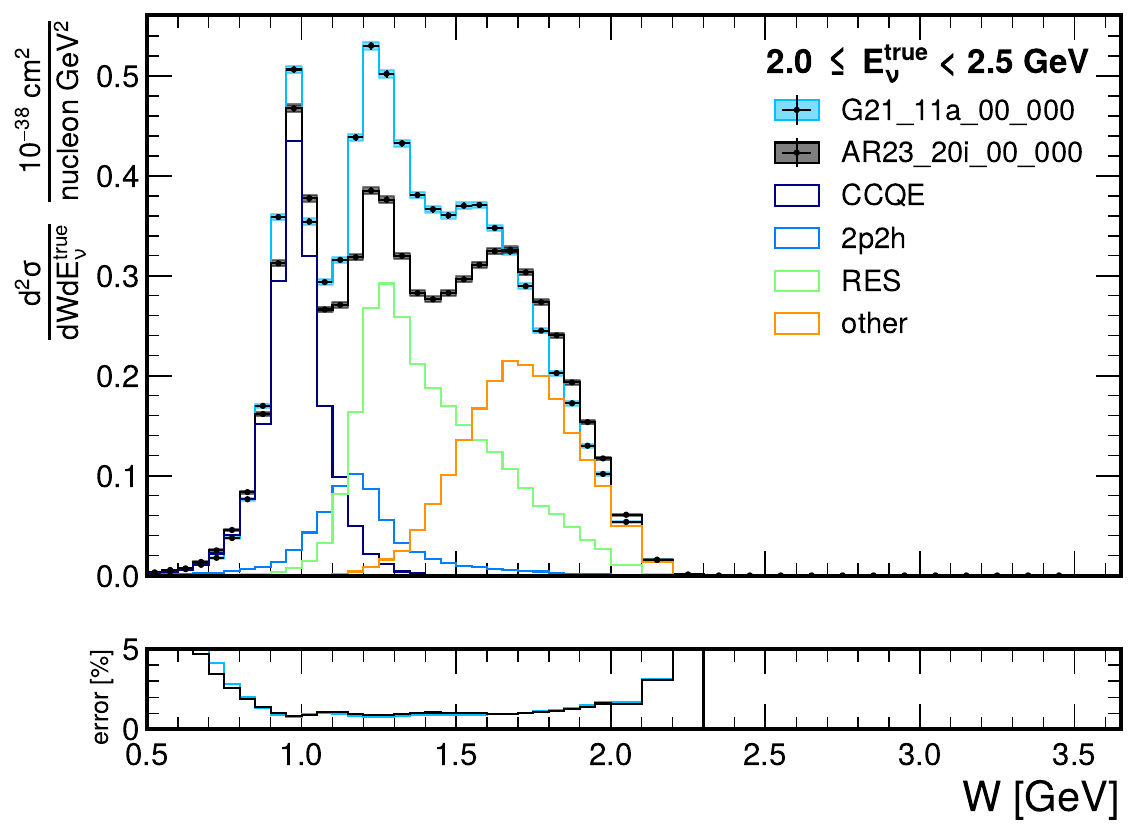}}
\subfloat{\includegraphics[width=0.34\textwidth]{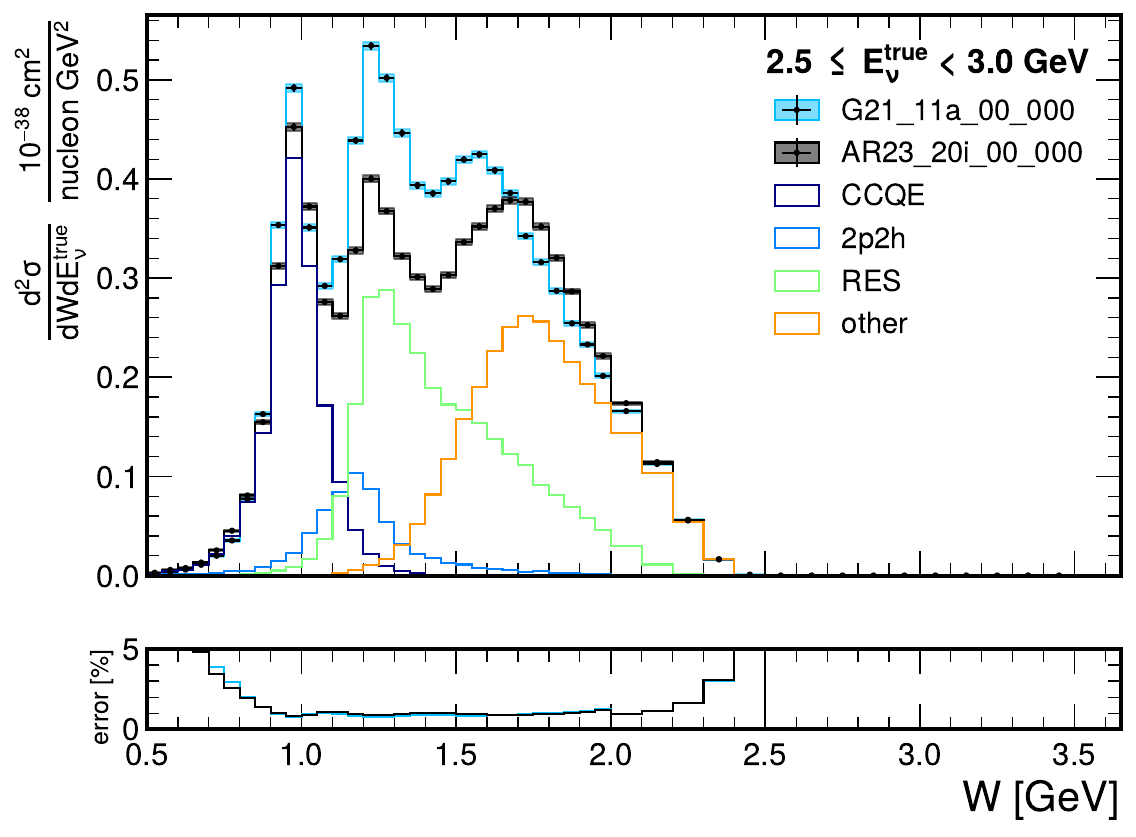}} \\
\subfloat{\includegraphics[width=0.34\textwidth]{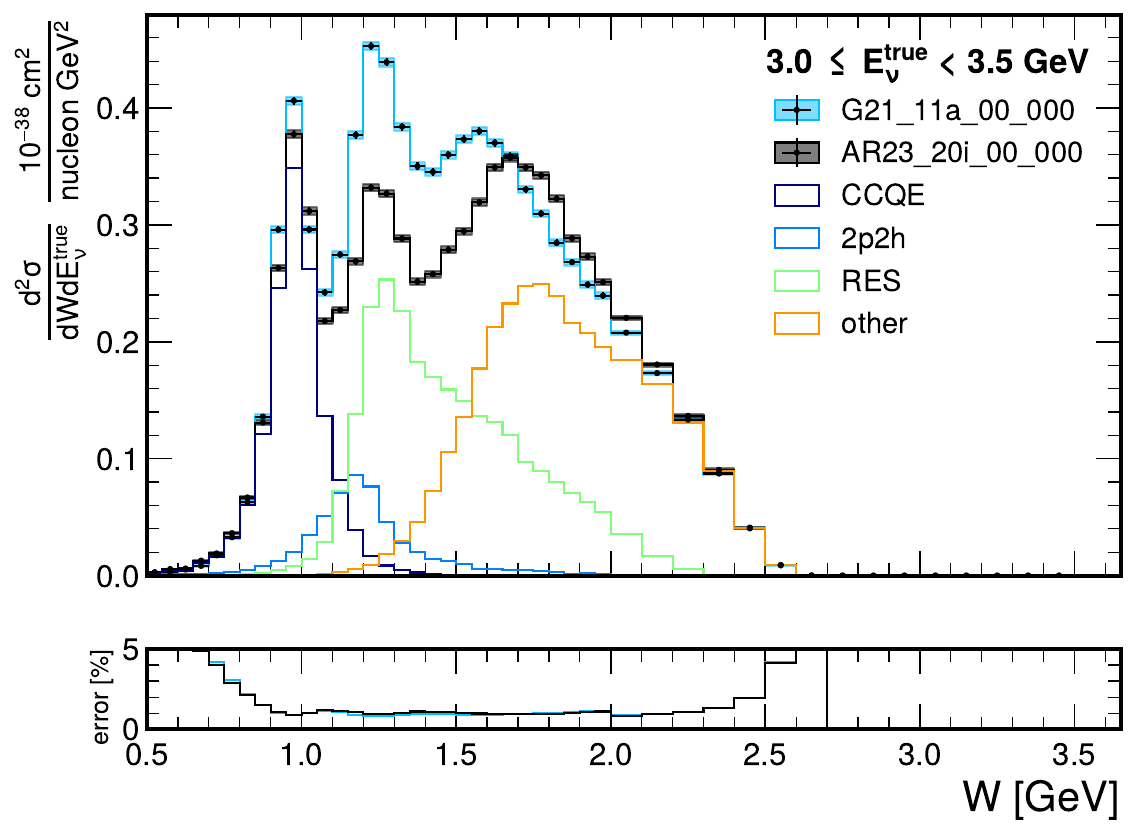}}
\subfloat{\includegraphics[width=0.34\textwidth]{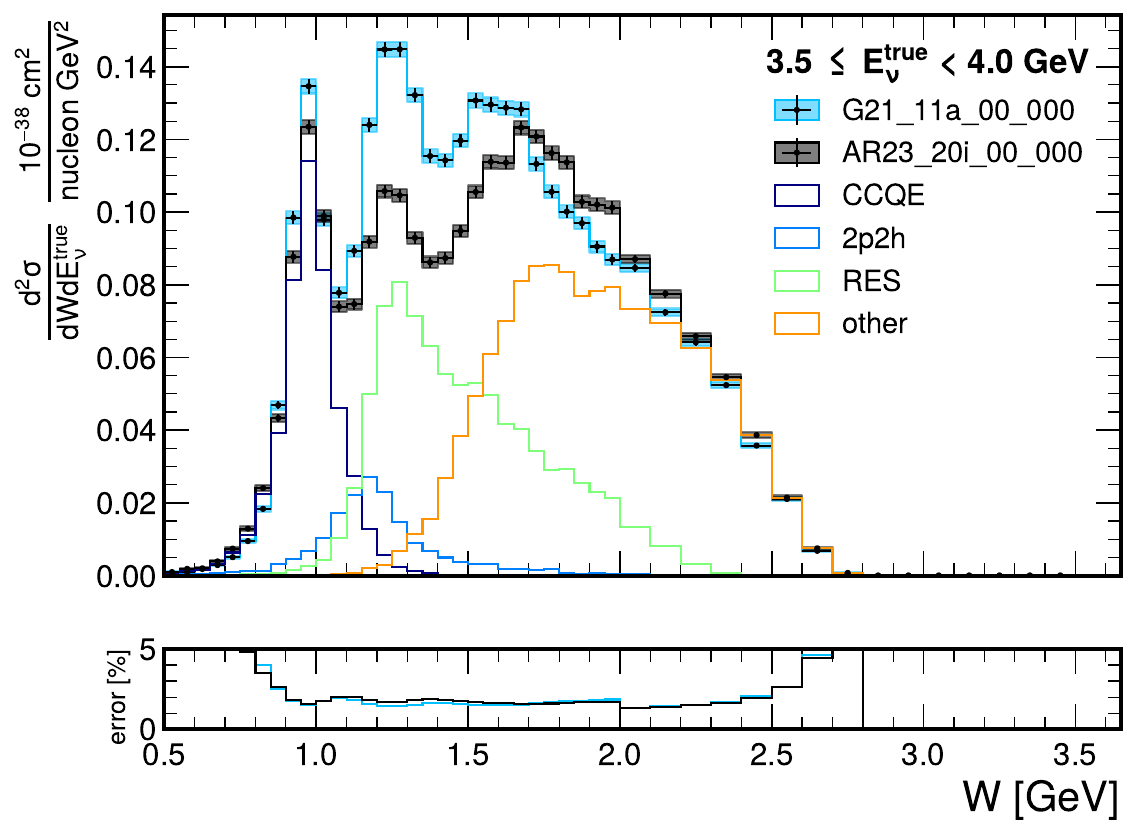}}
\subfloat{\includegraphics[width=0.34\textwidth]{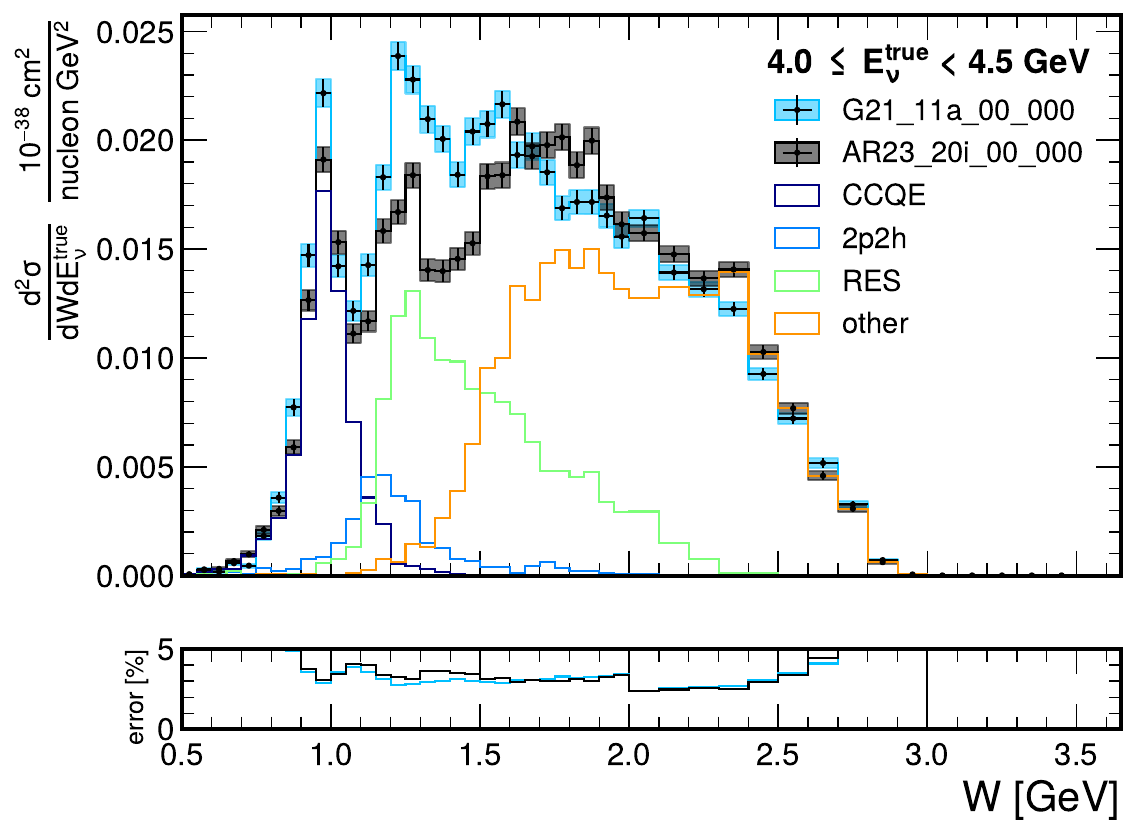}} \\
\subfloat{\includegraphics[width=0.34\textwidth]{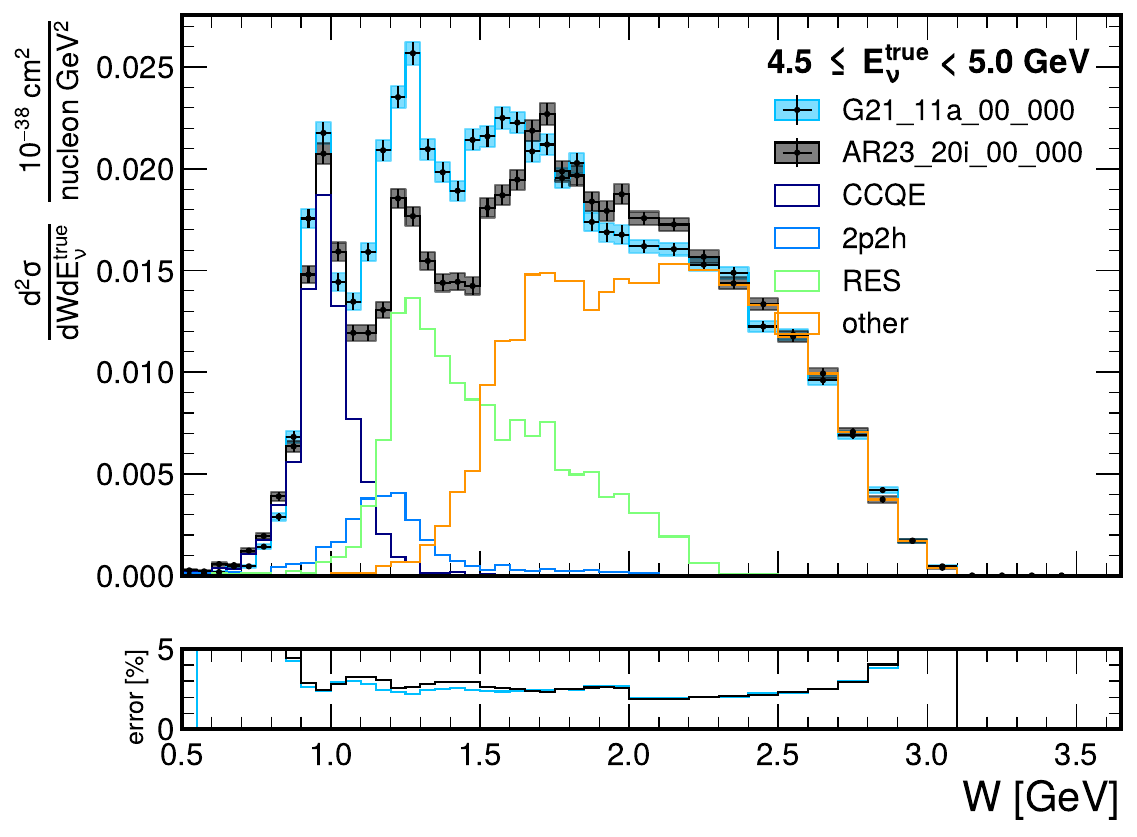}}

\caption{CC \numu double differential cross-section as a function of the invariant mass of the hadronic system, \wexp, and different neutrino energy, \enutrue, regions. The filled regions show the associated statistical uncertainty for the \argenie (black) and \susagenie (light blue) models. The breakdown by interaction mode is given for the \argenie model. Each figure is accompanied by the evolution of the associated statistical uncertainty on the measurement, shown underneath.}
\label{fig:nutag_escat_wexp}
\end{figure}

Note that this is an inclusive definition of the invariant hadronic mass, using only the neutrino energy and the four-momentum transfer in the calculation (i.e. it does not add all of the hadronic components, as they are not visible inside the detector), and assumes that the nucleon is at rest. With the tagging technique, this variable becomes an experimental observable, as \enu, \etransfer, and $Q^2$ are known on an event-by-event basis, and it provides insight notably on the mechanisms for hadron generation in neutrino interactions. The distribution of \wexp will exhibit peaks corresponding to the different baryons that were probed during the interaction -- first, a nucleon for QE and 2p2h interaction, then a $\Delta(1232)$ resonance for the lightest resonance in pion production processes, and then increasingly higher-order resonances. At DUNE energies, resonant pion production processes account for roughly 1/3 of the available CC interactions. The projected measurement is shown in \autoref{fig:nutag_escat_wexp}. The figure illustrates that as the mean value of neutrino energies increases, the phase space for interaction as a function of \wexp probes increasingly higher mass baryons, and the interactions become more inelastic. The associated statistical error on the cross-section is far below 5\% (often below 2\%) across all regions of interest.

\subsubsection{Other measurements with a tagged neutrino beam}

Whilst the projected measurements in this section give a sense of some of the measurements that could be made with a tagged neutrino beam, it is by no means exhaustive. An experiment with event-by-event knowledge of the neutrino energy would represent a paradigm shift for neutrino cross-section measurements, both from the perspective of mitigating systematic uncertainties in neutrino oscillation analyses and of using neutrinos as a weakly-interacting probe of the nucleus. Below, we highlight just a few of the additional possibilities for measurements with a tagged neutrino beam. 

The measurement in \autoref{sec:prism_studies} could be repeated without the statistically challenging linear combinations of NBOA fluxes. Instead, an inclusive \numu flux integrated tagging analysis could simply be reweighted on an event-by-event basis to model-independently correct the measurement to the shape of the \nue flux. The same concept could be applied to directly measure the \numu cross-section in the oscillated or unoscillated flux of neutrino oscillation experiments, allowing a precision check of the cross-sections relevant to those experiments. 

Reconstructing specifically the hadronic energy and momentum imbalance could be used for precision studies of final state interactions (effectively fixing $\omega$ and comparing it to the observed hadronic energy) or to construct spectral functions from neutrino scattering measurements in the same way it is done for electron scattering~\cite{JeffersonLabHallA:2022cit}, allowing a detailed evaluation of the nuclear initial state. 

Knowledge of the neutrino energy would also allow a detailed exploration of the very poorly understood resonant to deep inelastic transition region, which is of particular importance to experiments operating at the few $\si{GeV}$ energy scale such as DUNE. The Bjorken and Nachtmann variables could be precisely reconstructed, and the suppression of the cross-section with respect to calculations naively extrapolating standard parton distribution functions into the non-perturbative region could be precisely mapped. Hadronization could also be explored in much more detail, using the precise knowledge of four-momentum transfer allowed using a tagged beam.

\subsection{Summary of the nuSCOPE physics potential}
\label{sec:xsec_summary}

nuSCOPE offers the possibility to make a wide range of previously unattainable measurements that may be the key to allowing the next generation of long-baseline neutrino oscillation experiments to reach their ultimate precision. It directly confronts aspects of neutrino interaction physics expected to drive the leading sources of systematic uncertainty for neutrino oscillation experiments:
\begin{itemize}
    \item The energy dependence of the cross-section is essential for extrapolating between near and far detectors; we demonstrate how this can be directly measured in \autoref{sec:numu_ccinc} and \autoref{sec:nutag_ccinc}.
    \item The smearing of neutrino energy must be understood to interpret far detector event rates as an oscillation probability. Measurements in a tagged neutrino beam offer a unique opportunity to directly measure this, as demonstrated in \autoref{sec:nutag_energybias}.
    \item The muon to electron neutrino cross-section ratio is projected to be a dominant systematic uncertainty for measurements of CP-violation, we show that this can be directly constrained using a monitored beam in \autoref{sec:prism_studies}, with potential improvements possible using a tagged beam.
    \item NC processes are a key background at the experiment's far detectors; we show how these can be precisely constrained in \autoref{sec:ncpizero}. 
\end{itemize}

\noindent
More generally, collecting high-statistics data with a tagged neutrino beam across the energies relevant to next-generation neutrino oscillation experiments offers a unique way to directly calibrate their cross-section and the energy smearing. A library of measurements with a tagged beam could be leveraged to determine how simulated events for DUNE or Hyper-Kamiokande are expected to look in their near or far detectors, thanks to the event-by-event characterization of neutrino energy.
Measurements with a tagged neutrino could also be leveraged to explore nuclear physics with a neutrino probe at the GeV-energy scale in a previously impossible way, as explored in \autoref{sec:nutag_escat_measurements}.

Future work will focus on expanding these studies to include simulations using a water-based target tailored to supporting Hyper-Kamiokande's physics program, anti-neutrino beam running, lower energy beam configurations, sensitivity to BSM physics, and a detailed assessment of the detector performance required to realize the full nuSCOPE physics program.

\section{Conclusions}

nuSCOPE addresses a fundamental limitation in neutrino physics: the lack of precise knowledge of neutrino cross-sections at the GeV scale.
This uncertainty currently hinders the physics potential of long-baseline experiments like DUNE and Hyper-Kamiokande. It is based on two innovative techniques, whose proof-of-principle has been demonstrated in recent years by the NP06/ENUBET and NuTag Collaborations. nuSCOPE introduces a novel beam that monitors leptons at the single-particle level and can tag both the parent meson and the accompanying muon of the neutrinos interacting in the detector.

The proposed beamline leverages the CERN-SPS to produce a highly controlled neutrino beam.  
Various siting options within CERN’s infrastructure were analyzed, with TT61/TNC tunnels and ECN4 emerging as potential locations.
The proposal integrates liquid argon and water Cherenkov detectors, ensuring high-resolution event reconstruction. The instrumented decay tunnel provides real-time flux monitoring using tracking detectors and calorimeters to measure lepton production. The ability to measure neutrino energy on an event-by-event basis represents a groundbreaking improvement over conventional techniques.
With a percent level of accuracy in energy and flux normalization, neutrino-nucleus cross-section measurements could reach precision levels comparable to electron-nucleus cross-sections, reducing systematic errors in oscillation experiments.
The slow-extracted proton beam minimizes background noise, enabling more accurate event reconstruction.

While the current focus has been put on neutrino cross-sections, additional studies are being performed to address the potential of nuSCOPE in terms of BSM physics topics that could benefit from the presence of a high-granularity massive detector close to a proton and hadron beam dump associated with an unprecedented control over neutrino-originating decays.
After the successful R\&D phase and beam optimization, a new Collaboration is being formed to present the experiment proposal within a two-year timescale. The nuSCOPE project has the potential to revolutionize neutrino cross-section measurements, supporting the next generation of neutrino physics experiments. Leveraging CERN’s infrastructure, advanced detection techniques, and precision beam monitoring, this proposal marks a crucial step toward reducing systematic uncertainties in neutrino oscillation physics and advancing our understanding of neutrino interactions in nuclear media.

\bibliographystyle{spphys}
\bibliography{bibliography}   

\end{document}